\documentclass[12pt]{article}
\usepackage{amsmath,amssymb}
\usepackage{graphicx}
\usepackage[pdfencoding=auto]{hyperref}
\usepackage{float}
\usepackage{caption,subcaption}
\begin{document}

\title{Production of massive $W^{\pm}$ bosons and
fermion-antifermion pairs from vacuum in the de Sitter Universe}
\author{Cosmin Crucean and Amalia Dariana Fodor\\ \thanks{cosmin.crucean@e-uvt.ro, amalia.fodor97@e-uvt.ro}\\
{\small \it Faculty of Physics, West University of Timi\c soara,}\\
{\small \it V. Parvan Avenue 4 RO-300223 Timi\c soara,  Romania}}
\maketitle

\begin{abstract}
In this paper we study the charged electro-weak interactions in the de Sitter geometry. We develop the reduction formalism for the Proca field with the help of the solutions for the interacting fields. Perturbation theory is used for obtaining the definition of the transition amplitudes in the first order. We apply our formalism to the study of spontaneous vacuum emission of fermions, anti-fermions and $W^{\pm}$ bosons in the expanding de Sitter universe. Our results are generalizations of the Weinberg-Salam electro-weak theory to curved space-time, in the case of $W^{\pm}$ boson interaction with leptons. The probability and transition amplitude are found to be a quantities which depend on the Hubble parameter. Our analytical and graphical results prove that such perturbative processes are possible only for the large expansion conditions of the early Universe. The total probability and rate of transition are obtained for the case of large expansion and we use dimensional regularization and Pauli-Villars regularization for the momenta integrals. In the end we recover the Minkowski limit, where our probabilities vanish, thus confirming the well known fact that spontaneous vacuum emission of $W^{\pm}$ bosons and fermions is forbidden in flat space-time due to energy-momentum conservation.
\end{abstract}

\section{Introduction}

The theory of weak interactions allows us to understand fundamental quantum processes between quarks, leptons and massive bosons. One example of this is the process where a "down" quark transforms in an "up" quark by emitting a $W^-$ boson (or absorbing a $W^+$ boson). The massive boson is unstable and decays into an electron and an antineutrino, which enables neutrons to turn into protons \cite{12}. This process is known as $\beta^-$ decay and increases the atomic number of an atomic nucleus by one, while emitting an electron and an antineutrino. The inverse transformation turns an "up" quark into a "down" quark. There is also large number of observed $\beta$-decays in nuclei and all these imply the transformations of neutrons into protons and vice versa \cite{12}. All these processes are possible due to the existence and properties of the charged $W^{\pm}$ bosons. For these reasons it is a fundamental problem to understand the mechanisms which generate these bosons in our universe.

The charged massive bosons $W^{\pm}$ could exist as stable particles only in the early universe, as long as the background energy was greater than or equal to their rest energy. Subsequently these bosons decay into more light and stable particles, such as electrons and neutrinos. It is also well established that, in the very early stages of universe, the temperatures and densities were high enough to allow these bosons to exist in large numbers at thermal equilibrium \cite{w1,w2}. Space-time expansion could also be a mechanism of massive boson production. This effect can be revealed by using the perturbative methods of quantum field theory from flat space-time. The perturbative approach allows one to obtain the dependence of the transition rates on the Hubble parameter and explore the Minkowski limit, where processes which involve the spontaneous emission of particles from vacuum are forbidden. In this paper we explore the vacuum generation of massive $W^{\pm}$ bosons due to space-time expansion in the de Sitter geometry.

The electro-weak theory in Minkowski space-time was developed in order to understand the interactions between massive bosons and fermions. As far as free fields are concerned, the massive vector fields are described by the Proca equation, which was first proposed in \cite{PR1}. There are a few results that consider the study of the Proca field in curved space-times \cite{2,WT}, where the problem of fundamental solutions and propagators was approached. It is a well established fact that the massive charged $W^{\pm}$ bosons were generated in the early universe \cite{w1}, but the mechanisms that were involved deserve further studies. This is because the conditions of the early universe, where a large space-time expansion is present, are suitable for a perturbative treatment of the problems which involve particle production. The perturbative approach could help one study particle generation from field interactions \cite{24,25,26,27,28,29,30,31,32,b1,b2,cc,33,34,35}. The problem related to quantum fields in the presence of the large expansion conditions of the early universe was also studied in \cite{15,LL,LL1,23,24,25,26,27,28,29,30,31,32,b1,b2,33,34,35}, where perturbative methods were used. The theory of the free Dirac and Proca fields in the de Sitter geometry was studied in \cite{2,24}, and a model which investigates neutral current interactions in the de Sitter geometry was proposed in \cite{24,cc}. The Weinberg-Salam theory \cite{PR1,w1,12,3,4,5,6,7,8,9,10,11,cr} of electro-weak interactions in Minkowski space-time can be extended to the de Sitter geometry by using perturbations for defining the transition amplitudes, with the specification that the Feynman rules will be written down in the coordinate representation. This allows us to extend the definitions of the transition amplitudes from the Minkowski theory to curved space-time. In our study we will also consider, for the first time, the problem of spontaneous vacuum emission of the triplet $W^-$ boson, positron and neutrino, $vac \rightarrow W^-+e^++\nu$, as well as the process involving the positively charged boson $vac \rightarrow W^++e^-+\widetilde{\nu}$. The theory of electro-weak interactions between the charged $W^{\pm}$ bosons and fermions can be constructed by considering the lagrangian densities for the free Dirac and Proca fields, as well as the interaction lagrangian density, constructed as the electrodynamic coupling $-J^\mu A_\mu$, where the current represents the charged currents. The non-perturbative approach to the problem of massive boson generation is less studied in literature, and we mention here the results obtained in \cite{37}. Our approach is based on the fundamental results obtained in \cite{32,33,34,35}, where the ideas that space-time expansion could generate particles were first proposed. In our study we use a perturbative approach for defining the first order transition amplitudes corresponding to the de Sitter electro-weak theory with charged currents. This perturbative approach could be one of the mechanisms that explain the generation of massive $W^{\pm}$ bosons in the early universe. The perturbative approach to the problem of particle production in expanding geometries was also examined in \cite{17,18}, without taking into account the problem of massive boson generation.

The paper is organized as follows: in the second section we present the equations of interaction intermediated by charged currents in de Sitter space-time. In the third section we present the reduction formalism for the Proca field. The amplitudes of electro-weak theory with charged $W^{\pm}$ bosons in de Sitter geometry are defined in the fourth section. The fifth and sixth sections are dedicated to the amplitudes and probabilities corresponding to the process of spontaneous generation from vacuum of $W^{\pm}$ bosons, fermions and anti-fermions. In the seventh section we present the amplitude and probability for the case of longitudinal polarization. In the eighth section we compute the total probability by using dimensional regularization and Pauli-Villars regularization for the integrals over the final momenta. The rate of transition is obtained in the ninth section, where we also discuss the limit of a large expansion factor, relative to the particle masses. Our conclusions are summarized in section ten.

\section{Equations for interacting fields}
Let us begin with the de Sitter line element written in conformal form \cite{1}:
\begin{equation}\label{metr}
ds^2=dt^2-e^{2\omega t}d\vec{x}^2=\frac{1}{(\omega t_{c})^2}(dt_{c}^2-d\vec{x}^2)
\end{equation}

The conformal time is related to the proper time by $t_{c}=\frac{-e^{-\omega t}}{\omega}$, where $\omega$ is the Hubble parameter ($\omega>0$).

In our present study we propose for the first time a perturbative approach to the problem of massive charged $W^{\pm}$ boson production in early universe. Our computations are done in the de Sitter geometry, defined by a conformal chart with conformal time $t_{c}\in(-\infty,0)$, which covers the expanding portion of de Sitter space-time \cite{1}. For the line element (\ref{metr}) in the Cartesian gauge we have the non-vanishing tetrad components:
\begin{equation}
e^{0}_{\widehat{0}}=-\omega t_c  ;\,\,\,e^{i}_{\widehat{j}}=-\delta^{i}_{\widehat{j}}\,\omega t_c.
\end{equation}

The perturbative QED in de Sitter space-time was developed in \cite{24}, where it was shown that the generators of the covariant representation are the differential operators produced by the Killing vectors, which are associated with isometries, according to the generalized Carter and McLenagan formula \cite{CML}. These generators come from an algebra of conserved observables that commute with the operators of the field equations. With this method the quantum states can be defined on the entire manifold and be independent of local coordinates, so the vacuum state is unique and stable \cite{24}. This approach allows one to consider that the quantum states are prepared and measured by the same global apparatus which consists of the largest freely generated algebra which includes the conserved operators \cite{24}.
For developing the theory of electro-weak interactions between charged $W^{\pm}$ bosons and fermions we assume that the electro-weak transitions are measured by the same global apparatus which prepares all the quantum states, and this includes the $in$ and $out$ asymptotic free fields which are minimally coupled to gravity. This apparatus complies with an asymptotic prescription of frequency separation which assures the uniqueness and stability of  the vacuum state of the free fields \cite{2,22}. This goes back to the fact that our apparatus can record particle creation only in the presence of electromagnetic interactions, and the difference between our apparatus and the local detectors \cite{GH,DW} is that the latter can record particle production without electromagnetic interaction.

In this section we present the main steps for developing the interaction theory between massive charged bosons and fermions. We start with the action integral, which contains the lagrangian densities for the free Dirac field, the free charged Proca field and the interaction term from the electro-weak theory corresponding to charged currents. We also assume that the Dirac and Proca fields are minimally coupled to gravity.

\begin{equation}
S = \int d^{4}x \sqrt{-g(x)} \left( \mathcal{L}_{Proca} + \mathcal{L}_{Dirac} +  \mathcal{L}_{int}\nonumber \right)
\end{equation}

The explicit form of the action can be found bellow, where the action is expressed in terms of the tetrad-gauge invariant Lagrangian density, which determines the coupling between $W^{\pm}$ bosons and leptons, written in turn with regard to the point independent Dirac matrices $\gamma^{\hat\mu}$ and tetrad fields $e_{\hat{\alpha}}^{\mu}$ :
\begin{eqnarray}
S &=& \int d^{4}x \sqrt{-g(x)}\Bigg\{\frac{i}{2}\left[\bar{\psi}_e\gamma^{\hat{\alpha}}(D_{\hat{\alpha}}\psi_e)-(D_{\hat{\alpha}}\bar{\psi}_e)\gamma^{\hat{\alpha}}\psi_e\right]-m\bar{\psi}_e\psi_e\nonumber\\
&&+\frac{i}{2}\left[\bar{\psi}_\nu\gamma^{\hat{\alpha}}(D_{\hat{\alpha}}\psi_\nu)-(D_{\hat{\alpha}}\bar{\psi}_\nu)\gamma^{\hat{\alpha}}\psi_\nu\right]-\frac{1}{2}F_{\mu\nu}^+ F^{\mu\nu} + M_W^{2}A_{\mu}^+A^{\mu}\nonumber\\
 &&-\frac{g}{2\sqrt{2}}\bar{\psi}_e\gamma^{\hat{\alpha}}e_{\hat{\alpha}}^{\mu}\left(1-\gamma^{5}\right)\psi_{\nu} A_{\mu}-\frac{g}{2\sqrt{2}}\bar{\psi}_{\nu}\gamma^{\hat{\alpha}}e_{\hat{\alpha}}^{\mu}\left(1-\gamma^{5}\right)\psi_{e} A_{\mu}^+\Bigg\},
\end{eqnarray}
where $\psi_e$ describe the electron-positron field, $\psi_{\nu}$ is the neutrino-antineutrino field and $A_\mu$,$A_\mu^+$ describe the massive boson fields $W^{-}$ and $W^{+}$, with field strength defined as $F_{\mu\nu}=\partial_\mu A_{\nu}-\partial_\nu A_{\mu}$.
The coupling constant $g$ is expressed in terms of Fermi constant as $\frac{G_F}{\sqrt{2}}=\frac{g^2}{8M_W^2}$.

Term $D_{\hat{\alpha}}=\partial_{\hat{\alpha}}+\Gamma_{\hat{\alpha}}$ denotes the covariant derivatives in local frames. The covariant derivative depends on the spin connections $\Gamma_{\hat{\alpha}}=\Gamma_{\hat{\alpha}\hat{\mu}\hat{\nu}}S^{\hat{\mu}\hat{\nu}}$, which are given in terms of the basis generators $S^{\hat{\mu}\hat{\nu}}=i/4\{\gamma^{\hat{\mu}},\gamma^{\hat{\nu}}\}$ of the spinorial representation of the $SL(2,\textbf{C})$ group.

By using Euler-Lagrange equations we obtain a set of four interaction equations for the fields $\psi_{e},\psi_{e\nu},A_\mu,A_\mu^+$ :
\begin{eqnarray}\label{fi}
\begin{cases}
(i\gamma^{\hat{\alpha}}D_{\hat{\alpha}}-m)\psi_e(x)=\frac{g}{2\sqrt{2}}\sqrt{-g(x)}\left[\gamma^{\hat{\alpha}}e_{\hat{\alpha}}^{\mu}(1-\gamma^{5})\psi_\nu(x)A_{\mu}(x)\right], \\
i\gamma^{\hat{\alpha}}D_{\hat{\alpha}}\psi_\nu(x)=\frac{g}{2\sqrt{2}}\sqrt{-g(x)}\left[\gamma^{\hat{\alpha}}e_{\hat{\alpha}}^{\mu}(1-\gamma^{5})\psi_e(x)A_{\mu}(x)\right],\\
\partial_\rho \big( g^{\rho \alpha} g^{\mu \beta} \sqrt{-g(x)}F_{\alpha \beta}(x) \big) + \sqrt{-g} \, M_W^2 A^\mu(x)=\frac{g}{2\sqrt{2}}\sqrt{-g(x)}\left[\bar{\psi}_\nu(x)\gamma^{\hat{\alpha}}e_{\hat{\alpha}}^{\mu}(1-\gamma^{5})\psi_e(x)\right],\\
\partial_\rho \big( g^{\rho \alpha} g^{\mu \beta} \sqrt{-g(x)}F_{\alpha \beta}^+(x) \big) + \sqrt{-g} \, M_W^2 A^{\mu+}(x)=\frac{g}{2\sqrt{2}}\sqrt{-g(x)}\left[\bar{\psi}_e(x)\gamma^{\hat{\alpha}}e_{\hat{\alpha}}^{\mu}(1-\gamma^{5})\psi_\nu(x)\right]
\end{cases}
\end{eqnarray}

We mention that the Dirac operator in de Sitter metric has the expression \cite{22}:
\begin{equation}
i\gamma^{\hat{\alpha}}D_{\hat{\alpha}}-m=E_{D} = -i\omega t_{c} (\gamma^{0}\partial_{t_{c}} + \gamma^{i}\partial_{i})+\frac{3i\omega}{2}\gamma^{0}-m
\end{equation}

The Proca equation in de Sitter geometry reads:
\begin{equation}
\frac{1}{\sqrt{-g(x)}}\left[\delta_{\mu}^{\beta}\left(\partial_{\rho}\partial^{\rho} + \frac{\mu^{2}}{t_{c}^{2}}\right) - \partial_{\mu}\partial^{\beta} \right] A_{\beta} = 0,
\end{equation}
where we introduce the Proca operator:
\begin{equation}\label{procaop}
{(E_{P})}_{\mu}^{\beta}= \frac{1}{\sqrt{-g(x)}}\left[\delta_{\mu}^{\beta}\left(\partial_{\rho}\partial^{\rho} + \frac{\mu^{2}}{t_{c}^{2}}\right) - \partial_{\mu}\partial^{\beta}\right]
\end{equation}

In the case of the free Proca field, the equations for the temporal and spatial components of the vector potential in the de Sitter metric and chart $\{t_c,\vec x\}$ can be found by using the Lorentz condition \cite{2}:
\begin{equation}
\partial_{\mu}(\sqrt{-g}A^{\mu})=0
\end{equation}

The equations for the components of the vector potential are \cite{2} :
\begin{eqnarray}
\begin{cases}
-\Delta A_{0} + \frac{\mu^{2}}{t_{c}^{2}}A_{0} + \partial_{t_{c}}(\partial_{i}A_{i}) = 0 ,\\
\left(\partial_{t_{c}}^{2}-\Delta + \frac{\mu^{2}}{t_{c}^{2}}\right) A_{i} - \partial_{t_{c}}\left(\partial_{i}A_{0}\right) - \partial_{i}\left(\partial_{k}A_{k}\right) = 0,
\end{cases}
\end{eqnarray}
where $\mu=\frac{M_W}{\omega}$.

With the help of (\ref{procaop}), the last two equations from (\ref{fi}) can be rewritten in the form:
\begin{eqnarray}\label{pfi}
{(E_{P})}_{\mu}^{\beta}A^\mu=\frac{g}{2\sqrt{2}}\left[\bar{\psi}_\nu(x)\gamma^{\hat{\alpha}}e_{\hat{\alpha}}^{\beta}(1-\gamma^{5})\psi_e(x)\right],\nonumber\\
{(E_{P})}_{\mu}^{\beta}A^{\mu+}=\frac{g}{2\sqrt{2}}\left[\bar{\psi}_e(x)\gamma^{\hat{\alpha}}e_{\hat{\alpha}}^{\beta}(1-\gamma^{5})\psi_\nu(x)\right].
\end{eqnarray}

The Green functions of the Proca equation obey relation \cite{2,WT}:
\begin{eqnarray}\label{pg}
\eta^{\alpha\beta}\partial_{\alpha}[\partial_{\beta}G_{\mu\nu}(x,y)-\partial_{\mu}G_{\beta\nu}(x,y)] + \frac{\mu^{2}}{t_{c}^{2}}G_{\mu\nu}(x,y) = \\ \eta_{\mu\nu}\delta^{4}(x-y)
\end{eqnarray}

This can be written in terms of the Proca operator as:
\begin{equation}\label{gpro}
  \bigg[ \delta^\beta_\mu \bigg( \partial_\rho \partial^\rho + \frac{\mu^2}{t_c^2} \bigg) - \partial^\beta \partial_\mu \bigg] G_{\beta\nu} (x-y) = \eta_{\mu\nu} \delta^4 (x-y)
\end{equation}

The Green functions for the Dirac field will satisfy the following equations \cite{22}:
\begin{eqnarray}\label{dg}
E_{D}S(x-y) &=& -\frac{1}{\sqrt{-g}}\delta^{4}(x-y).
\end{eqnarray}

In order to develop the reduction formalism for the Proca field we will use the method of Green functions \cite{12,19,20} to write the solutions for the interacting field equations (\ref{pfi}):
\begin{eqnarray}\label{sp}
&&A_{\beta}(x) = \hat{A}_{\beta}(x) + \int d^{4}y \sqrt{-g(y)} G_{\beta\nu}(x,y)J^{\nu}\nonumber\\
&&=\hat{A}_{\beta}(x) + \frac{g}{2\sqrt{2}}\int d^{4}y \sqrt{-g(y)} G_{\beta\nu}(x,y)\bar{\psi}_\nu(y)\gamma^{\hat{\alpha}}e_{\hat{\alpha}}^{\beta}(1-\gamma^{5})\psi_e(y)
\end{eqnarray}

The validity of the above solution can be verified if one applies the Proca operator ${(E_{P})}_{\mu}^{\beta}$ to it. Finally the solution for the interacting fields (\ref{sp}) can be expressed as:
\begin{eqnarray}\label{sp}
&&A_{\beta}(x) = \hat{A}_{\beta}(x) +  \int d^{4}y \sqrt{-g(y)} G_{\beta\nu}(x,y) E_{\alpha}^{\nu}(y)A^{\alpha}(y)
\end{eqnarray}

We proceed further by defining the IN and OUT fields with the help of the retarded and advanced Green's functions, while specifying that these fields are defined up to a normalization constant denoted by $\sqrt{z_3}$:
\begin{equation}
\sqrt{z_3}A_{\beta}^{IN/OUT}(x) = \hat{A}_{\beta}(x) - \int d^{4}y \sqrt{-g(y)} G_{\beta\nu}^{A/R}(x,y) E_{\alpha}^{\nu}(y)A^{\alpha}(y)
\end{equation}

The above fields are asymptotically equal to free fields for $t\rightarrow \pm \infty$, where the advanced and retarded Green functions vanish. The IN/OUT fields defined above represent the basis for the reduction formalism.

\section{Reduction formalism for the Proca field}

We apply the same method of reduction as in the flat space-time field theory \cite{19}. The reduction of the Proca field from the OUT state can be constructed by starting with the amplitude that corresponds to a transition of a  Proca particle between IN/OUT states. We also denote the other particles by $\alpha,\beta$ in the matrix element:
\begin{equation}
\langle \,out \,\beta \, 1 (\vec{p^{\prime}},\lambda^{\prime}) \,|\, in\,\alpha \, 1 (\vec{p},\lambda)\,\rangle = \langle \, out \, \beta\,|\,a_{out}(\vec{p^{\prime}},\lambda^{\prime})\,|\,in\, \alpha \,1(\vec{p},\lambda)\, \rangle
\end{equation}

By using the properties of the creation and annihilation operators $a^{\dag}$ and $a$:
\begin{equation}
\left[a(\vec{p},\lambda),a^{\dag}(\vec{p^{\prime}},\lambda^{\prime})\right] =  \delta_{\lambda\lambda^{\prime}}\delta^{3}(\vec{p}-\vec{p^{\prime}}),
\end{equation}
we can rewrite the amplitude in the following way:
\begin{eqnarray}\label{aaa}
&&\langle \, out \, \beta \,|\, (a_{out}(\vec{p^{\prime}},\lambda^{\prime}) \,|\, in \, \alpha \,1 (\vec{p},\lambda)\, \rangle = \nonumber
\\ &&\langle \, out \, \beta \,|\, [a_{out}(\vec{p^{\prime}},\lambda^{\prime})- a_{in}(\vec{p^{\prime}},\lambda^{\prime})] \,|\, in \,\alpha \,1(\vec{p},\lambda)\,\rangle \nonumber
\\ &&+ \langle \, out \, \beta \,|\, a_{in}(\vec{p^{\prime}},\lambda^{\prime})a_{in}^{\dag}(\vec{p},\lambda)\,|\, in \, \alpha \, \rangle.
\end{eqnarray}

The field operator for the Proca field can be expanded in terms of the fundamental solutions for the Proca equation in the momentum-helicity basis \cite{2}:
\begin{equation}
A_{\beta} (x) = \sum_{\lambda} \int d^{3}p \,[ a(\vec{p},\lambda)f_{{\vec{p}\lambda}_{\beta}}(x) + b^{\dag}(\vec{p},\lambda)f_{{\vec{p}\lambda}_{\beta}}^{*}(x)],
\end{equation}
where $a(\vec{p},\lambda)$ are the operators for the particle while $b^{\dag}(\vec{p},\lambda)$ are the operators for the antiparticle.

The definitions for the creation and annihilation operators for the particle can be obtained as:
\begin{eqnarray}
&&{a^{\dag}}^{IN/OUT} = i \int f_{\vec{p}\lambda}^{\mu}(x) \stackrel{\leftrightarrow}{\partial_{t_c}} A_{\mu}^{IN/OUT\dag} (x) d^{3}x\\
&&a^{IN/OUT} = -i \int {f^{*}}_{\vec{p}\lambda}^{\mu}(x) \stackrel{\leftrightarrow}{\partial_{t_c}} A_{\mu}^{IN/OUT} (x) d^{3}x.
\end{eqnarray}

Then one can evaluate the difference from equation (\ref{aaa})
\begin{eqnarray}
&&a_{out} (\vec{p^{\prime}},\lambda^{\prime}) - a_{in} (\vec{p^{\prime}},\lambda^{\prime}) =
i \int d^{3}x {f^{*}}_{\vec{p^{\prime}}\lambda^{\prime}}^{\mu} (x) \stackrel{\leftrightarrow}{\partial_{t_c}} \left[A_{\mu}^{OUT}(x) - A_{\mu}^{IN}(x)\right]\nonumber\\
&&= \frac{i}{\sqrt{z_3}} \int d^{4}y \sqrt{-g(y)} \int d^{3}x {f^{*}}_{\vec{p^{\prime}}\lambda^{\prime}}^{\beta} (x) \stackrel{\leftrightarrow}{\partial_{t_c}} \left[G_{\beta\nu}^{A}(x,y) - G_{\beta\nu}^{R}(x,y)\right]({E_{P}})_{\alpha}^{\nu}(y)A^{\alpha}(y),
\nonumber\\
\end{eqnarray}
where the total commutator function is defined as:
\begin{eqnarray}
&&G_{\beta\nu}(x,y) = G_{\beta\nu}^{A}(x,y) - G_{\beta\nu}^{R}(x,y)\nonumber\\
&& = \sum_{\lambda} \int d^{3}p \left[{f_{\vec{p}\lambda}}_{\beta}(x){{f^{*}}_{\vec{p}\lambda}}_{\nu}(y) - {{f^{*}}_{\vec{p}\lambda}}_{\beta}(x){f_{\vec{p}\lambda}}_{\nu}(y)\right].
\end{eqnarray}

Then we solve the $x$ integral
\begin{equation}
i\int d^{3}x {f^{*}}_{\vec{p^{\prime}}\lambda^{\prime}}^{\beta}(x) \stackrel{\leftrightarrow}{\partial_{t_c}} G_{\beta\nu}(x,y) = i {f^{*}}_{\vec{p^{\prime}}\lambda^{\prime}}^{\beta}(y),
\end{equation}
and one obtains the final result for the operator difference:
\begin{equation}
a_{out} (\vec{p^{\prime}},\lambda^{\prime}) - a_{in} (\vec{p^{\prime}},\lambda^{\prime}) = i \int d^{4}y \sqrt{-g(y)} {f^{*}}_{\vec{p^{\prime}}\lambda^{\prime}}^{\beta}(y) ({E_{P}})_{\alpha}^{\nu}(y)A^{\alpha}(y).
\end{equation}

The final reduction formula for the Proca field in the OUT state is:
\begin{eqnarray}
&&\langle\, out\, \beta\, 1(\vec{p^{\prime}},\lambda^{\prime}) \,|\, in\,\alpha\, 1(\vec{p},\lambda)\, \rangle =\delta_{\lambda\lambda^{\prime}}\delta^{3}(\vec{p}-\vec{p^{\prime}}) \langle \,out \,\beta\,|\,in \,\alpha\, \rangle \nonumber \\
&&+ \frac{i}{\sqrt{z_3}} \int d^{4}y \sqrt{-g(y))} {{f^{*}}_{\vec{p^{\prime}}\lambda^{\prime}}}_{\nu}(y) (E_{P})_{\alpha}^{\nu}(y) \langle\, out\, \beta \, |\, A^{\alpha}(y)\,|\,in \, \alpha \,1 (\vec{p},\lambda)\,\rangle,
\nonumber\\
\end{eqnarray}
where the first term represents the process when the Proca particle do not interact with other particles at the IN/OUT transition and will be neglected when we compute the amplitude.

In the case of reduction from the IN state the amplitude reads
\begin{equation}
\langle \,out \,\beta \, 1 (\vec{p^{\prime}},\lambda^{\prime}) \,|\, in\,\alpha \, 1 (\vec{p},\lambda)\,\rangle = \langle \, out \, \beta\, 1(\vec{p}\,^{\prime},\lambda^{\prime}) \,|\, a_{in}^{\dag}(\vec{p},\lambda) \,|\, in\, \alpha \, \rangle.
\end{equation}

Following the same steps as in the case of reduction for the OUT particle we obtain the operator difference:

\begin{equation}
a_{in}^{\dag} (\vec{p},\lambda) - a_{out}^{\dag} (\vec{p},\lambda) = i \int d^{4}y \sqrt{-g(y)} {f_{\vec{p}\lambda}}_{\nu}(y) ({E_{P}})_{\alpha}^{\nu}(y)A^{\alpha}(y),
\end{equation}

and the final reduction formula for the Proca field in the IN state:
\begin{eqnarray}
\langle\, out\, \beta\, 1(\vec{p^{\prime}},\lambda^{\prime}) \,|\, in\,\alpha\, 1(\vec{p},\lambda)\, \rangle =\delta_{\lambda\lambda^{\prime}}\delta^{3}(\vec{p}-\vec{p^{\prime}}) \langle \,out \,\beta\,|\,in \,\alpha\, \rangle \nonumber \\
+ \frac{i}{\sqrt{z_3}} \int d^{4}y \sqrt{-g(y)} {f_{\vec{p}\lambda}}_{\nu}(y) (E_{P})_{\alpha}^{\nu}(y) \langle\, out\, \beta \, 1(\vec{p^{\prime}},\lambda^{\prime})\, |\, A^{\alpha}(y)\,|\,in \, \alpha \,\rangle.
\end{eqnarray}

When all particles from the IN and OUT state are reduced we obtain the vacuum average of a chronological product of field operators. This allows us to use perturbative methods for defining the amplitudes by making T-contractions between the cinematical part obtained from the reduction formalism and the dynamical part given by the different orders in the $S$ operator.

\section{The transition amplitude}

The process we want to study is the spontaneous vacuum emission of a $W^{-}$ boson, a positron and a neutrino, under the conditions of an expanding background:
\begin{equation}
vacuum \longrightarrow W^{-} + e^{+} + \nu.
\end{equation}

The lagrangian of interaction between leptons and charged bosons $W^{\pm}$ from the flat space-time theory \cite{12}, will be adapted to our de Sitter geometry and will be written in terms of the point independent Dirac matrices $\gamma^{\hat\alpha}$ and the tetrad fields $e_{\hat{\alpha}}^{\mu}$ :
\begin{equation}
 \mathcal{L}_{int}= -\frac{g}{2\sqrt{2}}\left[\bar{\psi}_{\nu} \gamma^{\hat{\alpha}} e_{\hat{\alpha}}^{\mu}  (1-\gamma^{5}) \psi_{e} A_{\mu}^+ + \bar{\psi}_{e} \gamma^{\hat{\alpha}} e_{\hat{\alpha}}^{\mu}  (1-\gamma^5) \psi_{(\nu)} A_\mu\right],
\end{equation}
where $g$ is the coupling constant.

The first step in finding out the amplitude is to compute the S-matrix elements using the reduction method. The IN state is empty, whereas in the OUT state there are one negatively charged boson, a positron and a neutrino.
We denote the momenta and polarizations of $W^{-}$, $e^{+}$ and $\nu$ by $(P,\lambda)$, $(p$'$,\sigma$'$)$ and $(p,\sigma)$, respectively. We also specify that the reduction formalism for fermions in de Sitter space-time can be found in \cite{24,rc}.

The reduction formula for the OUT particles is:
\begin{eqnarray}
&&\langle out 1(P,\lambda), 1(p,\sigma),\tilde{1}(p^{\prime},\sigma^{\prime})| in 0 \rangle = \bigg( \frac{i}{\sqrt{z_3}} \bigg) \bigg( \frac{i}{\sqrt{z_2}}\bigg)^2 \int d^{4}y_1 \sqrt{-g(y_1)} {f^{*}}_{\vec{P}\lambda}^\nu(y_1)\nonumber \\
&&\times(E_{P})_{\nu}^{\beta}(y_1)  \int d^{4}y_2 \sqrt{-g(y_2)} \bar{u}_{\vec{p}\sigma}(y_2) E_{D}(y_2) \int d^{4}y_3 \sqrt{-g(y_3)} v_{\vec{p^{\prime}}\sigma^{\prime}}(y_3)E_{D}(y_3)
\nonumber\\ &&\times \langle 0|T(A_{\beta}(y_1)\psi(y_2)\bar{\psi}(y_3))|0 \rangle. \nonumber \\
\end{eqnarray}

To compute the time-ordered product of interacting fields we make use of the scattering operator \cite{12,19,24}:
\begin{equation}
 S=T(exp(-i\int d^{4}x \sqrt{-g}\mathcal{L}_{int})).
 \end{equation}

The first order term in the series expansion of the $S$ operator is:
\begin{equation}\label{ss}
S^{(1)} = -\frac{ig}{2\sqrt{2}}\int d^{4}x \sqrt{-g(x)}\left[\bar{\psi}_{\nu} \gamma^{\hat{\alpha}} e_{\hat{\alpha}}^{\mu}  (1-\gamma^{5}) \psi_{e} A_{\mu}^+ + \bar{\psi}_{e} \gamma^{\hat{\alpha}} e_{\hat{\alpha}}^{\mu}  (1-\gamma^5) \psi_{(\nu)} A_\mu\right].
\end{equation}

The Green functions for the interacting fields can be expressed in terms of free fields:
\begin{equation}
\langle 0|T(\mathcal{A}_{\beta}(y_1)\Psi(y_2)\bar{\Psi}(y_3))|0 \rangle = \frac{\langle 0|T(A_{\beta}(y_1)\psi(y_2)\bar{\psi}(y_3)S)|0 \rangle}{\langle 0|S|0 \rangle}.
\end{equation}

In our case we use only the term corresponding to the $W^-$ boson from (\ref{ss}). All possible T-contractions give :
\begin{eqnarray}
\langle 0|T(A_{\beta}(y_1)\psi(y_2)\bar{\psi}(y_3)[\bar{\psi}_{(e^{+})} \gamma^{\hat{\alpha}} e_{\hat{\alpha}}^{\mu} (1-\gamma^5) \psi_{(\nu)} A_\mu](x))|0 \rangle =\\ \nonumber \frac{1}{i}D_{\beta\mu}(y_1-x)\frac{1}{i}S(y_2-x)\gamma^{\hat{\alpha}} e_{\hat{\alpha}}^{\mu} (1-\gamma^5)\left(\frac{-1}{i}\right)S(y_3-x),
\end{eqnarray}
where $D_{\beta\mu}(y-x)$ and $S(y-x)$ are the Feynman propagators.

After substituting the results of the contractions into the reduction formula, we make use of the relations with propagators (\ref{pg}), (\ref{dg}) and compute the following delta integrals:
\begin{eqnarray}
&&\int d^{4}y_1  {f^{*}}_{\vec{P}\lambda,\mu}(y_1) \delta^4(y_1 -x) = {f^{*}}_{\vec{P}\lambda,\mu}(x) ;\\
&&\int d^{4}y_2  \bar{u}_{\vec{p}\sigma}(y_2) \delta^4(y_2 -x) = \bar{u}_{\vec{p}\sigma}(x); \\
&&\int d^{4}y_3  v_{\vec{p^{\prime}}\sigma^{\prime}}(y_3)\delta^4(y_3 -x) = v_{\vec{p^{\prime}}\sigma^{\prime}}(x).
\end{eqnarray}

The amplitude for the process of spontaneous vacuum emission of a $W^{-}$ boson, a positron and a neutrino then is:
\begin{equation}\label{ampl}
A_{i\rightarrow f}= \frac{ig}{2\sqrt{2}} \int d^{4}x \sqrt{-g(x)} \, \bar{u}_{\vec{p}\sigma}(x) \gamma^{\hat{\alpha}} e_{\hat{\alpha}}^{\mu}(1-\gamma^5) v_{\vec{p^{\prime}}\sigma^{\prime}}(x){f^{*}}_{\vec{P}\lambda,\mu} (x),
\end{equation}
where ${f^{*}}_{\vec{P}\lambda,\mu} (x), \bar{u}_{\vec{p}\sigma}(x)$ and $ v_{\vec{p^{\prime}}\sigma^{\prime}}(x)$ are the field equation solutions in the de Sitter geometry, written in the momentum-helicity basis. The amplitude corresponding to the vacuum emission of a $W^{+}$ boson, an electron and an anti-neutrino is then:
\begin{equation}\label{ampl5}
A_{i\rightarrow f}'= \frac{ig}{2\sqrt{2}} \int d^{4}x \sqrt{-g(x)} \, \bar{u}_{\vec{p\,'}\sigma'}(x) \gamma^{\hat{\alpha}} e_{\hat{\alpha}}^{\mu}(1-\gamma^5) v_{\vec{p}\sigma}(x){f}_{\vec{P}\lambda,\mu} (x),
\end{equation}
where in this case $\bar{u}_{\vec{p\,'}\sigma'}(x)$ describes the electron while $v_{\vec{p}\sigma}(x)$ describes the anti-neutrino. We mention that the probabilities corresponding to the above mentioned processes are equal.

\section{The amplitude for $\lambda = \pm 1$}

We can split the amplitude in terms of the polarization $\lambda$ of $W^{-}$, which can take the values $0$ and $\pm 1$. Firstly we are going to compute for $\lambda=\pm 1$:
\begin{equation}\label{amplt}
A_{i\rightarrow f}= \frac{ig}{2\sqrt{2}} \int d^{4}x \sqrt{-g(x)} \, \bar{u}_{\vec{p}\sigma}(x) \gamma^{\hat{i}} e_{\hat{i}}^{j}(1-\gamma^5) v_{\vec{p^{\prime}}\sigma^{\prime}}(x){f^{*}}_{\vec{P}\lambda,j} (x).
\end{equation}

The solutions of the Proca equation in de Sitter space-time for $\lambda=\pm 1$ contain only the spatial part, while the temporal component of the Proca solution is $0$, \cite{2}:
\begin{eqnarray}
{f}_{\vec{P}\lambda=\pm 1,0}^{*} (x)&=& 0 \\ \nonumber
{f}_{\vec{P}\lambda=\pm 1,j}^{*} (x)&=& \frac{\sqrt{\pi} e^{-\pi K/2}}{2(2\pi)^{3/2}} \sqrt{-t_{c}} \mathcal{H}_{-iK}^{(1)}(-P t_{c}) e^{-i\vec{P}\vec{x}} {\epsilon}^{*}_{j}(\vec{P},\lambda=\pm 1),
\end{eqnarray}
where $\mathcal{H}_{-iK}^{(2)}(-P t_{c})$ are Hankel functions of first kind, $K=\sqrt{\left(\frac{M_W}{\omega}\right)^{2}-\frac{1}{4}}$ and $\vec{\epsilon}\,(\vec{P},\lambda)$ are the polarization vectors. For $\lambda=\pm 1$ these vectors are orthogonal with the momentum
$\vec{P}\cdot\vec{\epsilon}\,({P},\lambda=\pm1)=0$.

The solutions with well determined momentum and helicity for the Dirac equation in de Sitter geometry were obtained in \cite{22}. In the case of zero mass fermions, the solution for the neutrino was obtained in \cite{22}:
\begin{equation}
u_{\vec{p}\sigma}^{\dag}(x) = \left(\frac{-\omega t_c}{2\pi}\right)^{3/2} \left(\left(\frac{1}{2}-\sigma\right)\xi_{\sigma}^{\dag}(x),0 \right)e^{-i\vec{p}\vec{x} + ipt_c},
\end{equation}
considering that $\bar{u}=u^{\dag}\gamma^{0}$.

For the positron we have the solution with respect to the helicity basis \cite{22}:
\begin{eqnarray}
v_{\vec{p^{\prime}}\sigma^{\prime}}(x) = i\sqrt{\frac{\pi p^{\prime}}{\omega}}\left(\frac{1}{2\pi}\right)^{3/2}\left(
                                                                                      \begin{array}{c}
                                                                                        -\sigma^{\prime} e^{-\pi k/2} \mathcal{H}_{\nu_{-}}^{(2)}\left(\frac{p^{\prime}}{\omega}e^{-\omega t}\right)\eta_{\sigma^{\prime}}(\vec{p^{\prime}}) \\
                                                                                        \frac{1}{2} e^{\pi k/2}\mathcal{H}_{\nu_{+}}^{(2)}\left(\frac{p^{\prime}}{\omega}e^{-\omega t}\right)\eta_{\sigma^{\prime}}(\vec{p^{\prime}}) \\
                                                                                      \end{array}
                                                                                    \right) e^{-i\vec{p^{\prime}}\vec{x}-2\omega t},
\end{eqnarray}
where $\mathcal{H}^{(1)}_{\nu}(z), \mathcal{H}^{(2)}_{\nu}(z)$ are Hankel functions of the first and second kind and $k=\frac{m_{e}}{\omega},\nu_{\pm}=\frac{1}{2}\pm ik$.

The helicity spinors satisfy the relation:
\begin{equation}\label{pa}
\vec{\sigma}\vec{p}\,\xi_{\sigma}(\vec{p}\,)=2p\sigma\xi_{\sigma}(\vec{p}\,)
\end{equation}
with $\sigma=\pm1/2$, where $\vec{\sigma}$ are the Pauli matrices, and $p=\mid\vec{p}\mid$ is the modulus of the momentum vector, while $\eta_{\sigma}(\vec{p}\,)= i\sigma_2
[\xi_{\sigma}(\vec{p}\,)]^{*}$.

By substituting the solutions into the amplitude we get:
\begin{eqnarray}
A_{i\rightarrow f} &=& \frac{-g}{2\sqrt{2}} \int d^{4}x \sqrt{-g(x)} \left(\frac{-\omega t_c}{2\pi}\right)^{3/2} \left(-2\sigma^i\left(\frac{1}{2}-\sigma\right)\xi_{\sigma}^{\dag}(\vec{p}),0 \right)e^{-i\vec{p}\vec{x} + ipt_c}\nonumber \\
  &&\times \sqrt{\frac{\pi p^{\prime}}{\omega}}\left(\frac{1}{2\pi}\right)^{3/2}
  \left(
   \begin{array}{c}
   -\sigma^{\prime} e^{-\pi k/2} \mathcal{H}_{\nu_{-}}^{(2)}\left(\frac{p^{\prime}}{\omega}e^{-\omega t}\right)\eta_{\sigma^{\prime}}(\vec{p^{\prime}}) \\
   \frac{1}{2} e^{\pi k/2}\mathcal{H}_{\nu_{+}}^{(2)}\left(\frac{p^{\prime}}{\omega}e^{-\omega t}\right)\eta_{\sigma^{\prime}}(\vec{p^{\prime}}) \\
   \end{array}
   \right) e^{-i\vec{p^{\prime}}\vec{x}-2\omega t} \nonumber \\
   &&\times e_{\hat{i}}^{j}\frac{\sqrt{\pi} e^{-\pi K/2}}{2(2\pi)^{3/2}} \sqrt{-t_{c}} \mathcal{H}_{-iK}^{(2)}(-P t_{c}) e^{-i\vec{P}\vec{x}} {\epsilon}^{*}_{j}(\vec{P},\lambda=\pm 1),
\end{eqnarray}
where we have used the chiral basis for representing the Dirac matrices and  have replaced term $\gamma^0\gamma^{\hat{i}}e_{\hat{i}}^{j}(1-\gamma^5)$.

The conformal map we are working in is defined by $t_c=-\frac{1}{\omega}e^{-\omega t}$. The metric in this map is $\sqrt{-g} = e^{3\omega t}$ and the tetrad coefficients are $e_{\hat{i}}^{j} = -e^{-\omega t}\delta_{ij}$.

We can group some of the time-dependent terms:
\begin{equation}
\sqrt{-g}(-\omega t_c)^{3/2}e_{\hat{i}}^{j} \frac{1}{\sqrt{\omega}}e^{-2\omega t}\sqrt{t_{c}}= -\frac{1}{\omega}e^{-2\omega t}\delta_{ij}.
\end{equation}

We can also directly compute the spatial integral:
\begin{equation}
\int d^{3}x e^{-i(\vec{p}+\vec{p^{\prime}}+\vec{P})\vec{x}} = (2\pi)^{3}\delta^{3}(\vec{p}+\vec{p^{\prime}}+\vec{P}),
\end{equation}
which takes care of momentum conservation in this process.

By substituting all of the above results and multiplying the matrices and vectors the transition amplitude becomes:
\begin{eqnarray}\label{amge}
&&A_{i\rightarrow f}= -\frac{g}{2\sqrt{2}}\frac{\pi \sqrt{p^{\prime}} e^{-\pi K/2}e^{-\pi k/2}}{(2\pi)^{3/2}}\delta^{3}(\vec{p}+\vec{p^{\prime}}+\vec{P})\nonumber\\
&&\times\int_{-\infty}^{+\infty} d t \frac{e^{-2\omega t}}{\omega}e^{ip^{\prime}t_{c}}\left(\frac{1}{2}-\sigma\right)\sigma^{\prime}
\cdot\mathcal{H}_{\nu_{-}}^{(2)}\left(\frac{p^{\prime}}{\omega}e^{-\omega t}\right)\mathcal{H}_{-iK}^{(2)}(-P t_{c})\nonumber\\
&&\times\xi_{\sigma}^{\dag}(\vec{p}\,)\sigma^{i}\eta_{\sigma^{\prime}}(\vec{p^{\prime}}){\epsilon}^{*}_{i}(\vec{P},\lambda=\pm 1).
\end{eqnarray}

In order to solve the integral, we have to change the integration variable from $t$ to $z=-t_c=\frac{1}{\omega}e^{-\omega t}$.
The amplitude in terms of the new variable $z$ is:
\begin{eqnarray}\label{amp1}
A_{i\rightarrow f}= -\frac{g}{2\sqrt{2}}\frac{\pi \sqrt{p^{\prime}} e^{-\pi K/2}e^{-\pi k/2}}{(2\pi)^{3/2}}\delta^{3}(\vec{p}+\vec{p^{\prime}}+\vec{P}) \left(\frac{1}{2}-\sigma\right)\sigma^{\prime} \cdot \nonumber \\ \int_{0}^{+\infty} dz \cdot z e^{-ipz}\mathcal{H}_{\nu_{-}}^{(2)}(p^{\prime}z)\mathcal{H}_{-iK}^{(2)}(P z)\xi_{\sigma}^{\dag}(\vec{p})\sigma^{i}\eta_{\sigma^{\prime}}(\vec{p^{\prime}}){\epsilon}^{*}_{i}(\vec{P},\lambda=\pm 1).
\end{eqnarray}

The next step is to solve the integral with respect to $z$.
To use relation (\ref{appell}) from the Appendix, we need to transform the Hankel functions and the exponential from (\ref{amp1}) into Bessel functions. The following relations and properties allow us to do so \cite{AS,21}:
\begin{eqnarray}
&&e^{-z} = \sqrt{\frac{2z}{\pi}}\mathcal{K}_{\frac{1}{2}}(z);\label{etok} \\
&&\mathcal{K}_{\nu}(iz) = \mathcal{H}_{\nu}^{(2)}(z)\frac{\pi}{2i}e^{-i\nu_\frac{\pi}{2}};\label{ktoh} \\
&&\mathcal{H}_{\nu}^{(2)}(z) = \frac{2i}{\pi}e^{i\nu_\frac{\pi}{2}}\mathcal{K}_{\nu}(iz);\label{htok}\\
&&\mathcal{H}_{\nu}^{(2)}(z) = \frac{e^{i\pi\nu}\mathcal{J}_{\nu}(z)-\mathcal{J}_{-\nu}(z)}{i \sin({\pi\nu})}.\label{htoj}
\end{eqnarray}

From the exponential factor in the amplitude integral (\ref{amp1}) we want to obtain a Hankel function by firstly using (\ref{etok}) and then (\ref{ktoh}), meaning we get:
\begin{equation}
e^{-ipz}=\sqrt{\frac{2ipz}{\pi}}\frac{\pi}{2i}e^{-i\frac{\pi}{4}}\mathcal{H}_{\frac{1}{2}}^{(2)}(pz).
\end{equation}

We further transform the Hankel function in the above relation into Bessel functions using (\ref{htoj}) \cite{AS,21}:
\begin{equation}\label{expneutrino}
\sqrt{\frac{2ipz}{\pi}}\frac{\pi}{2i}e^{-i\frac{\pi}{4}}\mathcal{H}_{\frac{1}{2}}^{(2)}(pz) = \sqrt{\frac{2ipz}{\pi}}\frac{\pi}{2i}e^{-i\frac{\pi}{4}}\left[\mathcal{J}_{\frac{1}{2}}(pz)-\frac{1}{i}\mathcal{J}_{-\frac{1}{2}}(pz)\right],
\end{equation}

The Hankel function from the $v_{\vec{p^{\prime}},\sigma^{\prime}}(x)$ solution also becomes, by virtue of (\ref{htoj}) \cite{AS,21}:
\begin{equation}\label{hankelpositron}
\mathcal{H}_{\frac{1}{2}-ik}^{(2)}(p^{\prime}z)=\frac{ie^{\pi k}\mathcal{J}_{\frac{1}{2}-ik}(p^{\prime}z)-\mathcal{J}_{-\frac{1}{2}+ik}(p^{\prime}z)}{i\, \cosh(\pi k)}.
\end{equation}

The Hankel function which depends on the momentum of the $W^{-}$ boson becomes a modified Bessel $\mathcal{K}$ function \cite{AS,21}:
\begin{equation}\label{hankelboson}
\mathcal{H}_{-iK}^{(2)}(Pz)=\frac{2i}{\pi}e^{\frac{K\pi}{2}} \mathcal{K}_{-iK}(iPz).
\end{equation}

Now we may substitute (\ref{expneutrino}),(\ref{hankelpositron}) and (\ref{hankelboson}) into the temporal integral and we obtain:
\begin{eqnarray}
&\int_{0}^{\infty} dz\quad z e^{-ipz}\mathcal{H}_{\nu_{-}}^{(2)}(p^{\prime}z)\mathcal{H}_{-i\it{K}}^{(2)}(P z) = \sqrt{\frac{2ip^{\prime}}{\pi}}e^{-i\frac{\pi}{4}}e^{\mathcal{K}\frac{\pi}{2}}\nonumber\\
&\times\int_{0}^{\infty}dz\quad z\sqrt{z}\left[\mathcal{J}_{\frac{1}{2}}(pz)-\frac{1}{i}\mathcal{J}_{-\frac{1}{2}}(pz)\right]
\left[\frac{ie^{\pi k}\mathcal{J}_{\frac{1}{2}-ik}(p^{\prime}z)-\mathcal{J}_{-\frac{1}{2}+ik}(p^{\prime}z)}{i\,\cosh(\pi k)}\right]\mathcal{K}_{-iK}(iPz).\nonumber\\
\end{eqnarray}

The final form of the temporal integral is:
\begin{equation}
\int_{0}^{\infty} dz\cdot z e^{-ipz}\mathcal{H}_{\nu_{-}}^{(2)}(p^{\prime}z)\mathcal{H}_{-i\it{K}}^{(2)}(P z) =\frac{\sqrt{\frac{2ip^{\prime}}{\pi}}e^{-i\frac{\pi}{4}}e^{K\frac{\pi}{2}}}{i\, cosh(\pi k)}[T_{1}-T_{2}-T_{3}+T_{4}],
\end{equation}
where the terms $T_{1},T_{2},T_{3}$ and $T_{4}$ denote the following integrals:
\begin{eqnarray}
T_{1}&=&\int_{0}^{\infty} dz\, z\sqrt{z}\,\mathcal{J}_{\frac{1}{2}}(pz)ie^{\pi k}\mathcal{J}_{\frac{1}{2}-ik}(p^{\prime}z)\mathcal{K}_{-iK}(iPz); \\
T_{2}&=&\int_{0}^{\infty} dz\, z\sqrt{z}\,\mathcal{J}_{\frac{1}{2}}(pz)\mathcal{J}_{-\frac{1}{2}+ik}(p^{\prime}z)\mathcal{K}_{-iK}(iPz);\\
T_{3}&=&\int_{0}^{\infty} dz\, z\sqrt{z}\,e^{\pi k}\mathcal{J}_{-\frac{1}{2}}(pz)\mathcal{J}_{\frac{1}{2}-ik}(p^{\prime}z)\mathcal{K}_{-iK}(iPz); \\
T_{4}&=&\int_{0}^{\infty} dz\, z\sqrt{z}\,\frac{1}{i}\mathcal{J}_{-\frac{1}{2}}(pz)\mathcal{J}_{-\frac{1}{2}+ik}(p^{\prime}z)\mathcal{K}_{-iK}(iPz).
\end{eqnarray}

We can then rewrite the amplitude:
\begin{eqnarray}
A_{i\rightarrow f}= -\frac{g}{2}\frac{\sqrt{\pi} \sqrt{pp^{\prime}}\sqrt{i} e^{-\pi k/2}}{(2\pi)^{3/2}}\delta^{3}(\vec{p}+\vec{p^{\prime}}+\vec{P}) \left(\frac{1}{2}-\sigma\right)\sigma^{\prime} \frac{e^{-i\frac{\pi}{4}}}{i\,\cosh(\pi k)}\nonumber \\ \times [T_{1}-T_{2}-T_{3}+T_{4}] \xi_{\sigma}^{\dag}(\vec{p})\sigma^{i}\eta_{\sigma^{\prime}}(\vec{p^{\prime}}){\epsilon}^{*}_{i}(\vec{P},\lambda=\pm 1).
\end{eqnarray}

Terms $T_{1},T_{2},T_{3}$ and $T_{4}$ may now each be integrated using formula (\ref{appell}) from Appendix, taking note of the fact that in this case $Q=\frac{5}{2}$.

After integration the results are:
\begin{eqnarray}\label{ttt}
T_{1}=\frac{ie^{\pi k}\sqrt{2}p^{\frac{1}{2}}{p^{\prime}}^{\frac{1}{2}-ik}(iP)^{-\frac{7}{2}+ik}}{\Gamma\left(\frac{3}{2}\right)\Gamma\left(\frac{3}{2}-ik\right)} \Gamma\left(\frac{\frac{7}{2}-i(k-K)}{2}\right)\Gamma\left(\frac{\frac{7}{2}-i(k+K)}{2}\right)\nonumber \\
\times\mathcal{F}_{4}\left(\frac{\frac{7}{2}-i(k-K)}{2},\frac{\frac{7}{2}-i(k+K)}{2},\frac{3}{2},\frac{3}{2}-ik,\frac{p^{2}}{P^{2}},\frac{{p^{\prime}}^{2}}{P^{2}}\right);\\
T_{2}=\frac{\sqrt{2}p^{\frac{1}{2}}{p^{\prime}}^{-\frac{1}{2}+ik}(iP)^{-\frac{5}{2}-ik}}{\Gamma\left(\frac{3}{2}\right)\Gamma\left(\frac{1}{2}+ik\right)} \Gamma\left(\frac{\frac{5}{2}+i(k+K)}{2}\right)\Gamma\left(\frac{\frac{5}{2}+i(k-K)}{2}\right)\nonumber \\
\times\mathcal{F}_{4}\left(\frac{\frac{5}{2}+i(k+K)}{2},\frac{\frac{5}{2}+i(k-K)}{2},\frac{3}{2},\frac{1}{2}+ik,\frac{p^{2}}{P^{2}},\frac{{p^{\prime}}^{2}}{P^{2}}\right);\\
T_{3}=\frac{e^{\pi k}\sqrt{2}p^{-\frac{1}{2}}{p^{\prime}}^{\frac{1}{2}-ik}(iP)^{-\frac{5}{2}+ik}}{\Gamma\left(\frac{1}{2}\right)\Gamma\left(\frac{3}{2}-ik\right)} \Gamma\left(\frac{\frac{5}{2}-i(k-K)}{2}\right)\Gamma\left(\frac{\frac{5}{2}-i(k+K)}{2}\right)\nonumber \\
\times\mathcal{F}_{4}\left(\frac{\frac{5}{2}-i(k-K)}{2},\frac{\frac{5}{2}-i(k+K)}{2},\frac{1}{2},\frac{3}{2}-ik,\frac{p^{2}}{P^{2}},\frac{{p^{\prime}}^{2}}{P^{2}}\right);\\
T_{4}=\frac{\frac{1}{i}\sqrt{2}p^{-\frac{1}{2}}{p^{\prime}}^{-\frac{1}{2}+ik}(iP)^{-\frac{3}{2}-ik}}{\Gamma\left(\frac{1}{2}\right)\Gamma\left(\frac{1}{2}+ik\right)} \Gamma\left(\frac{\frac{3}{2}+i(k+K)}{2}\right)\Gamma\left(\frac{\frac{3}{2}+i(k-K)}{2}\right)\nonumber \\
\times\mathcal{F}_{4}\left(\frac{\frac{3}{2}+i(k+K)}{2},\frac{\frac{3}{2}+i(k-K)}{2},\frac{1}{2},\frac{1}{2}+ik,\frac{p^{2}}{P^{2}},\frac{{p^{\prime}}^{2}}{P^{2}}\right),
\end{eqnarray}

Each of the four terms is expressed in terms of gamma Euler functions $\Gamma$  and the $\mathcal{F}_{4}$ Appell's function, which is defined by the double series \cite{AS,21}:
\begin{equation}
\mathcal{F}_{4}(\alpha,\beta,\gamma,\gamma';x,y)=\sum_{m=0}^{\infty} \sum_{n=0}^{\infty} \frac{\Gamma(\alpha+m+n)\Gamma(\beta+m+n)\Gamma(\gamma)\Gamma(\gamma')}{\Gamma(\alpha)\Gamma(\beta)\Gamma(\gamma +m)\Gamma(\gamma' +n)} \cdot\frac{x^{m}y^{n}}{m!n!}.
\end{equation}

There is little research on Appell's functions $\mathcal{F}_{4}$ and there are just a few known cases where they can be converted into other kinds of hypergeometric functions, with the mention that this does not apply to our case. For this reason we will use in our further analysis the double series definition without any approximation.

The behaviour of the amplitude will be determined by the behaviour of the four terms which depend on Appell's function. The parameters of interest are $K=\sqrt{\left(\frac{M_{W}}{\omega}\right)^{2}-\frac{1}{4}}$ and $k=\frac{m_{e}}{\omega}$, where $M_{W}$ is the mass of the $W$ boson, $m_{e}$ is the mass of the positron and $\omega$ is the expansion parameter.

We want to see the behaviour of the amplitude as it relates to the ratios of the particle masses and the expansion parameter, and so we want to graphically represent each of the four terms $T_{1},T_{2},T_{3}$ and $T_{4}$ in some way. We must also choose values for momenta $p', p, P$, which are present explicitly in the amplitude, as well as present in Appell's functions as arguments $x=p^{2}/P^{2}$ and $y={p^{\prime}}^{2}/P^{2}$.

In order to add up all relevant factors we denote:
\begin{eqnarray}
A = \frac{\sqrt{i}\sqrt{pp^{\prime}}e^{-\frac{\pi i}{4}}{e^{-\pi\frac{k}{2}}}}{i\,\cosh(\pi k)} \, T_{1};\,\,\,B=\frac{\sqrt{i}\sqrt{pp^{\prime}}e^{-\frac{\pi i}{4}}{e^{-\pi\frac{k}{2}}}}{i\,\cosh(\pi k)}\,T_2;\nonumber\\
C = \frac{\sqrt{i}\sqrt{pp^{\prime}}e^{-\frac{\pi i}{4}}{e^{-\pi\frac{k}{2}}}}{i\,\cosh(\pi k)} \, T_{3};\,\,\,D=\frac{\sqrt{i}\sqrt{pp^{\prime}}e^{-\frac{\pi i}{4}}{e^{-\pi\frac{k}{2}}}}{i\,\cosh(\pi k)}\,T_4.
\end{eqnarray}

The amplitude in terms of the new notation is:
\begin{eqnarray}
A_{i\rightarrow f}= -\frac{g}{2}\frac{\sqrt{\pi}}{(2\pi)^{3/2}}\delta^{3}(\vec{p}+\vec{p'}+\vec{P}) \left(\frac{1}{2}-\sigma\right)\sigma' \nonumber \\ \times [A-B-C+D] \xi_{\sigma'}^{\dag}(\vec{p'})\sigma^{i}\eta_{\sigma}(\vec{p}\,){\epsilon}^{*}_{i}(\vec{P},\lambda=\pm 1).
\end{eqnarray}

We will plot the real and imaginary parts of the functions that define the amplitude for a clear picture of their behaviour, in terms of parameters $\frac{m_{e}}{\omega},\,\frac{M_{W}}{\omega}$, for fixed values of momenta.

\begin{figure}[H]
\includegraphics[scale=0.35]{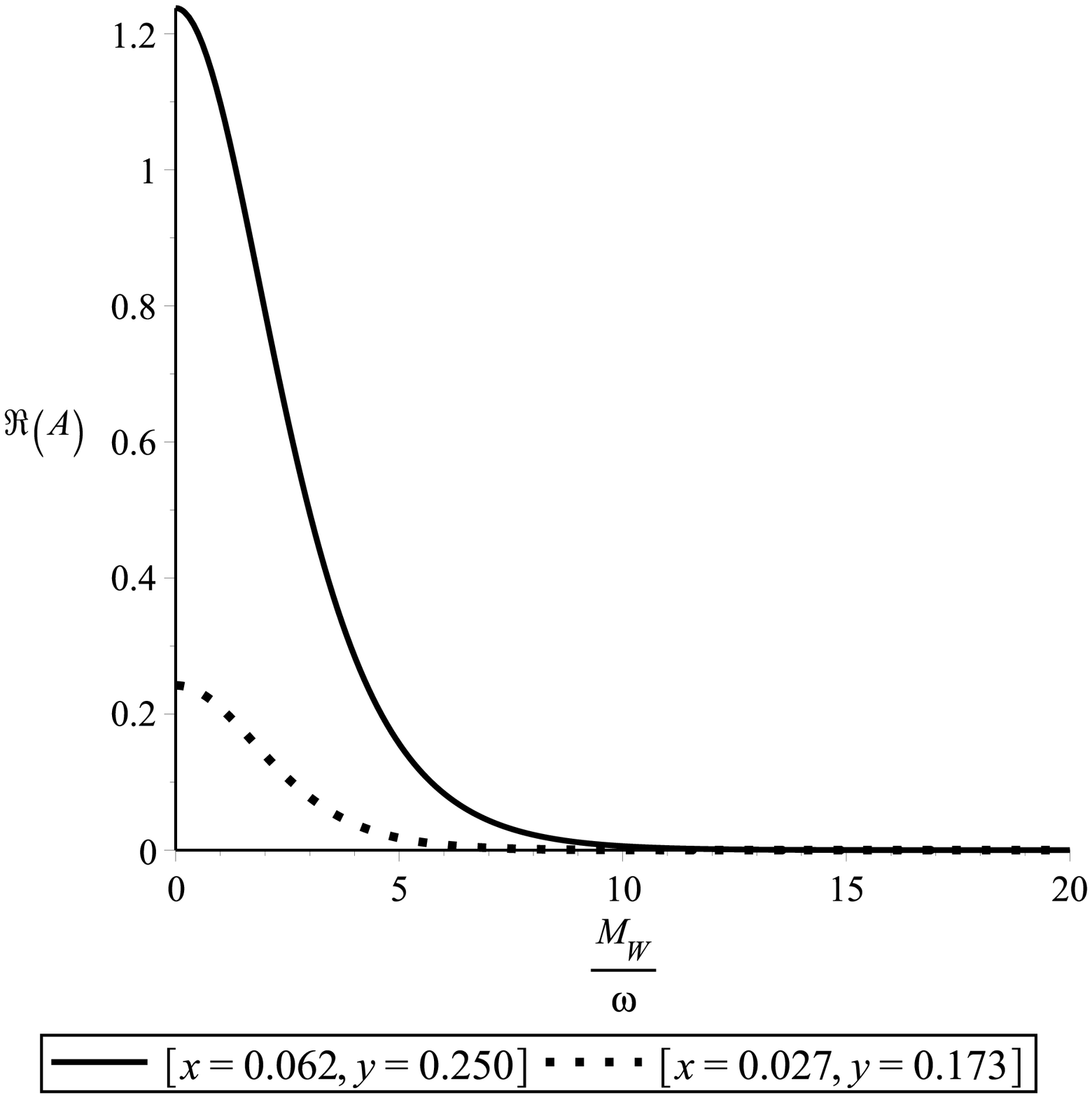}
\includegraphics[scale=0.35]{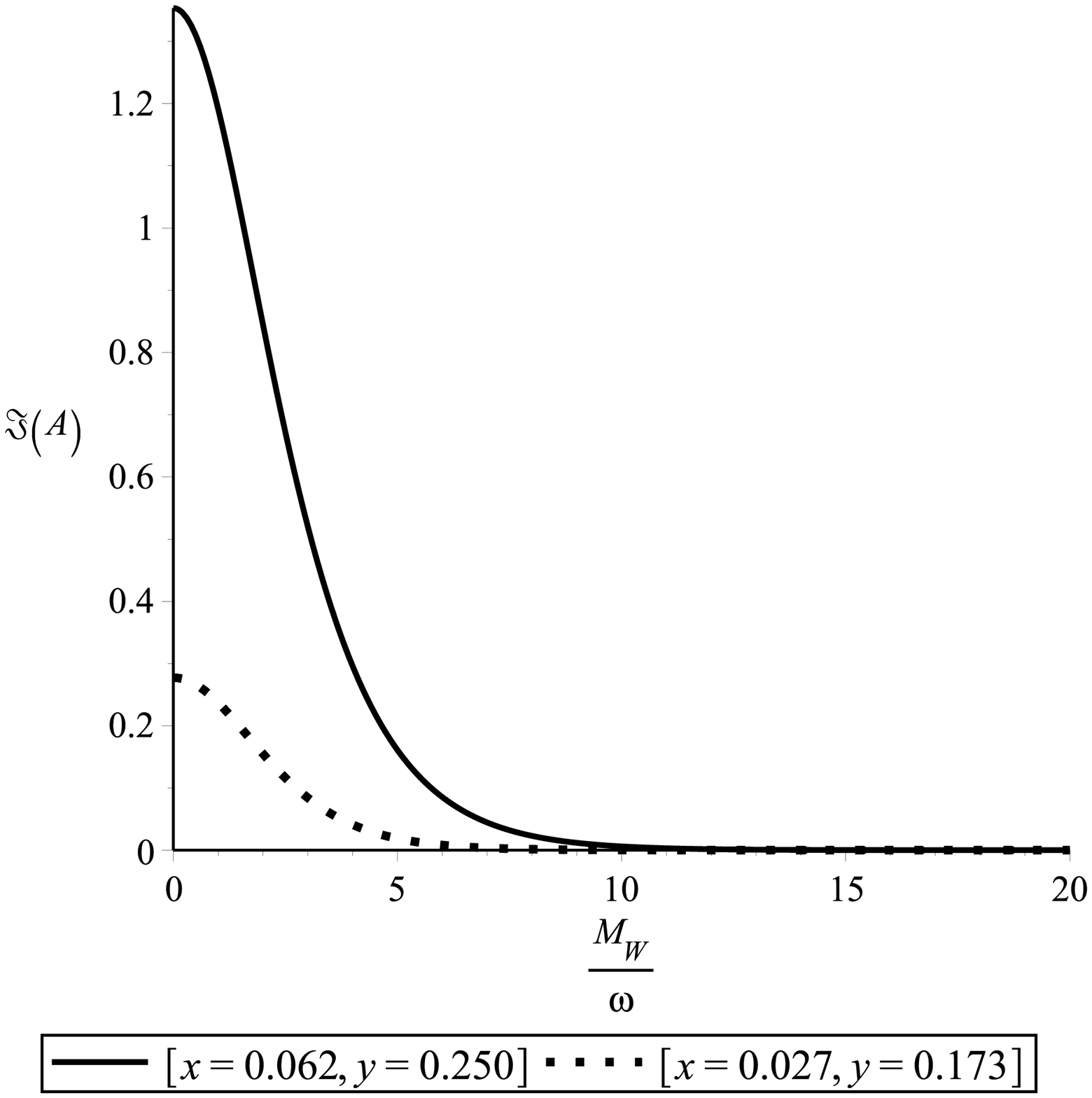}
\caption{The real and imaginary parts of $A$ with respect to $\frac{M_{W}}{\omega}$ for $k=0.1$ and momenta values $p' = 0.2, p = 0.1, P = 0.4$ for the straight line and $p' = 0.25, p = 0.1, P = 0.6$ for the dotted line.}
\label{A1W}
\end{figure}

\begin{figure}[H]
\includegraphics[scale=0.35]{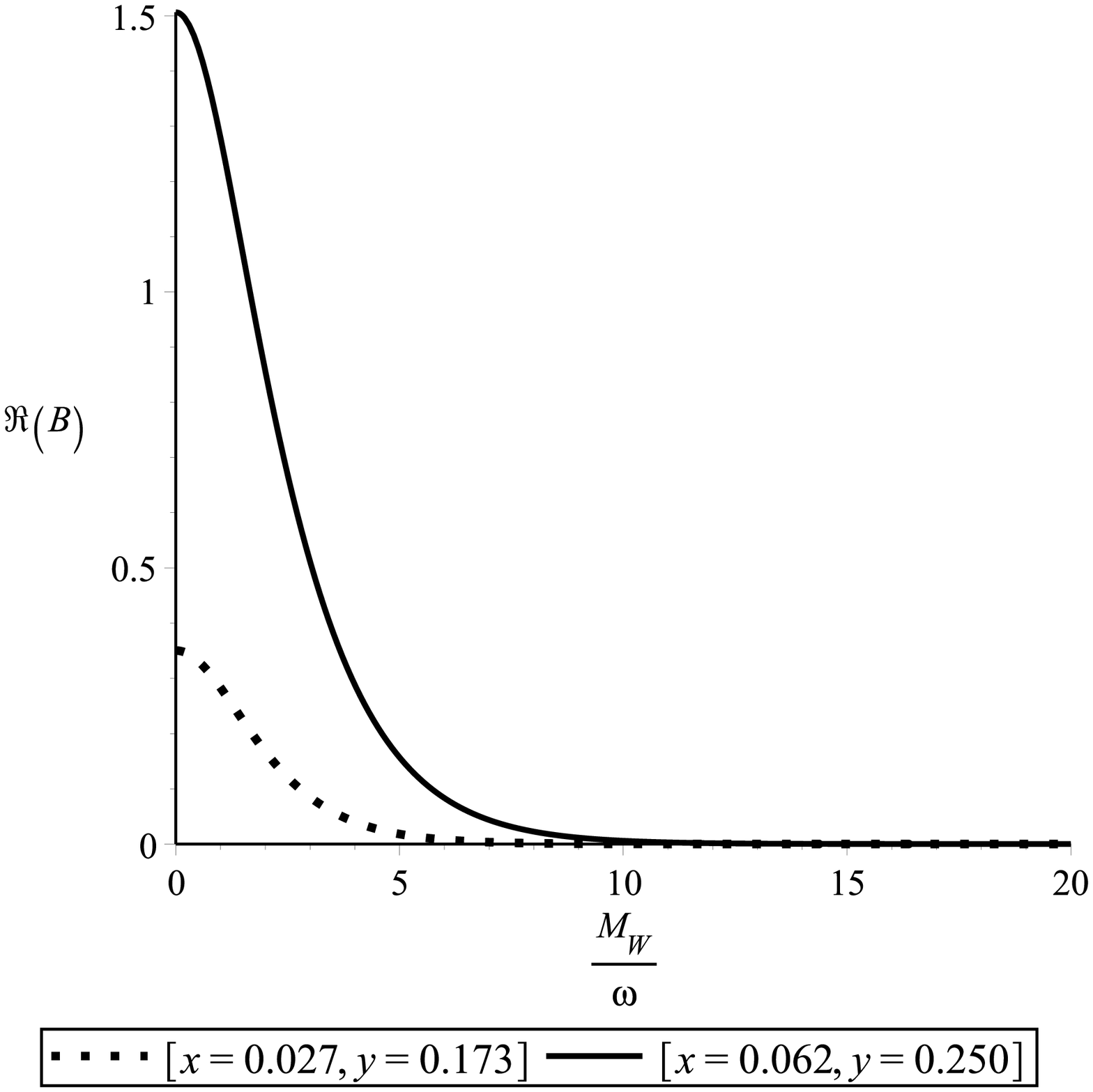}
\includegraphics[scale=0.35]{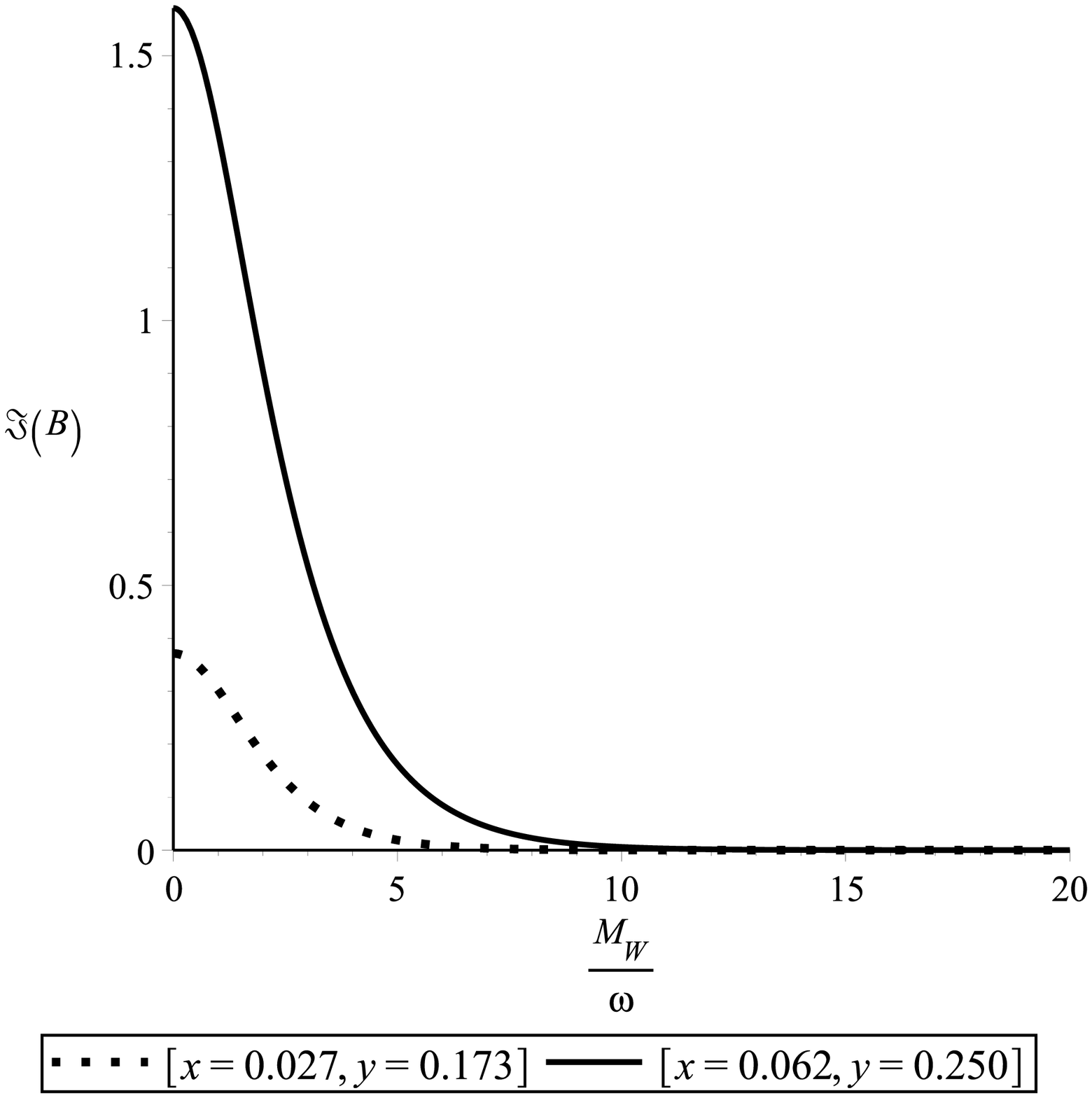}
\caption{The real and imaginary parts of $B$ with respect to $\frac{M_{W}}{\omega}$ for $k=0.1$ and momenta values $p' = 0.2, p = 0.1, P = 0.4$ for the straight line and $p' = 0.25, p = 0.1, P = 0.6$ for the dotted line.}
\label{B1W}
\end{figure}

\begin{figure}[H]
\includegraphics[scale=0.35]{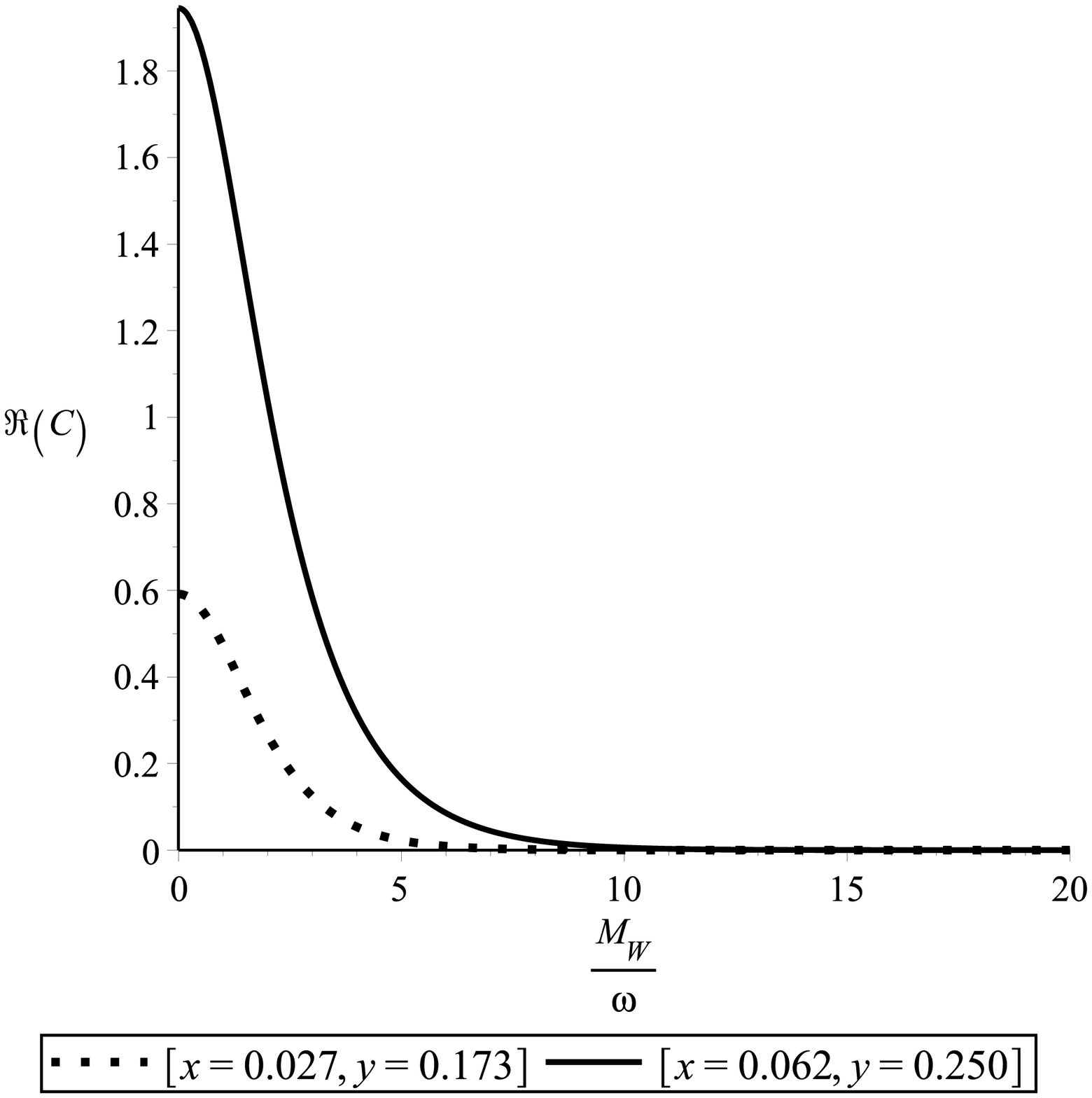}
\includegraphics[scale=0.35]{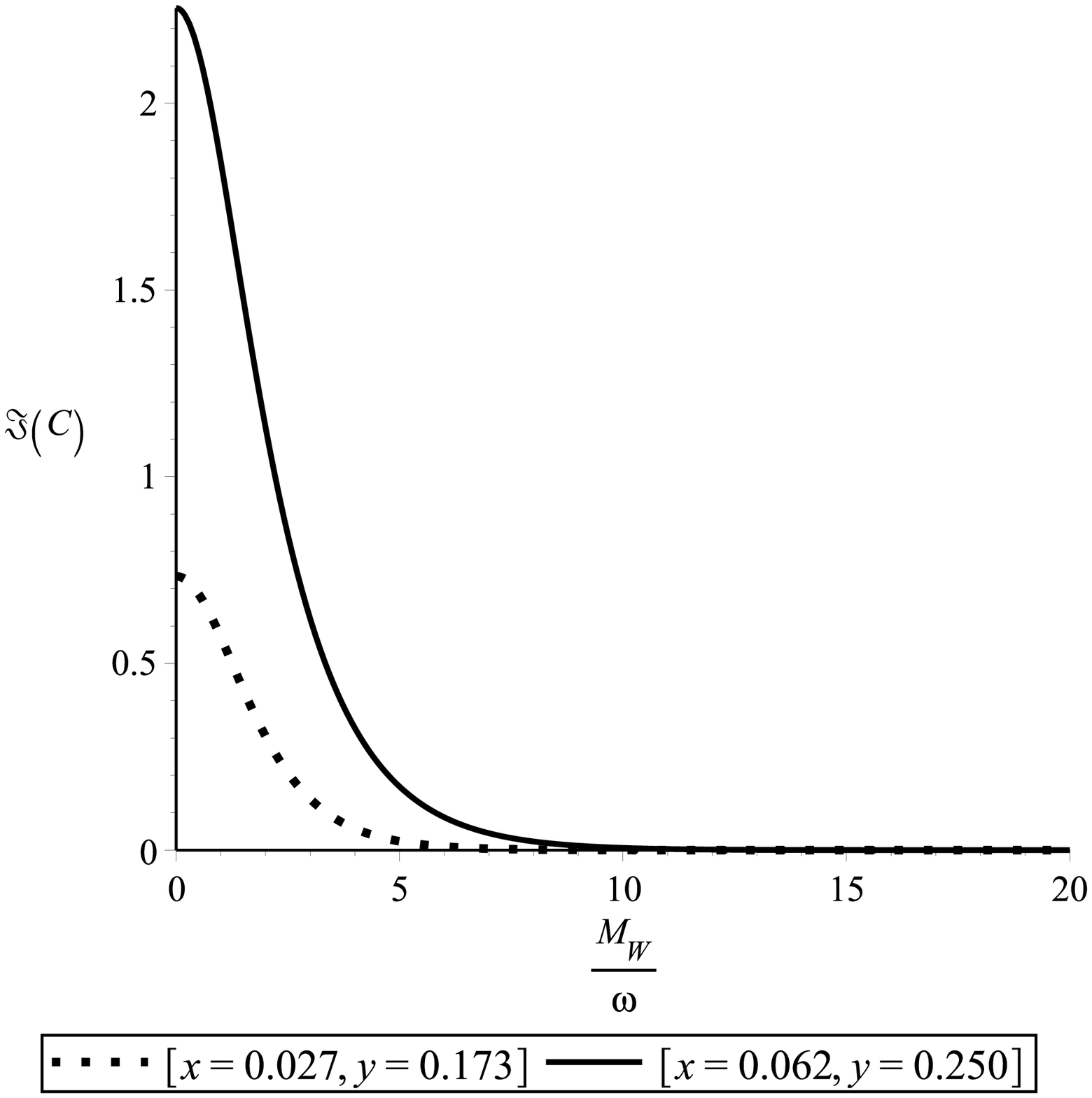}
\caption{The real and imaginary parts of $C$ with respect to $\frac{M_{W}}{\omega}$ for $k=0.1$ and momenta values $p' = 0.2, p = 0.1, P = 0.4$ for the straight line and $p' = 0.25, p = 0.1, P = 0.6$ for the dotted line.}
\label{C1W}
\end{figure}

\begin{figure}[H]
\includegraphics[scale=0.35]{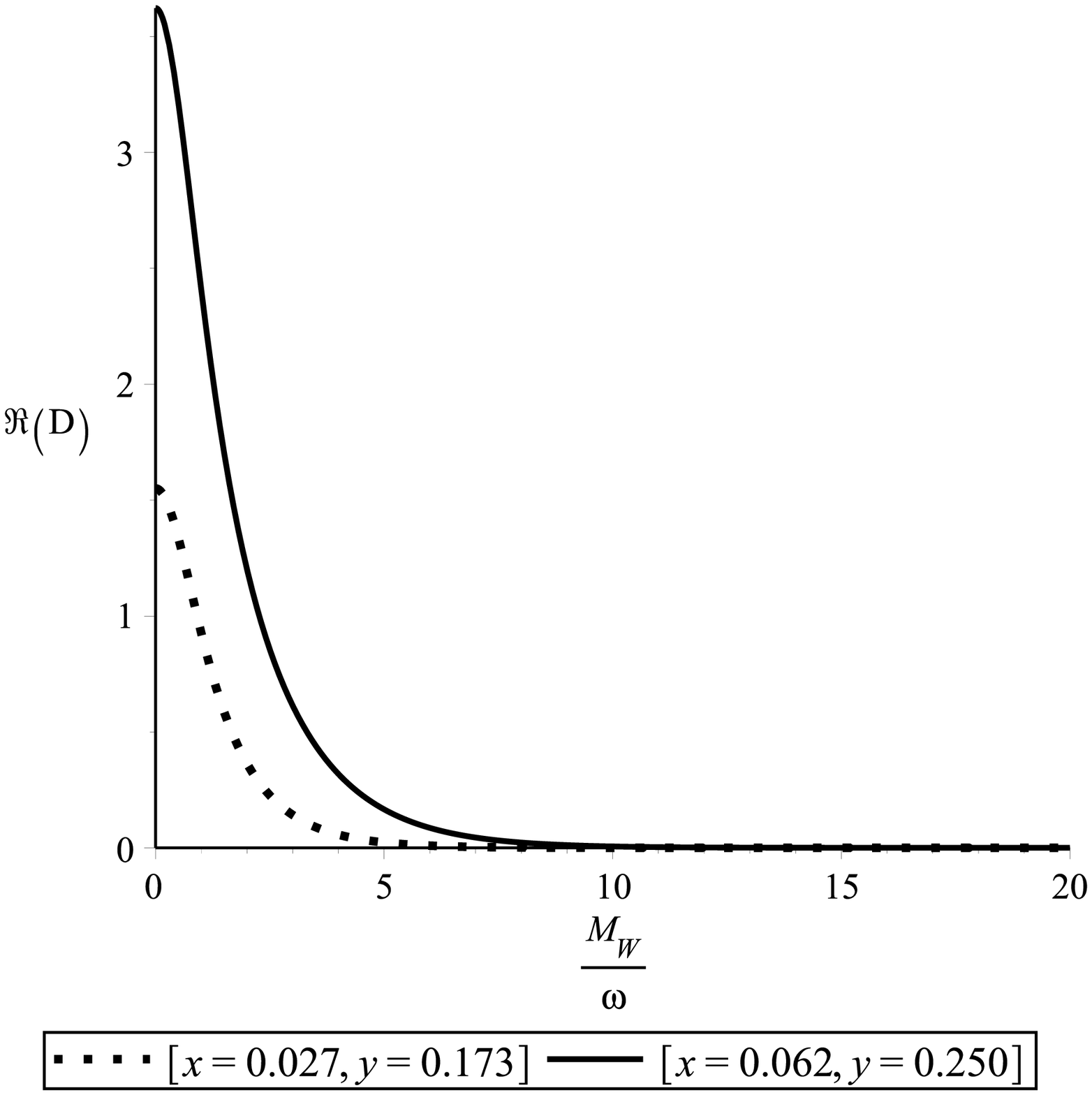}
\includegraphics[scale=0.35]{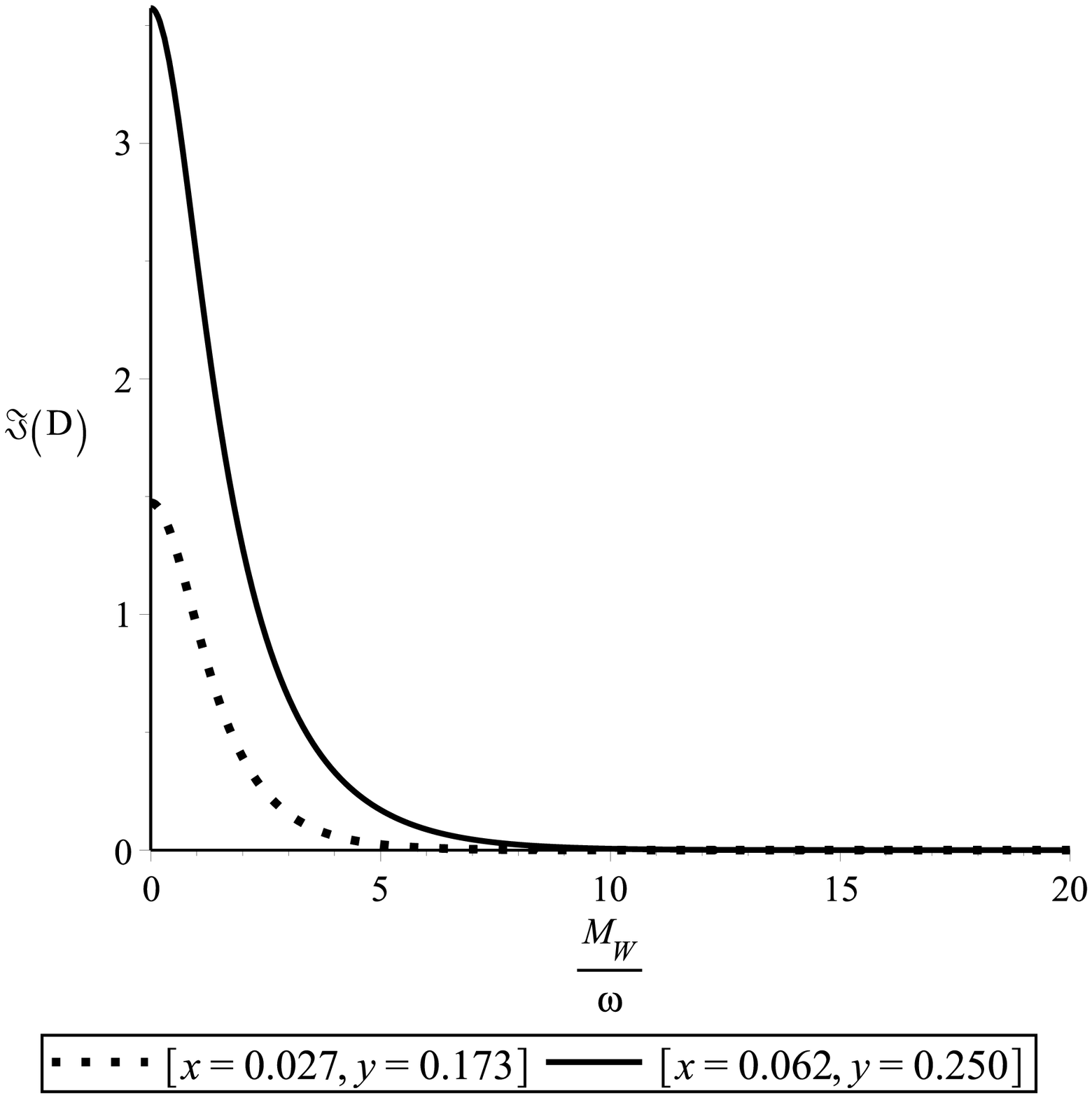}
\caption{The real and imaginary parts of $D$ with respect to $\frac{M_{W}}{\omega}$ for $k=0.1$ and momenta values $p' = 0.2, p = 0.1, P = 0.4$ for the straight line and $p' = 0.25, p = 0.1, P = 0.6$ for the dotted line.}
\label{D1W}
\end{figure}

\begin{figure}[H]
\includegraphics[scale=0.35]{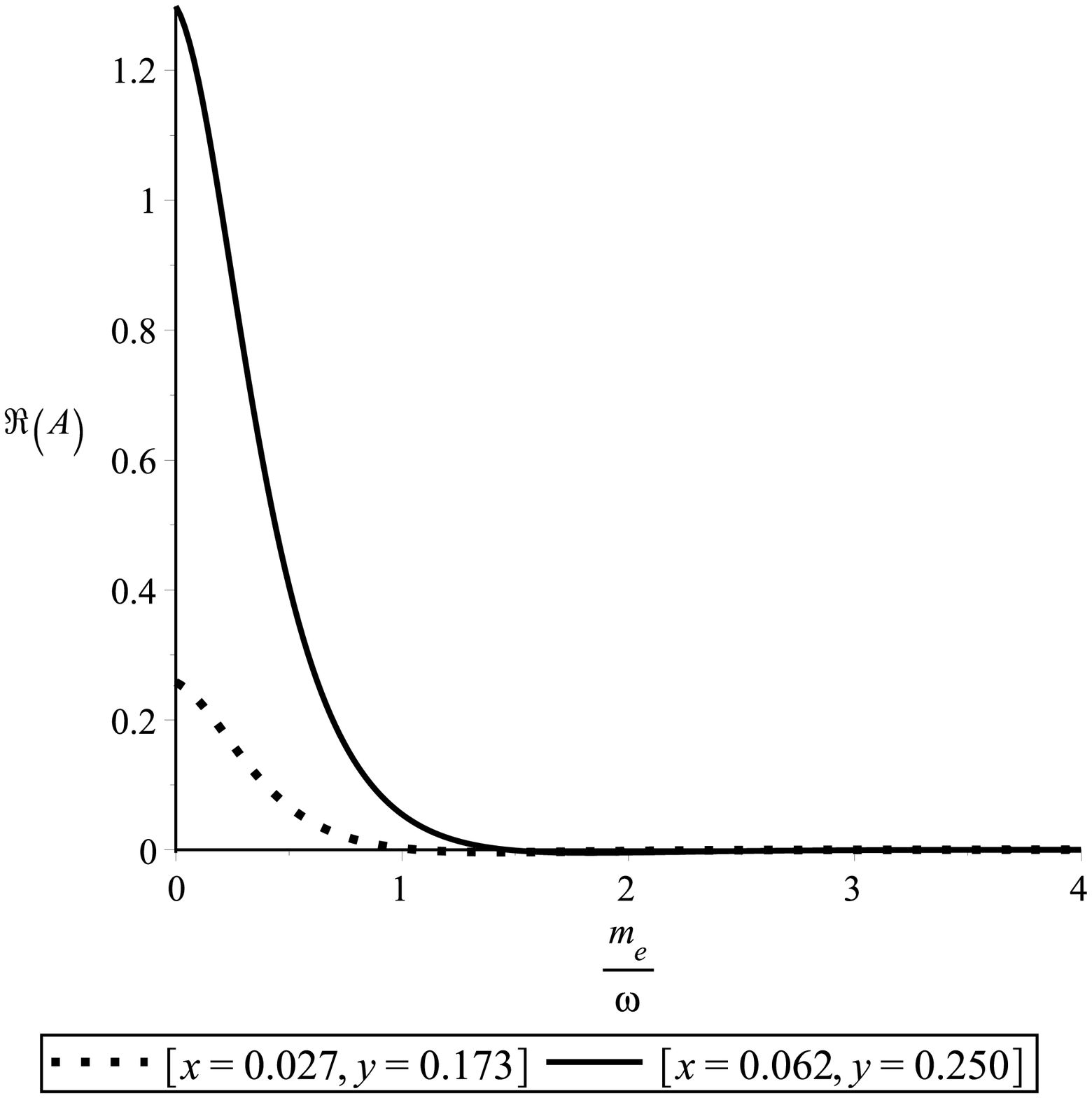}
\includegraphics[scale=0.35]{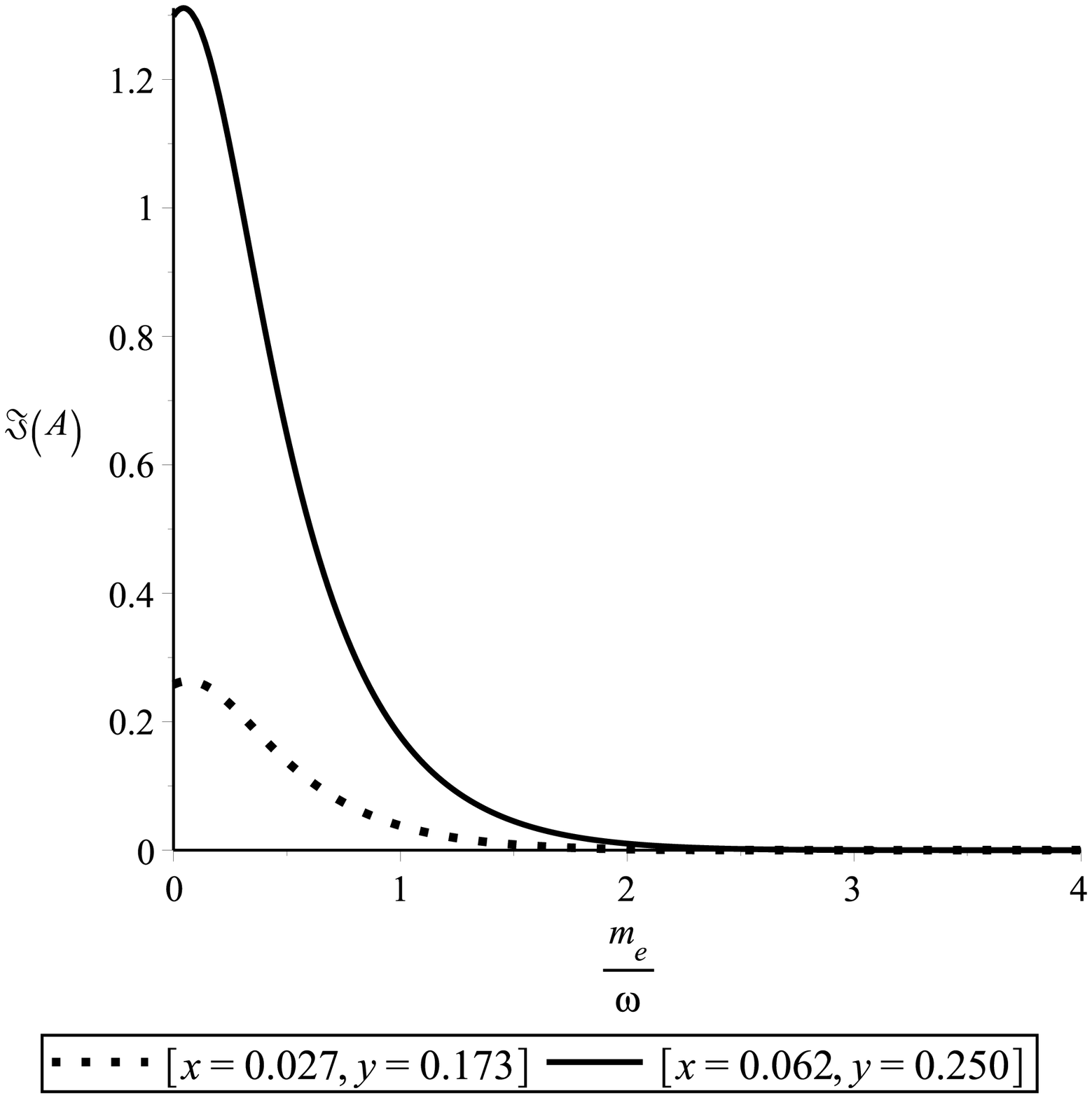}
\caption{The real and imaginary parts of $A$ with respect to $\frac{m_{e}}{\omega}$ for $\frac{M_{W}}{\omega}=0.6$ and momenta values $p' = 0.2, p = 0.1, P = 0.4$ for the straight line and $p' = 0.25, p = 0.1, P = 0.6$ for the dotted line.}
\label{A1m}
\end{figure}

\begin{figure}[H]
\includegraphics[scale=0.35]{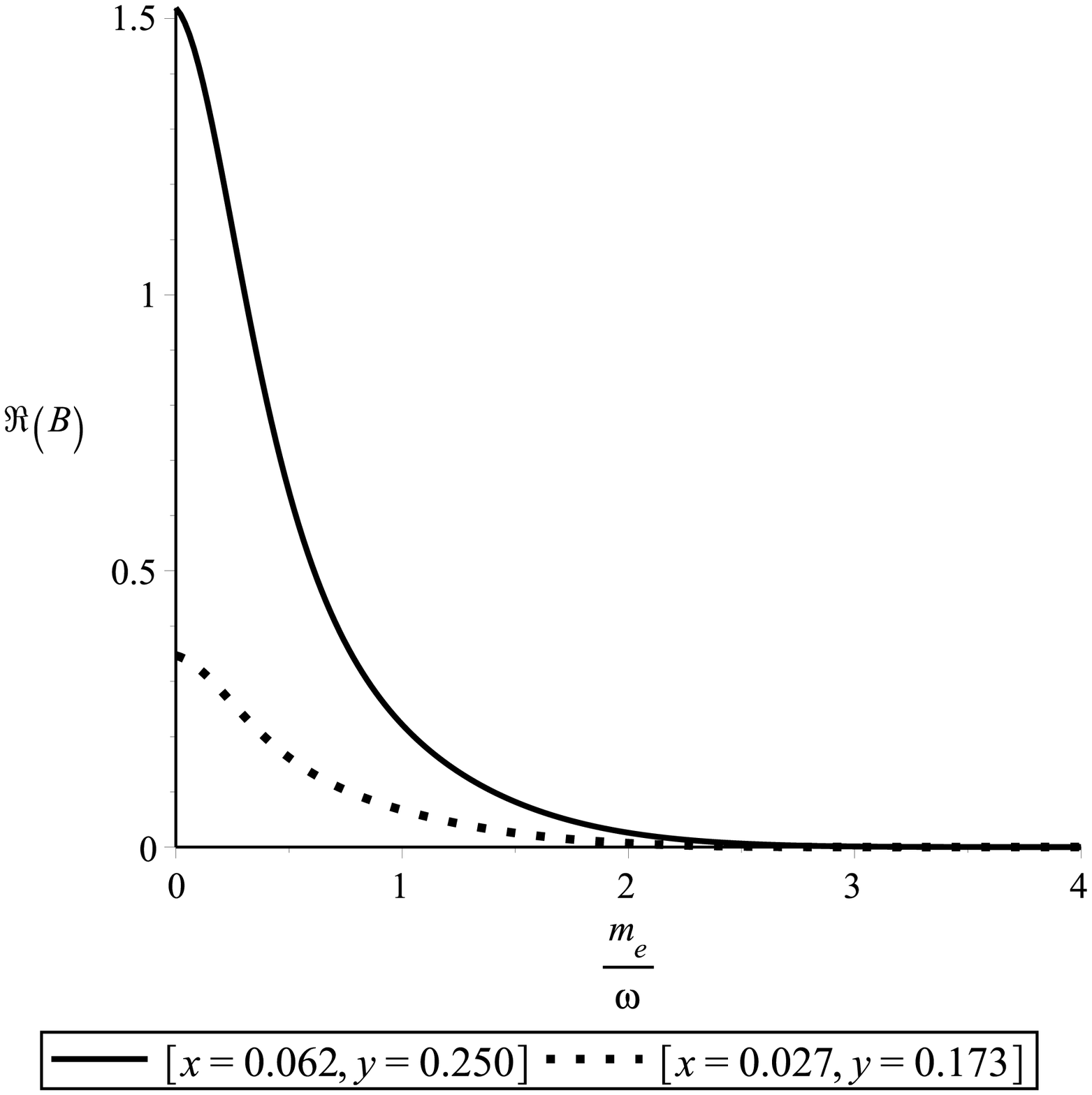}
\includegraphics[scale=0.35]{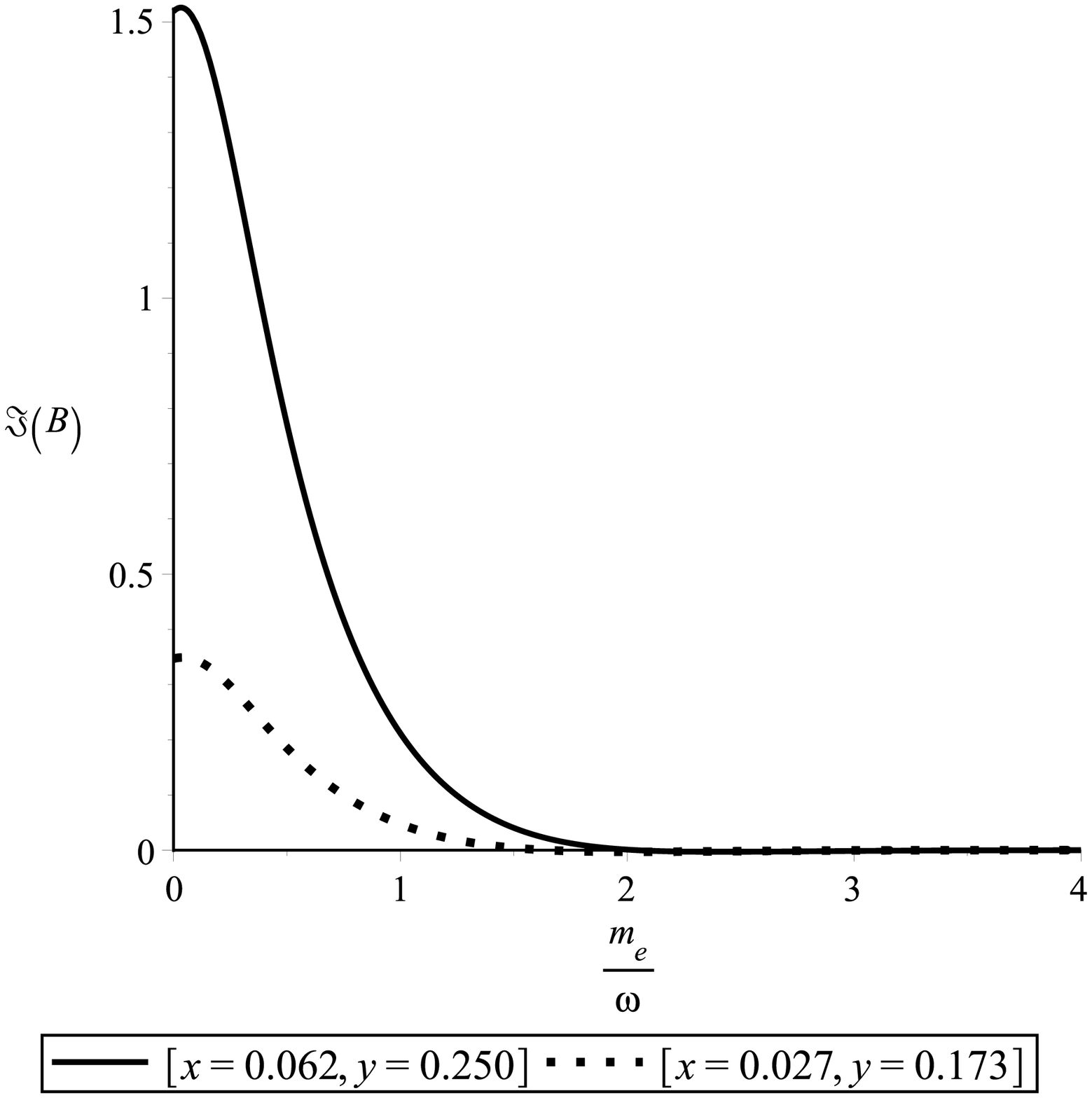}
\caption{The real and imaginary parts of $B$ with respect to $\frac{m_{e}}{\omega}$ for $\frac{M_{W}}{\omega}=0.6$ and momenta values $p' = 0.2, p = 0.1, P = 0.4$ for the straight line and $p' = 0.25, p = 0.1, P = 0.6$ for the dotted line.}
\label{B1m}
\end{figure}

\begin{figure}[H]
\includegraphics[scale=0.35]{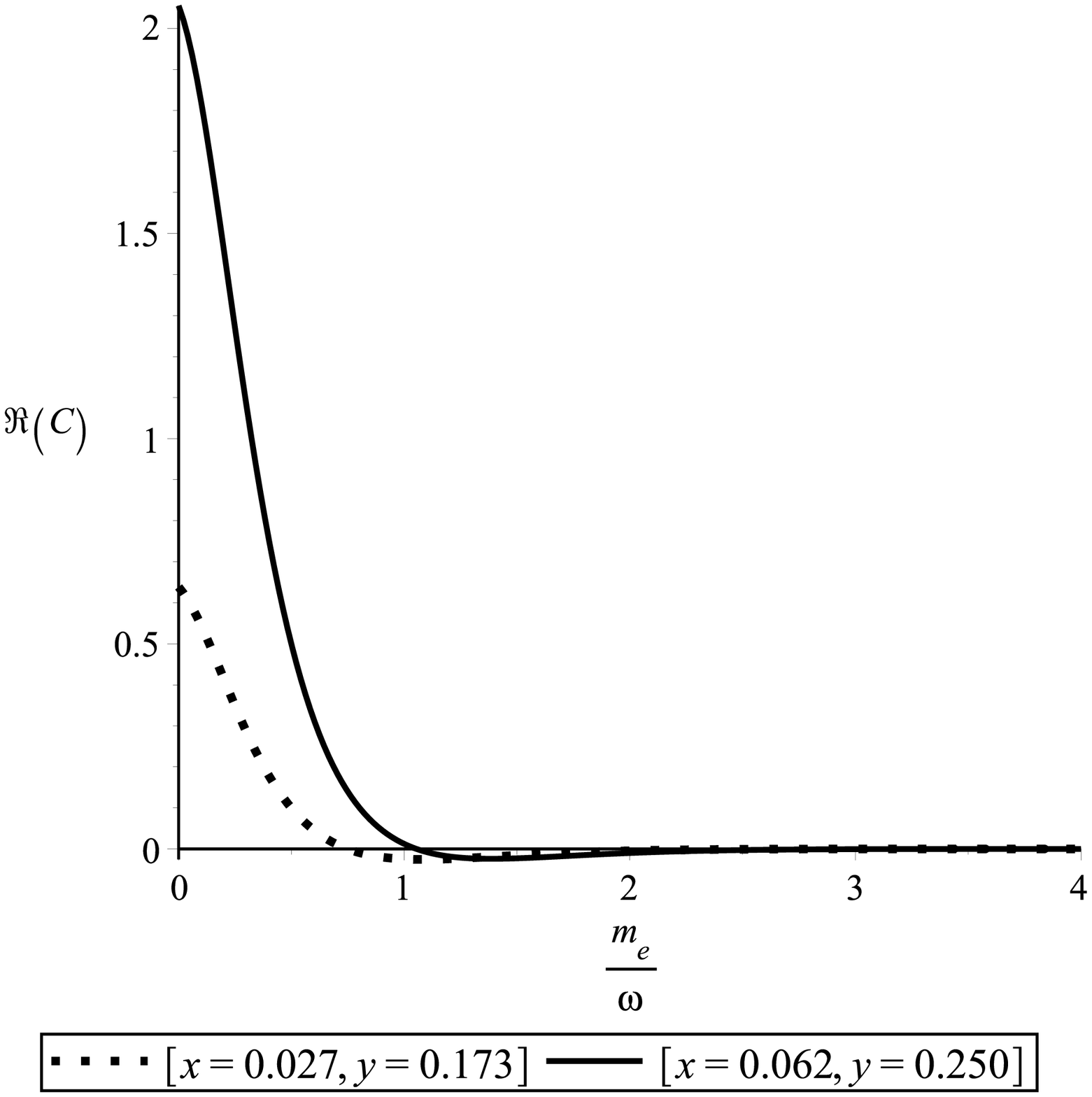}
\includegraphics[scale=0.35]{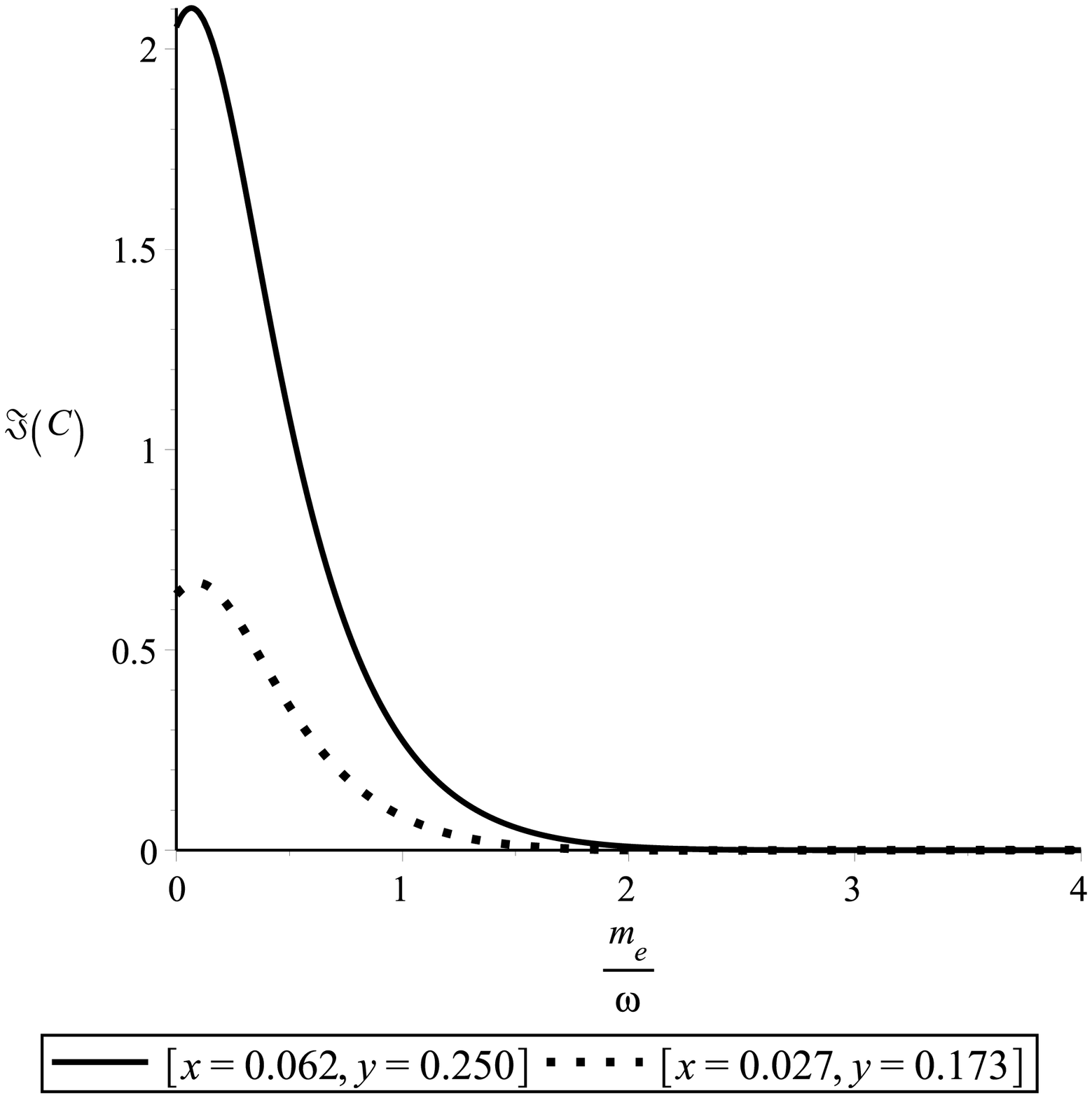}
\caption{The real and imaginary parts of $C$ with respect to $\frac{m_{e}}{\omega}$ for $\frac{M_{W}}{\omega}=0.6$ and momenta values $p' = 0.2, p = 0.1, P = 0.4$ for the straight line and $p' = 0.25, p = 0.1, P = 0.6$ for the dotted line.}
\label{C1m}
\end{figure}

\begin{figure}[H]
\includegraphics[scale=0.35]{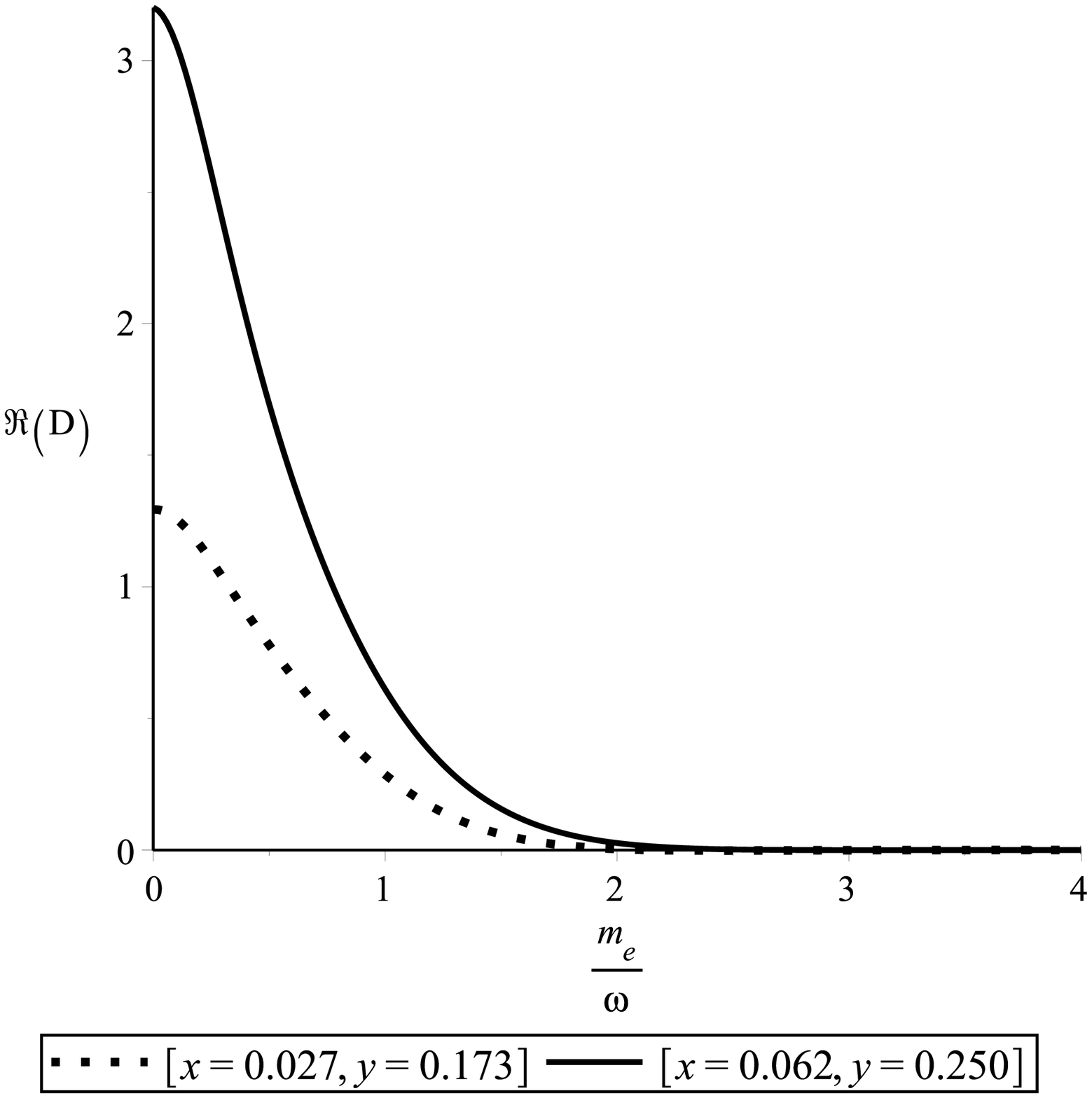}
\includegraphics[scale=0.35]{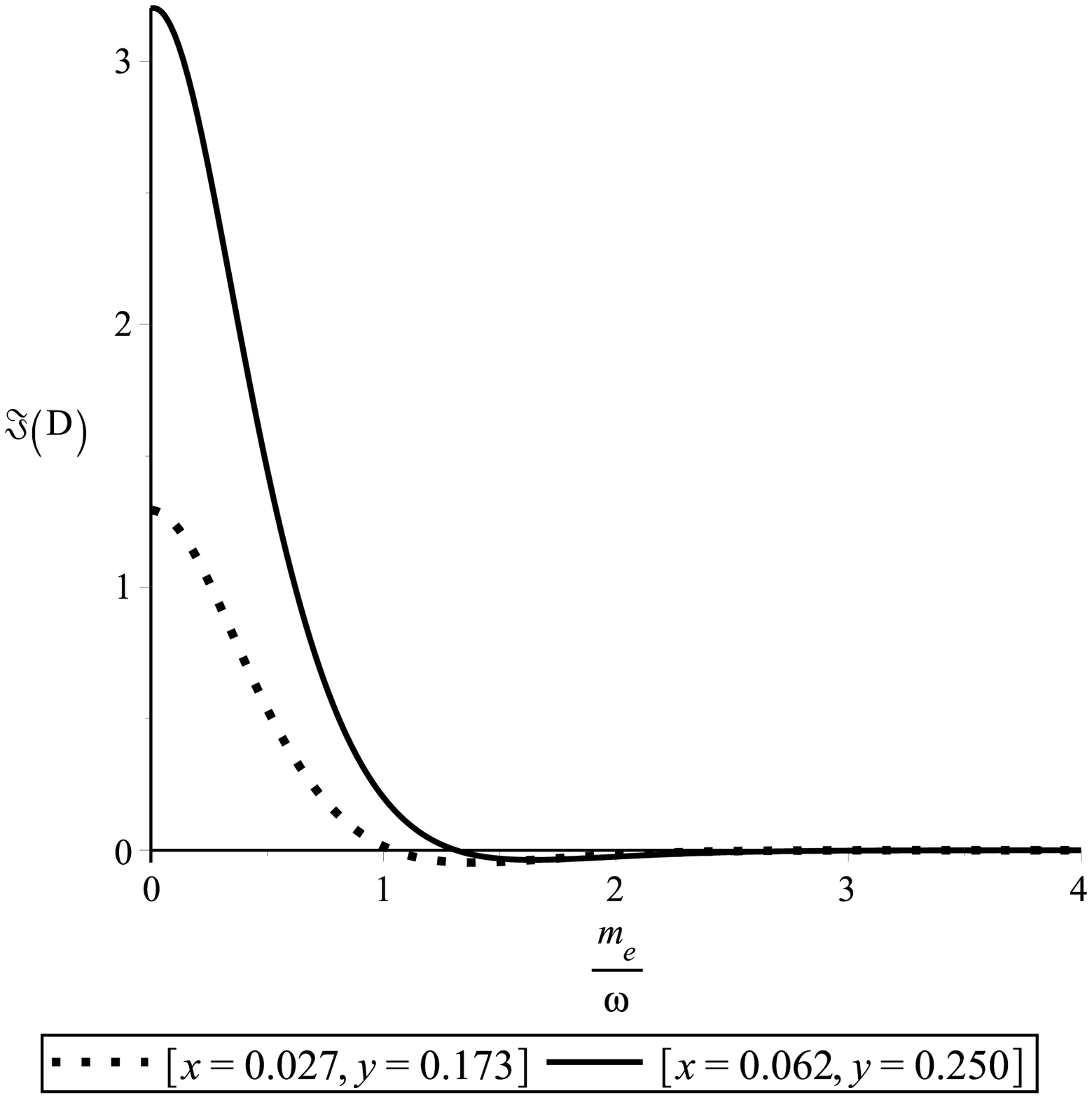}
\caption{The real and imaginary parts of $D$ with respect to $\frac{m_{e}}{\omega}$ for $\frac{M_{W}}{\omega}=0.6$ and momenta values $p' = 0.2, p = 0.1, P = 0.4$ for the straight line and $p' = 0.25, p = 0.1, P = 0.6$ for the dotted line.}
\label{D1m}
\end{figure}

Figures (\ref{A1W}) to (\ref{D1W}) show terms $A, B, C, D$ of the amplitude with respect to $\frac{M_{W}}{\omega}$, while figures (\ref{A1m}) to (\ref{D1m}) show terms $A, B, C, D$ with respect to $\frac{m_{e}}{\omega}$.

The amplitude of the particle production process is significantly higher when $\frac{M_{W}}{\omega}\rightarrow 0$, and wanes as the rest energy of the boson becomes much larger than the energy of the expanding background. This shows that this kind of process could have only taken place in the most early period of the universe, when space-time expansion was exceptionally high. The same observations are valid for the amplitude as it relates to $\frac{m_{e}}{\omega}$. The graphical analysis also proves that as $\frac{M_{W}}{\omega}\rightarrow \infty$ and $\frac{m_{e}}{\omega}\rightarrow \infty$, the transition amplitude vanishes and we recover the Minkowski limit, where this process is forbidden by energy and momentum conservation laws.

\section{The probability for $\lambda = \pm1$}

We can obtain the probability of this process by taking the absolute square value of the amplitude and by summing over all possible helicities, with the observation that in the case of neutrino the helicity is fixed:
\begin{equation}
P_{i\rightarrow f} = \frac{1}{4} \sum_{\sigma'\lambda}|A_{i\rightarrow f}A_{i\rightarrow f}^{*}|.
\end{equation}

This results in:
\begin{eqnarray}
P_{i\rightarrow f} &=&\frac{1}{4} \sum_{\sigma'\lambda} \frac{g^{2}}{4}\frac{\pi}{(2\pi)^{3}}\delta^{3}(\vec{p}+\vec{p'}+\vec{P}) \left(\frac{1}{2}-\sigma\right)^{2}\sigma'^{2} \nonumber \\
&&\times \left| F_{1} F_{1}^{*} \right|
\left|\xi_{\sigma}^{\dag}(\vec{p}\,)\sigma^{i}\eta_{\sigma'}(\vec{p'}\,){\epsilon}^{*}_{i}(\vec{P},\lambda=\pm 1)\right|^{2},
\end{eqnarray}
where we have denoted $F_{1} = (A-B-C+D)$.

We can assess the behaviour of the probability by looking at the absolute value of the product $F_{1}F_{1}^{*}$, which again contains all relevant terms to the evolution of the probability.

\begin{figure}[H]
\includegraphics[scale=0.4]{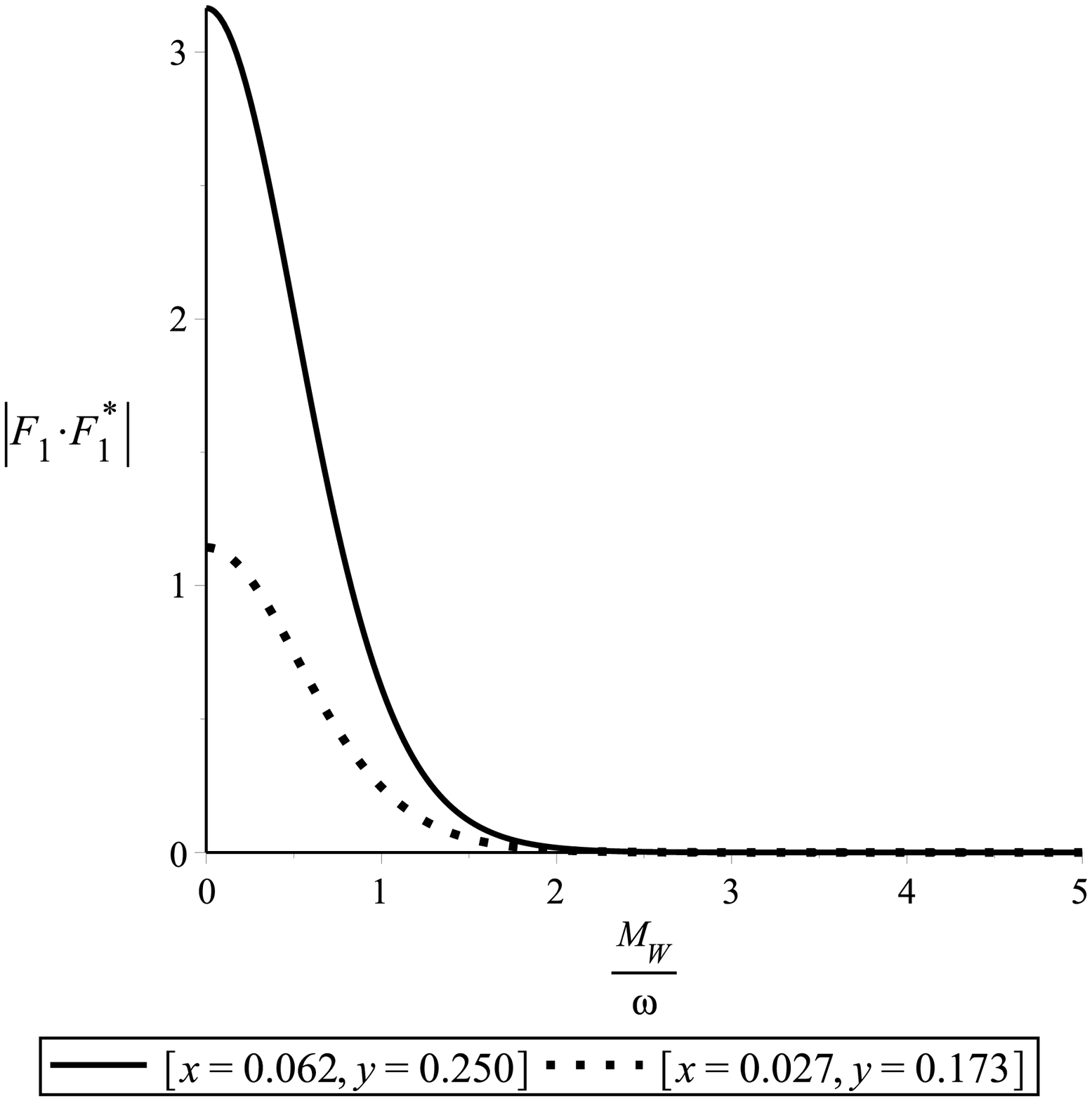}
\includegraphics[scale=0.4]{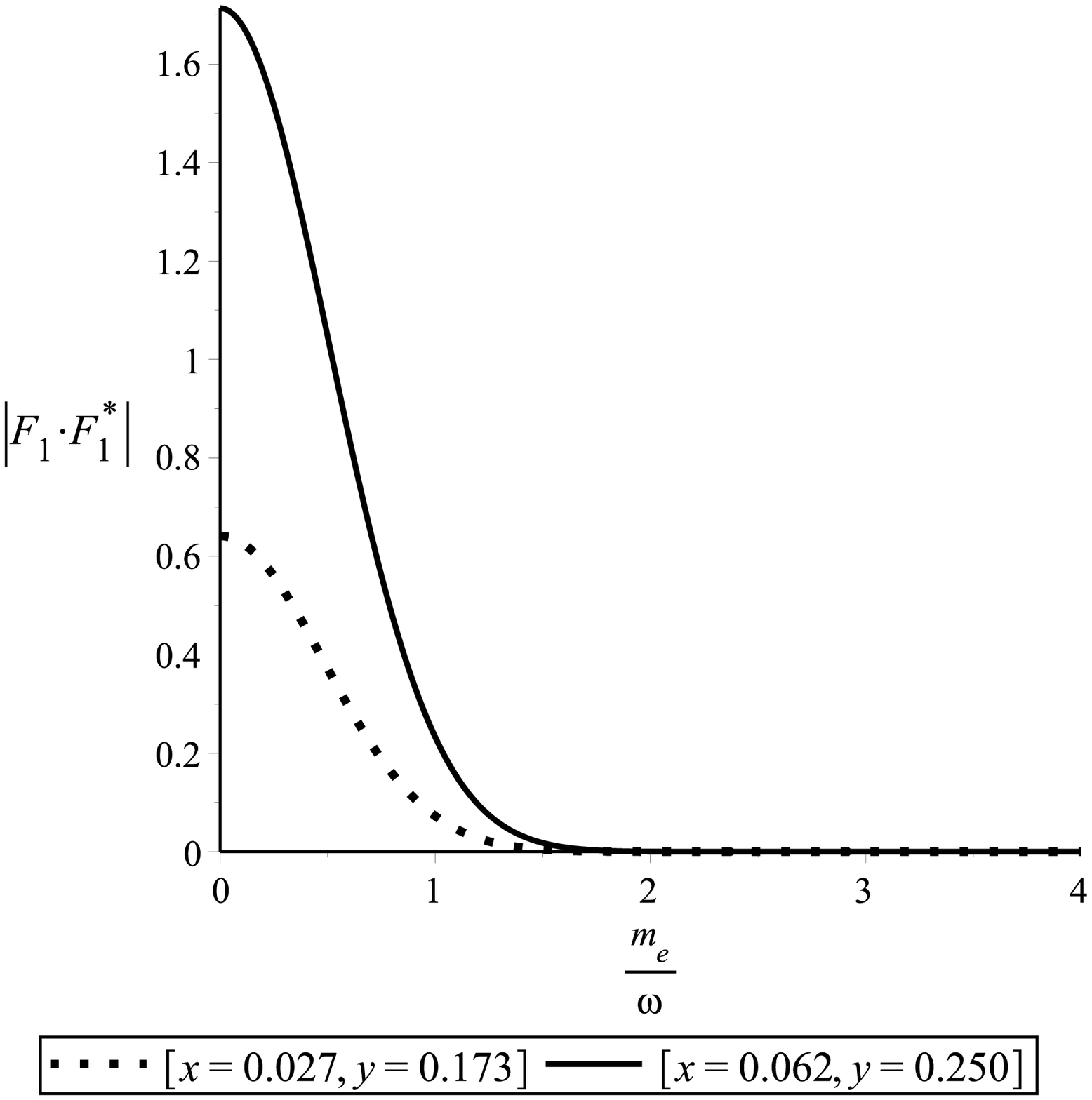}
\caption{The absolute value of $F_{1}$ for $\frac{m_{e}}{\omega}=0.1$ on the left and $\frac{M_{W}}{\omega}=0.6$ on the right, and momenta values $p' = 0.2, p = 0.1, P = 0.4$ for the straight line and $p' = 0.25, p = 0.1, P = 0.6$ for the dotted line.}\label{F1}
\end{figure}

From the graphical analysis of the probability it is clear that the phenomenon of spontaneous generation from vacuum of the massive $W$ bosons and fermions has nonvanishing probability only when the expansion factor exceeds the particle masses, which in our notations refers to small values of parameters $\frac{m_{e}}{\omega},\,\frac{M_{W}}{\omega}$. In the limit of small values of ratios $\frac{m_{e}}{\omega},\,\frac{M_W}{\omega}$ it is, in principle, possible to compute the total probability and total transition rate, which are the important physical quantities that are proportional with the density number of produced particles. For large values of $\frac{m_{e}}{\omega},\,\frac{M_{W}}{\omega}$, the probabilities vanish as expected in the Minkowski limit.

The computation of total probability is related to the dependence of the Appell hypergeometric functions on the momenta, which does not allow us to extract an analytical result when integrating over the final momenta. A solution may be to work with the functions $A,B,C,D$ where we neglect the Appell hypergeometric functions. The numerical and graphical analysis reveals that we can neglect Appell's functions since graphs of the probability in terms of parameters $\frac{m_{e}}{\omega},\,\frac{M_W}{\omega}$ are approximatively the same without them. We intend to approach this subject in a future study. In this way one may obtain the total probability dependence on the parameters $\frac{m_e}{\omega},\,\frac{M_W}{\omega}$.

\subsection{The Minkowski limit}

Let us analyse the Minkowski limit of our results. First our graphical analysis proves that the amplitudes and probability is vanishing for $\frac{m_{e}}{\omega}, \frac{M_W}{\omega} \rightarrow \infty$, which represents the Minkowski limit in our computations.

We start by considering one of the terms:
\begin{align}
A = \frac{\sqrt{i}e^{-\frac{\pi i}{4}}{e^{\pi\frac{k}{2}}}}{\cosh(\pi k)}\frac{\sqrt{2}{pp^{\prime}}^{1-ik}(iP)^{-\frac{7}{2}+ik}}{\Gamma\left(\frac{3}{2}\right)\Gamma\left(\frac{3}{2}-ik\right)}\, \Gamma\left(\frac{\frac{7}{2}-i(k-\frac{M_W}{\omega})}{2}\right)\Gamma\left(\frac{\frac{7}{2}-i(k+\frac{M_W}{\omega})}{2}\right)\nonumber \\
\times\mathcal{F}_{4}\left(\frac{\frac{7}{2}-i(k-\frac{M_W}{\omega})}{2},\frac{\frac{7}{2}-i(k+\frac{M_W}{\omega})}{2},\frac{3}{2},\frac{3}{2}-ik,\frac{p^{2}}{P^{2}},\frac{{p^{\prime}}^{2}}{P^{2}}\right).
\end{align}

Its behaviour is determined mainly by $\cosh^{-1}(\pi k)$ and the Gamma functions. Although the $\mathcal{F}_{4}$ functions diverge to infinity as the expansion parameter approaches zero, the Gamma functions and $cosh^{-1}(\pi k)$ factors decrease much more rapidly to zero, and make it so that term $A$ (and by extension $B$, $C$ and $D$) converges to zero as $\omega \rightarrow 0$, as seen below and also in figures (\ref{A1W}) to (\ref{D1m}).
\begin{figure}[H]
\includegraphics[scale=0.35]{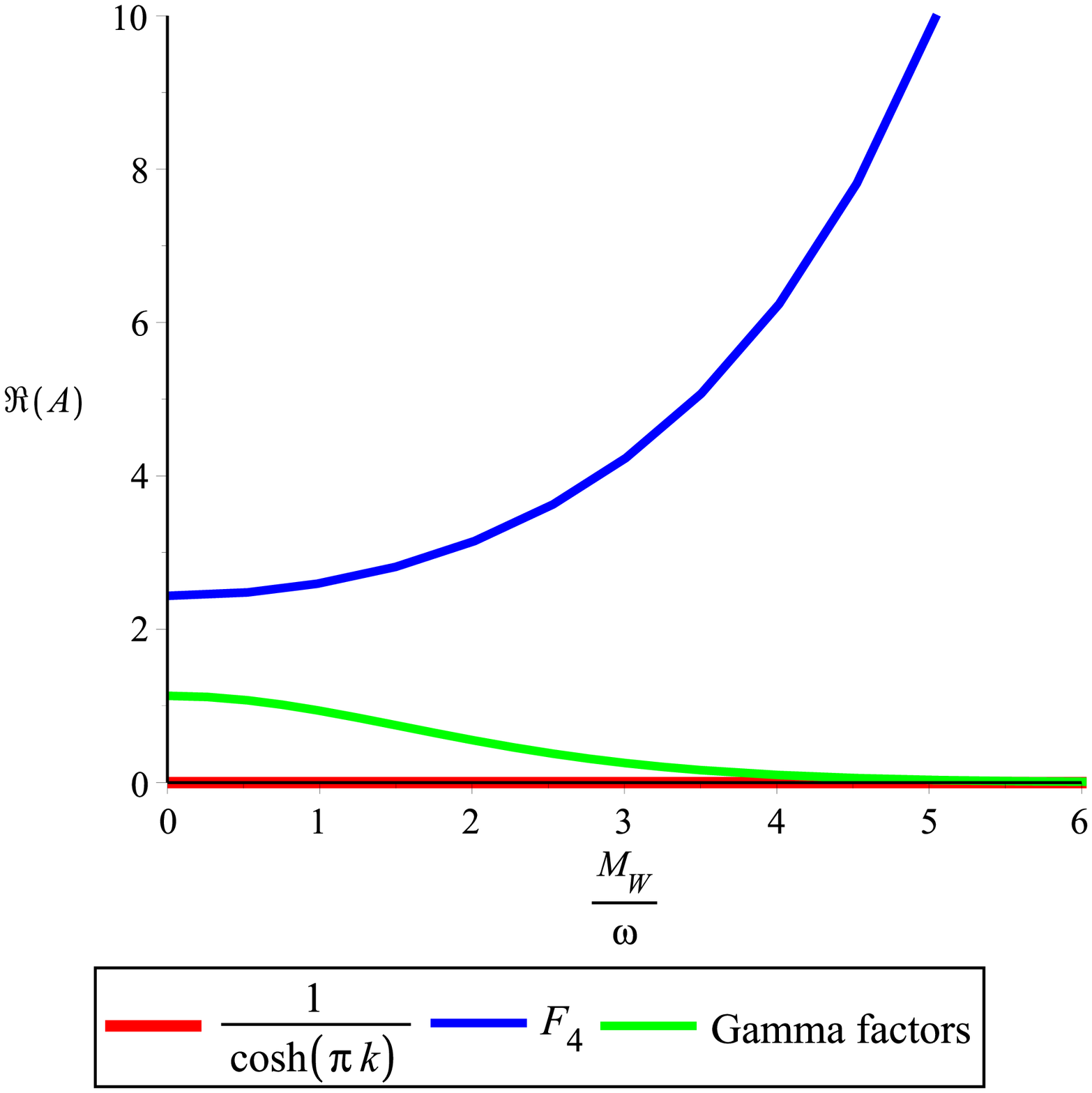}
\includegraphics[scale=0.35]{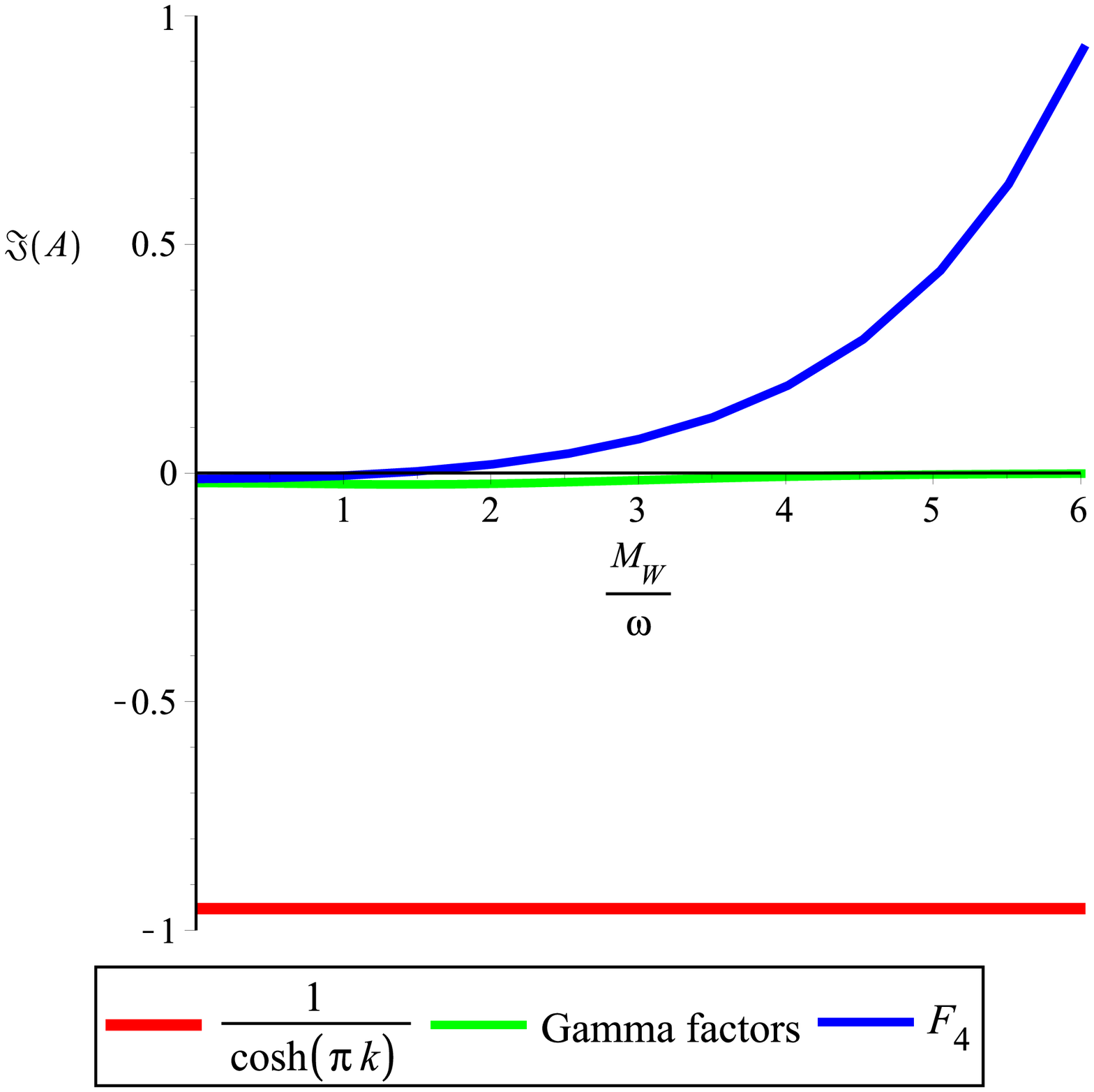}
\label{Abosmink}
\caption{The real and imaginary parts of $A$ with respect to $\frac{M_{W}}{\omega}$ for $k=0.1$ and momenta values $p' = 0.2, p = 0.1, P = 0.4$.}
\end{figure}

\begin{figure}[H]
\includegraphics[scale=0.35]{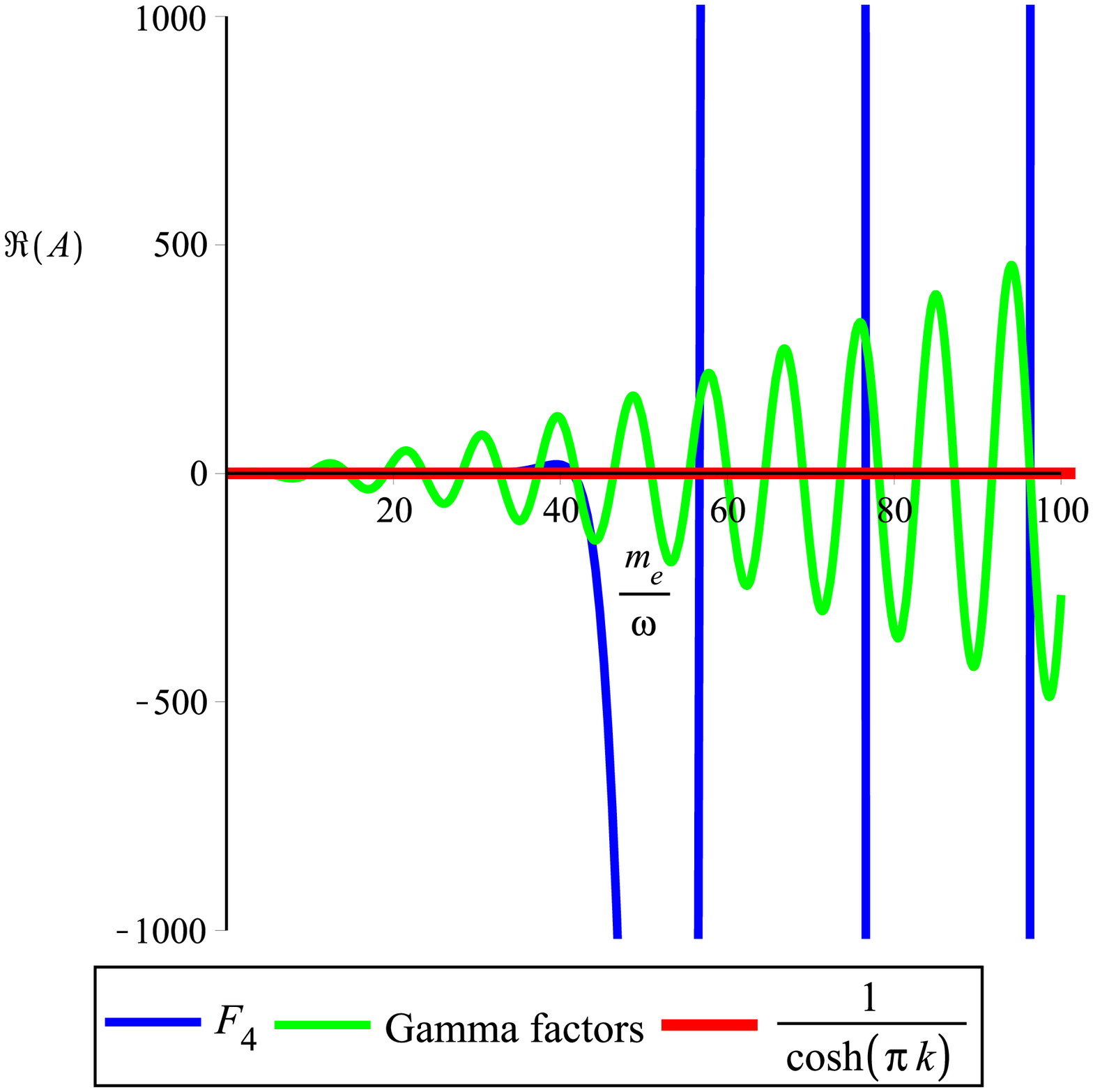}
\includegraphics[scale=0.35]{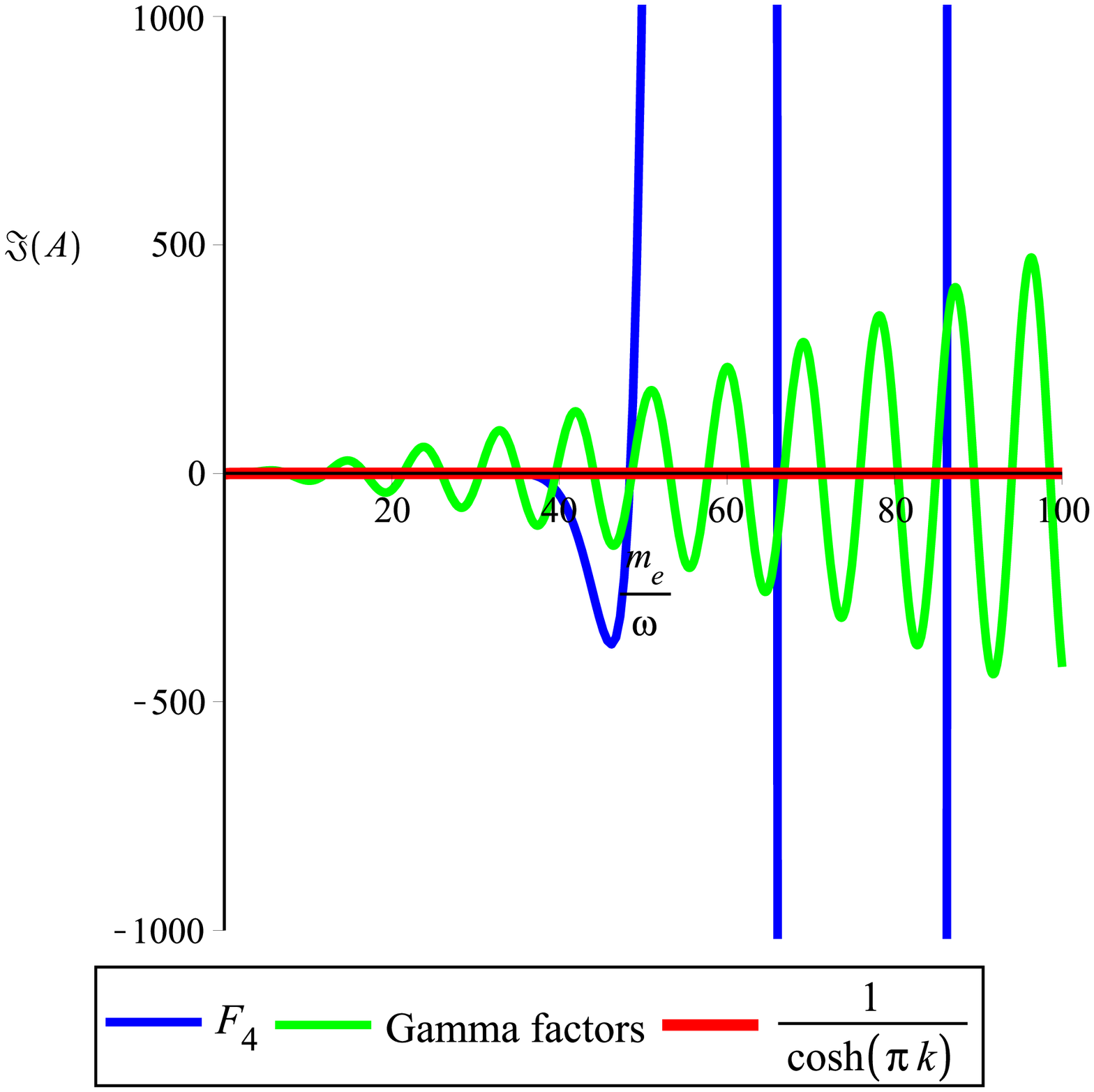}
\label{Aposmink}
\caption{The real and imaginary parts of $A$ with respect to $\frac{m_{e}}{\omega}$ for $\frac{M_{W}}{\omega}=0.6$ and momenta values $p' = 0.2, p = 0.1, P = 0.4$.}
\end{figure}

The $A,B,C,D$ functions that define the amplitude will be analysed for large values of parameters $\frac{m_{e}}{\omega}, \frac{M_{W}}{\omega}$. The parameter $K$ can be approximated as $K=\sqrt{\frac{M_W^2}{\omega^2}-\frac{1}{4}}\simeq \frac{M_W}{\omega}$ for $M_W\gg\omega$.

The Gamma functions can be further approximated by using the Stirling formula for $z \rightarrow \infty $ \cite{AS,21}:
\begin{equation}
\Gamma(z) \simeq e^{-z}z^{z}\left(\frac{2\pi}{z}\right)^{1/2}.
\end{equation}

We will transform the following functions for large arguments:
\begin{equation}
\Gamma\left(\frac{3}{2} - ik\right) = e^{-\frac{3}{2}+ik}\left(\frac{3}{2}-ik\right)^{1-ik}\sqrt{2\pi},
\end{equation}
and use
\begin{equation}
i^{-\frac{7}{2}+ik} = e^{-\frac{7i\pi}{4}}e^{-\frac{\pi k }{2}}.
\end{equation}

This will allow us to write down the gamma Euler functions as:
\begin{eqnarray}
&&\Gamma\left(\frac{7}{4}-\frac{i(k+\frac{M_W}{\omega})}{2}\right)\Gamma\left(\frac{7}{4}-\frac{i(k-\frac{M_W}{\omega})}{2}\right)=\frac{8\pi e^{-\frac{7}{2}+ik}}{\sqrt{49-4k^2-28ik+4\frac{M_W^2}{\omega^2}}}\nonumber\\
&&\times\left(\frac{7}{4} - \frac{i(k+\frac{M_W}{\omega})}{2}\right)^{\frac{7}{4}-\frac{i(k+\frac{M_W}{\omega})}{2}}
\left(\frac{7}{4} - \frac{i(k-\frac{M_W}{\omega})}{2}\right)^{\frac{7}{4}-\frac{i(k-\frac{M_W}{\omega})}{2}}.
\end{eqnarray}

Then for $M_{W}/\omega \gg 1$ and $m_{e}/\omega \gg 1$ we have:
\begin{align}
A \simeq \frac{8\sqrt{i\pi} e^{-2-2\pi i}p (p^{\prime})^{1-ik} (P)^{-\frac{7}{2}+ik}}{e^{\pi k}\Gamma\left(\frac{3}{2}\right)\left(\frac{3}{2}-ik\right)^{1-ik}}
\frac{\left(\frac{7}{4} - \frac{i(k+\frac{M_W}{\omega})}{2}\right)^{\frac{7}{4}-\frac{i(k+\frac{M_W}{\omega})}{2}}\left(\frac{7}{4} - \frac{i(k-\frac{M_W}{\omega})}{2}\right)^{\frac{7}{4}-\frac{i(k-\frac{M_W}{\omega})}{2}}}{\sqrt{49-4k^2-28ik+4\frac{M_W^2}{\omega^2}}}\nonumber\\
\times\mathcal{F}_{4}\left(\frac{\frac{7}{2}-i(k-\frac{M_W}{\omega})}{2},\frac{\frac{7}{2}-i(k+\frac{M_W}{\omega})}{2},\frac{3}{2},\frac{3}{2}-ik,\frac{p^{2}}{P^{2}},\frac{{p^{\prime}}^{2}}{P^{2}}\right).
\end{align}

The rest of the functions $B,C,D$ have similar expressions:
\begin{align}
B\simeq \frac{8\sqrt{i\pi} e^{-2-3\pi i/2}p (p^{\prime})^{ik} (P)^{-\frac{5}{2}-ik}}{ie^{\pi k}\Gamma\left(\frac{3}{2}\right)\left(\frac{1}{2}+ik\right)^{ik}}
\frac{\left(\frac{5}{4} + \frac{i(k+\frac{M_W}{\omega})}{2}\right)^{\frac{5}{4}+\frac{i(k+\frac{M_W}{\omega})}{2}}\left(\frac{5}{4} + \frac{i(k-\frac{M_W}{\omega})}{2}\right)^{\frac{5}{4}+\frac{i(k-\frac{M_W}{\omega})}{2}}}{\sqrt{25-4k^2+20ik+4\frac{M_W^2}{\omega^2}}}\nonumber\\
\times\mathcal{F}_{4}\left(\frac{\frac{5}{2}+i(k+\frac{M_W}{\omega})}{2},\frac{\frac{5}{2}+i(k-\frac{M_W}{\omega})}{2},\frac{3}{2},\frac{1}{2}+ik,\frac{p^{2}}{P^{2}},\frac{{p^{\prime}}^{2}}{P^{2}}\right);
\end{align}

\begin{align}
C \simeq \frac{8\sqrt{i\pi}e^{-1-3\pi i/2}(p^{\prime})^{1-ik} (P)^{-\frac{5}{2}+ik}}{ie^{\pi k}\Gamma\left(\frac{1}{2}\right)\left(\frac{3}{2}-ik\right)^{1-ik}}
\frac{\left(\frac{5}{4} - \frac{i(k-\frac{M_W}{\omega})}{2}\right)^{\frac{5}{4}-\frac{i(k-\frac{M_W}{\omega})}{2}}\left(\frac{5}{4} - \frac{i(k+\frac{M_W}{\omega})}{2}\right)^{\frac{5}{4}-\frac{i(k+\frac{M_W}{\omega})}{2}}}{\sqrt{25-4k^2-20ik+4\frac{M_W^2}{\omega^2}}}\nonumber\\
\times\mathcal{F}_{4}\left(\frac{\frac{5}{2}-i(k-\frac{M_W}{\omega})}{2},\frac{\frac{5}{2}-i(k+\frac{M_W}{\omega})}{2},\frac{1}{2},\frac{3}{2}-ik,\frac{p^{2}}{P^{2}},\frac{{p^{\prime}}^{2}}{P^{2}}\right);
\end{align}

\begin{align}
D\simeq-\frac{8\sqrt{i\pi} e^{-1-i\pi}(p^{\prime})^{ik} (P)^{-\frac{3}{2}-ik}}{e^{\pi k}\Gamma\left(\frac{1}{2}\right)\left(\frac{1}{2}+ik\right)^{ik}}
\frac{\left(\frac{3}{4} + \frac{i(k+\frac{M_W}{\omega})}{2}\right)^{\frac{3}{4}+\frac{i(k+\frac{M_W}{\omega})}{2}}\left(\frac{3}{4} + \frac{i(k-\frac{M_W}{\omega})}{2}\right)^{\frac{3}{4}+\frac{i(k-\frac{M_W}{\omega})}{2}}}{\sqrt{9-4k^2+12ik+4\frac{M_W^2}{\omega^2}}}\nonumber\\
\times\mathcal{F}_{4}\left(\frac{\frac{3}{2}+i(k+\frac{M_W}{\omega})}{2},\frac{\frac{3}{2}+i(k-\frac{M_W}{\omega})}{2},\frac{1}{2},\frac{1}{2}+ik,\frac{p^{2}}{P^{2}},\frac{{p^{\prime}}^{2}}{P^{2}}\right).
\end{align}

We obtain that in the limit $\frac{M_{W}}{\omega} \gg 1,\, \frac{m_{e}}{\omega} \gg 1$, terms $A,B,C,D$ decrease to zero like the functions $e^{-\pi m_{e}/\omega}$ and $\frac{1}{M_W/\omega}$ multiplied by factors at imaginary powers. It is also important to mention that the behaviour of these functions in terms of $m_{e}/\omega, M_W/\omega$ is mainly determined by the factors which multiply the Appell hypergeometric functions. These factors cancel any divergent behaviour of Appell's functions in the considered limit. The transition probabilitiy behaves like $e^{-2\pi m_{e}/\omega}$ and $\frac{1}{(M_{W}/\omega)^2}$, assuring the Minkowski limit when the ratios between particle masses and expansion factor become infinite. In the Minkowski theory, the amplitude and probability vanish due to the simultaneous conservation of momentum and energy \cite{12,19,20}.

\section{Transition amplitude and probability for $\lambda = 0$}

The amplitude defined in equation (\ref{ampl}) contains both spatial and temporal contributions in the case of longitudinal polarization, because the solution of the Proca equation has a nonvanishing temporal component in this case \cite{2}:
\begin{equation}
{f_{\vec{p}\lambda}}_{0}(\vec{x}) = \frac{\sqrt{\pi}e^{-K\pi/2}\omega P}{2(2\pi)^{3/2}M_W}(-t_c)^{3/2}\mathcal{H}_{iK}^{(1)}(-Pt_c)e^{i\vec{P}\vec{x}}.
\end{equation}

Moreover, for $\lambda=0$ the spatial component of the solution also has two terms \cite{2}:
\begin{equation}
{f_{\vec{p}\lambda}}_{i} (\vec{x}) = \frac{i\sqrt{\pi}e^{-k\pi/2}\omega P}{2(2\pi)^{3/2}M_W}\left[\left(\frac{1}{2}+iK\right)\frac{\sqrt{-t_c}}{P}\mathcal{H}_{iK}^{(1)}(-Pt_c) - t_{c}^{3/2}\mathcal{H}_{1+iK}^{(1)}(-Pt_c)\right]\epsilon_{i}(\vec{K},\lambda)e^{i\vec{P}\vec{x}}.
\end{equation}

In this case the amplitude reads:
\begin{eqnarray}
A_{i\rightarrow f}= \frac{ig}{2\sqrt{2}} \int d^{4}x \sqrt{-g(x)}  \bar{u}_{\vec{p}\sigma}(x) \gamma^{\hat{\alpha}} e_{\hat{\alpha}}^{\mu}(1-\gamma^5) v_{\vec{p^{\prime}}\sigma^{\prime}}(x){f^{*}_{\vec{P}\lambda}}_{\mu} (x)\nonumber\\
=\frac{ig}{2\sqrt{2}} \int d^{4}x \sqrt{-g(x)}  \bar{u}_{\vec{p}\sigma}(x) \gamma^{\hat{0}} e_{\hat{0}}^{0}(1-\gamma^5) v_{\vec{p^{\prime}}\sigma^{\prime}}(x){f^{*}_{\vec{P}\lambda}}_{0} (x)\nonumber\\
+\frac{ig}{2\sqrt{2}} \int d^{4}x \sqrt{-g(x)}  \bar{u}_{\vec{p}\sigma}(x) \gamma^{\hat{i}} e_{\hat{i}}^j(1-\gamma^5) v_{\vec{p^{\prime}}\sigma^{\prime}}(x){f^{*}_{\vec{P}\lambda}}_{j} (x).
\end{eqnarray}

Because of the nature of the Proca solutions \cite{2}, we are going to split the amplitude into three terms, for convenience:
\begin{equation}
A_{i\rightarrow f}(\lambda=0) = A_{i\rightarrow f}^{(1)} (\lambda = 0)+ A_{i\rightarrow f}^{(2)} (\lambda = 0) +  A_{i\rightarrow f}^{(3)} (\lambda = 0).
\end{equation}

The first term will contain the temporal component and we obtain something very similar to the case of $\lambda =\pm 1$:
\begin{eqnarray}
A_{i\rightarrow f}^{(1)} (\lambda = 0) &=& -\frac{g}{2\sqrt{2}}\frac{\omega P}{M_{W}}\frac{ \pi \sqrt{p^{\prime}} e^{-\pi K/2}e^{-\pi k/2}}{(2\pi)^{3/2}}\delta^{3}(\vec{p}+\vec{p^{\prime}}+\vec{P}) \left(\frac{1}{2}-\sigma\right)\sigma^{\prime} \nonumber \\ &&\times\int_{0}^{+\infty} dz \cdot z^{2} e^{-ipz}\mathcal{H}_{\nu_{-}}^{(2)}(p^{\prime}z)\mathcal{H}_{-iK}^{(2)}(P z)\xi_{\sigma}^{\dag}(\vec{p}\,)\eta_{\sigma^{\prime}}(\vec{p^{\prime}}).
\end{eqnarray}

This term can also be rewritten after transforming the Hankel functions into Bessel functions \cite{AS}:
\begin{eqnarray}
A_{i\rightarrow f}^{(1)} (\lambda = 0) &=& -\frac{g}{2}\frac{\sqrt{\pi} P\sqrt{pp^{\prime}}\sqrt{i} e^{-\pi k/2}}{(2\pi)^{3/2}}\frac{\omega}{M_{W}}\delta^{3}(\vec{p}+\vec{p^{\prime}}+\vec{P}) \left(\frac{1}{2}-\sigma\right)\sigma^{\prime} \frac{e^{-i\frac{\pi}{4}}}{i\, \cosh(\pi k)}\nonumber \\ &&\times [T_{5}-T_{6}-T_{7}+T_{8}] \xi_{\sigma}^{\dag}(\vec{p}\,)\eta_{\sigma^{\prime}}(\vec{p^{\prime}}),
\end{eqnarray}
where the terms $T_{5},T_{6},T_{7}$ and $T_{8}$ denote the following integrals:
\begin{eqnarray}
T_{5}&=&\int_{0}^{\infty} dz\, z^2\,\sqrt{z}\,\mathcal{J}_{\frac{1}{2}}(pz)ie^{\pi k}\mathcal{J}_{\frac{1}{2}-ik}(p^{\prime}z)\mathcal{K}_{-iK}(iPz); \\
T_{6}&=&\int_{0}^{\infty} dz\, z^2\,\sqrt{z}\,\mathcal{J}_{\frac{1}{2}}(pz)\mathcal{J}_{-\frac{1}{2}+ik}(p^{\prime}z)\mathcal{K}_{-iK}(iPz);\\
T_{7}&=&\int_{0}^{\infty} dz\, z^2\,\sqrt{z}\,e^{\pi k}\mathcal{J}_{-\frac{1}{2}}(pz)\mathcal{J}_{\frac{1}{2}-ik}(p^{\prime}z)\mathcal{K}_{-iK}(iPz); \\
T_{8}&=&\int_{0}^{\infty} dz\, z^2\,\sqrt{z}\,\frac{1}{i}\mathcal{J}_{-\frac{1}{2}}(pz)\mathcal{J}_{-\frac{1}{2}+ik}(p^{\prime}z)\mathcal{K}_{-iK}(iPz).
\end{eqnarray}

After integration using formula (\ref{appell}) from Appendix the results are:
\begin{eqnarray}
T_{5}=\frac{ie^{\pi k}{2}^{3/2}p^{\frac{1}{2}}{p^{\prime}}^{\frac{1}{2}-ik}(iP)^{-\frac{9}{2}+ik}}{\Gamma\left(\frac{3}{2}\right)\Gamma\left(\frac{3}{2}-ik\right)} \Gamma\left(\frac{\frac{9}{2}-i(k-K)}{2}\right)\Gamma\left(\frac{\frac{9}{2}-i(k+K)}{2}\right)\nonumber \\
\times \mathcal{F}_{4}\left(\frac{\frac{9}{2}-i(k-K)}{2},\frac{\frac{9}{2}-i(k+K)}{2},\frac{3}{2},\frac{3}{2}-ik,\frac{p^{2}}{P^{2}},\frac{{p^{\prime}}^{2}}{P^{2}}\right);\\
T_{6}=\frac{{2}^{3/2}p^{\frac{1}{2}}{p^{\prime}}^{-\frac{1}{2}+ik}(iP)^{-\frac{7}{2}-ik}}{\Gamma\left(\frac{3}{2}\right)\Gamma\left(\frac{1}{2}+ik\right)} \Gamma\left(\frac{\frac{7}{2}+i(k+K)}{2}\right)\Gamma\left(\frac{\frac{7}{2}+i(k-K)}{2}\right)\nonumber \\
\times\mathcal{F}_{4}\left(\frac{\frac{7}{2}+i(k-K)}{2},\frac{\frac{7}{2}+i(k-K)}{2},\frac{3}{2},\frac{1}{2}+ik,\frac{p^{2}}{P^{2}},\frac{{p^{\prime}}^{2}}{P^{2}}\right);\\
T_{7}=\frac{e^{\pi k}{2}^{3/2}p^{-\frac{1}{2}}{p^{\prime}}^{\frac{1}{2}-ik}(iP)^{-\frac{7}{2}+ik}}{\Gamma\left(\frac{1}{2}\right)\Gamma\left(\frac{3}{2}-ik\right)} \Gamma\left(\frac{\frac{7}{2}-i(k-K)}{2}\right)\Gamma\left(\frac{\frac{7}{2}-i(k+K)}{2}\right)\nonumber \\
\times\mathcal{F}_{4}\left(\frac{\frac{7}{2}-i(k-K)}{2},\frac{\frac{7}{2}-i(k+K)}{2},\frac{1}{2},\frac{3}{2}-ik,\frac{p^{2}}{P^{2}},\frac{{p^{\prime}}^{2}}{P^{2}}\right);\\
T_{8}=\frac{\frac{1}{i}2^{3/2}p^{-\frac{1}{2}}{p^{\prime}}^{-\frac{1}{2}+ik}(iP)^{-\frac{5}{2}-ik}}{\Gamma\left(\frac{1}{2}\right)\Gamma\left(\frac{1}{2}+ik\right)} \Gamma\left(\frac{\frac{5}{2}+i(k+K)}{2}\right)\Gamma\left(\frac{\frac{5}{2}+i(k-K)}{2}\right)\nonumber \\
\times\mathcal{F}_{4}\left(\frac{\frac{5}{2}+i(k+K)}{2},\frac{\frac{5}{2}+i(k-K)}{2},\frac{1}{2},\frac{1}{2}+ik,\frac{p^{2}}{P^{2}},\frac{{p^{\prime}}^{2}}{P^{2}}\right).
\end{eqnarray}

The second term and third terms of the amplitude sum over the spatial components of the Proca solution:
\begin{eqnarray}
A_{i\rightarrow f}^{(2)} (\lambda = 0) &=& -\frac{g}{2\sqrt{2}}\frac{\omega P}{M_{W}}\frac{ i\pi \sqrt{p^{\prime}} e^{-\pi K/2}e^{-\pi k/2}}{(2\pi)^{3/2}}\delta^{3}(\vec{p}+\vec{p^{\prime}}+\vec{P}) \left(\frac{1}{2}-\sigma\right)\sigma^{\prime}\frac{\left(\frac{1}{2}-iK\right)}{P} \cdot \nonumber \\ &&\times\int_{0}^{+\infty} dz \,z\, e^{-ipz}\mathcal{H}_{\nu_{-}}^{(2)}(p^{\prime}z)\mathcal{H}_{-iK}^{(2)}(P z)\xi_{\sigma}^{\dag}(\vec{p}\,)\sigma^{i}\eta_{\sigma^{\prime}}(\vec{p^{\prime}}){\epsilon}^{*}_{i}(\vec{P},\lambda= 0).\nonumber\\
\end{eqnarray}

This second term is actually the amplitude for $\lambda=\pm 1$ up to a an imaginary factor and the ratio of the expansion parameter to the boson mass:
\begin{eqnarray}
A_{i\rightarrow f}^{(2)} &=& -i\left(\frac{1}{2} - iK\right)\frac{g}{2}\frac{\omega}{M_{W}}\frac{\sqrt{\pi} \sqrt{pp^{\prime}}\sqrt{i} e^{-\pi k/2}}{(2\pi)^{3/2}}\delta^{3}(\vec{p}+\vec{p^{\prime}}+\vec{P}) \left(\frac{1}{2}-\sigma\right)\sigma^{\prime} \frac{e^{-i\frac{\pi}{4}}}{i\,\cosh(\pi k)}\nonumber \\ &&\times [T_{1}-T_{2}-T_{3}+T_{4}] \xi_{\sigma}^{\dag}(\vec{p}\,)\sigma^{i}\eta_{\sigma^{\prime}}(\vec{p^{\prime}}){\epsilon}^{*}_{i}(\vec{P},\lambda=\pm 0),
\end{eqnarray}
where $T_1,T_2,T_3,T_4$ are defined in equation (\ref{ttt}).

The third term is very close to the first term of $A_{i\rightarrow f}(\lambda=0)$, with the exception of the Hankel function dependent on $P$:
\begin{eqnarray}
A_{i\rightarrow f}^{(3)} (\lambda = 0) &=& \frac{g}{2\sqrt{2}}\frac{\omega P}{M_{W}}\frac{i \pi \sqrt{p^{\prime}} e^{-\pi K/2}e^{-\pi k/2}}{(2\pi)^{3/2}}\delta^{3}(\vec{p}+\vec{p^{\prime}}+\vec{P}) \left(\frac{1}{2}-\sigma\right)\sigma^{\prime}  \nonumber \\ &&\times\int_{0}^{+\infty} dz \cdot z^{2} e^{-ipz}\mathcal{H}_{\nu_{-}}^{(2)}(p^{\prime}z)\mathcal{H}_{1-iK}^{(2)}(P z)\xi_{\sigma}^{\dag}(\vec{p}\,)\sigma^{i}\eta_{\sigma^{\prime}}(\vec{p^{\prime}}){\epsilon}^{*}_{i}(\vec{P},\lambda= 0).\nonumber
\end{eqnarray}

Written with respect to Bessel functions \cite{AS}, $A_{i\rightarrow f}^{(3)} (\lambda = 0)$ becomes:
\begin{eqnarray}
A_{i\rightarrow f}^{(3)} (\lambda = 0) &=& \frac{g}{2}\frac{\sqrt{\pi} P\sqrt{pp^{\prime}}i\sqrt{i} e^{-\pi k/2}}{(2\pi)^{3/2}}\frac{\omega}{M_{W}}\delta^{3}(\vec{p}+\vec{p^{\prime}}+\vec{P}) \left(\frac{1}{2}-\sigma\right)\sigma^{\prime} \frac{e^{i\frac{\pi}{4}}}{i\, \cosh(\pi k)}\nonumber \\&& \times [T_{9}-T_{10}-T_{11}+T_{12}] \xi_{\sigma}^{\dag}(\vec{p}\,)\sigma^{i}\eta_{\sigma^{\prime}}(\vec{p^{\prime}}){\epsilon}^{*}_{i}(\vec{P},\lambda= 0),
\end{eqnarray}
where the terms $T_{9},T_{10},T_{11}$ and $T_{12}$ denote the following integrals:
\begin{eqnarray}
T_{9}&=&\int_{0}^{\infty} dz\, z^2\,\sqrt{z}\,\mathcal{J}_{\frac{1}{2}}(pz)ie^{\pi k}\mathcal{J}_{\frac{1}{2}-ik}(p^{\prime}z)\mathcal{K}_{1-iK}(iPz); \\
T_{10}&=&\int_{0}^{\infty} dz\, z^2\,\sqrt{z}\,\mathcal{J}_{\frac{1}{2}}(pz)\mathcal{J}_{-\frac{1}{2}+ik}(p^{\prime}z)\mathcal{K}_{1-iK}(iPz);\\
T_{11}&=&\int_{0}^{\infty} dz\, z^2\,\sqrt{z}\,e^{\pi k}\mathcal{J}_{-\frac{1}{2}}(pz)\mathcal{J}_{\frac{1}{2}-ik}(p^{\prime}z)\mathcal{K}_{1-iK}(iPz); \\
T_{12}&=&\int_{0}^{\infty} dz\, z^2\,\sqrt{z}\,\frac{1}{i}\mathcal{J}_{-\frac{1}{2}}(pz)\mathcal{J}_{-\frac{1}{2}+ik}(p^{\prime}z)\mathcal{K}_{1-iK}(iPz).
\end{eqnarray}

The explicit form of these terms can be found bellow and was obtained by using formula (\ref{appell}):
\begin{eqnarray}
T_{9}=\frac{ie^{\pi k}{2}^{3/2}p^{\frac{1}{2}}{p^{\prime}}^{\frac{1}{2}-ik}(iP)^{-\frac{9}{2}+ik}}{\Gamma\left(\frac{3}{2}\right)\Gamma\left(\frac{3}{2}-ik\right)} \Gamma\left(\frac{\frac{7}{2}-i(k-K)}{2}\right)\Gamma\left(\frac{\frac{11}{2}-i(k+K)}{2}\right)\nonumber \\
\times \mathcal{F}_{4}\left(\frac{\frac{7}{2}-i(k-K)}{2},\frac{\frac{11}{2}-i(k+K)}{2},\frac{3}{2},\frac{3}{2}-ik,\frac{p^{2}}{P^{2}},\frac{{p^{\prime}}^{2}}{P^{2}}\right);\\
T_{10}=\frac{{2}^{3/2}p^{\frac{1}{2}}{p^{\prime}}^{-\frac{1}{2}+ik}(iP)^{-\frac{7}{2}-ik}}{\Gamma\left(\frac{3}{2}\right)\Gamma\left(\frac{1}{2}+ik\right)} \Gamma\left(\frac{\frac{5}{2}+i(k+K)}{2}\right)\Gamma\left(\frac{\frac{9}{2}+i(k-K)}{2}\right)\nonumber \\
\times\mathcal{F}_{4}\left(\frac{\frac{5}{2}+i(k-K)}{2},\frac{\frac{9}{2}+i(k-K)}{2},\frac{3}{2},\frac{1}{2}+ik,\frac{p^{2}}{P^{2}},\frac{{p^{\prime}}^{2}}{P^{2}}\right);\\
T_{11}=\frac{e^{\pi k}{2}^{3/2}p^{-\frac{1}{2}}{p^{\prime}}^{\frac{1}{2}-ik}(iP)^{-\frac{7}{2}+ik}}{\Gamma\left(\frac{1}{2}\right)\Gamma\left(\frac{3}{2}-ik\right)} \Gamma\left(\frac{\frac{5}{2}-i(k-K)}{2}\right)\Gamma\left(\frac{\frac{9}{2}-i(k+K)}{2}\right)\nonumber \\
\times\mathcal{F}_{4}\left(\frac{\frac{5}{2}-i(k-K)}{2},\frac{\frac{9}{2}-i(k+K)}{2},\frac{1}{2},\frac{3}{2}-ik,\frac{p^{2}}{P^{2}},\frac{{p^{\prime}}^{2}}{P^{2}}\right);\\
T_{12}=\frac{\frac{1}{i}2^{3/2}p^{-\frac{1}{2}}{p^{\prime}}^{-\frac{1}{2}+ik}(iP)^{-\frac{5}{2}-ik}}{\Gamma\left(\frac{1}{2}\right)\Gamma\left(\frac{1}{2}+ik\right)} \Gamma\left(\frac{\frac{3}{2}+i(k+K)}{2}\right)\Gamma\left(\frac{\frac{7}{2}+i(k-K)}{2}\right)\nonumber \\
\times\mathcal{F}_{4}\left(\frac{\frac{3}{2}+i(k+K)}{2},\frac{\frac{7}{2}+i(k-K)}{2},\frac{1}{2},\frac{1}{2}+ik,\frac{p^{2}}{P^{2}},\frac{{p^{\prime}}^{2}}{P^{2}}\right).
\end{eqnarray}

We can see that for $\lambda = 0$ there is a much stronger dependence on the expansion factor as a consequence of all terms being multiplied by the ratio between $\omega$ and $M_{W}$.

The probability for this process, when computed with the longitudinal modes for the $W^-$ boson, is defined as the square modulus of the amplitude summed over $\sigma'$, since $\lambda, \sigma$ are fixed:
\begin{eqnarray}
&&P(\lambda=0)=\frac{1}{2}\sum_{\sigma'}|A_{i\rightarrow f} (\lambda = 0)|^2\nonumber\\
&&=\frac{1}{2}\sum_{\sigma'}|(A_{i\rightarrow f}^{(1)} + A_{i\rightarrow f}^{(2)}  +  A_{i\rightarrow f}^{(3)} )(A_{i\rightarrow f}^{(1)} + A_{i\rightarrow f}^{(2)}  +  A_{i\rightarrow f}^{(3)} )^*|.
\end{eqnarray}

To assess this probability we again have to take into account all terms which depend on the expansion parameter. In this case they are:
\begin{eqnarray}
F_{2}&=& -\frac{\omega}{M_{W}}\frac{P\sqrt{p p^{\prime}}\sqrt{i}e^{-\pi k/2}e^{-i \pi /4}}{i\, \cosh(\pi k)}[T_{5}-T_{6}-T_{7}+T_{8}] \nonumber \\
&&-i\left(\frac{1}{2}-iK\right)\frac{\omega}{M_{W}}\frac{\sqrt{pp^{\prime}}\sqrt{i}e^{-\pi k/2}e^{-i\pi/4}}{i\, \cosh(\pi k)}[T_{1}-T_{2}-T_{3}+T_{4}] \nonumber \\
&&+\frac{\omega}{M_{W}}\frac{P\sqrt{pp^{\prime}}i\sqrt{i}e^{-\pi k/2}e^{i\pi/4}}{i\, \cosh(\pi k)}[T_{9}-T_{10}-T_{11}+T_{12}] .\nonumber \\
\end{eqnarray}

As before, the relevant term for the probability is the absolute value of the product $F_{2}F_{2}^{*}$.

\begin{figure}[H]
\includegraphics[scale=0.41]{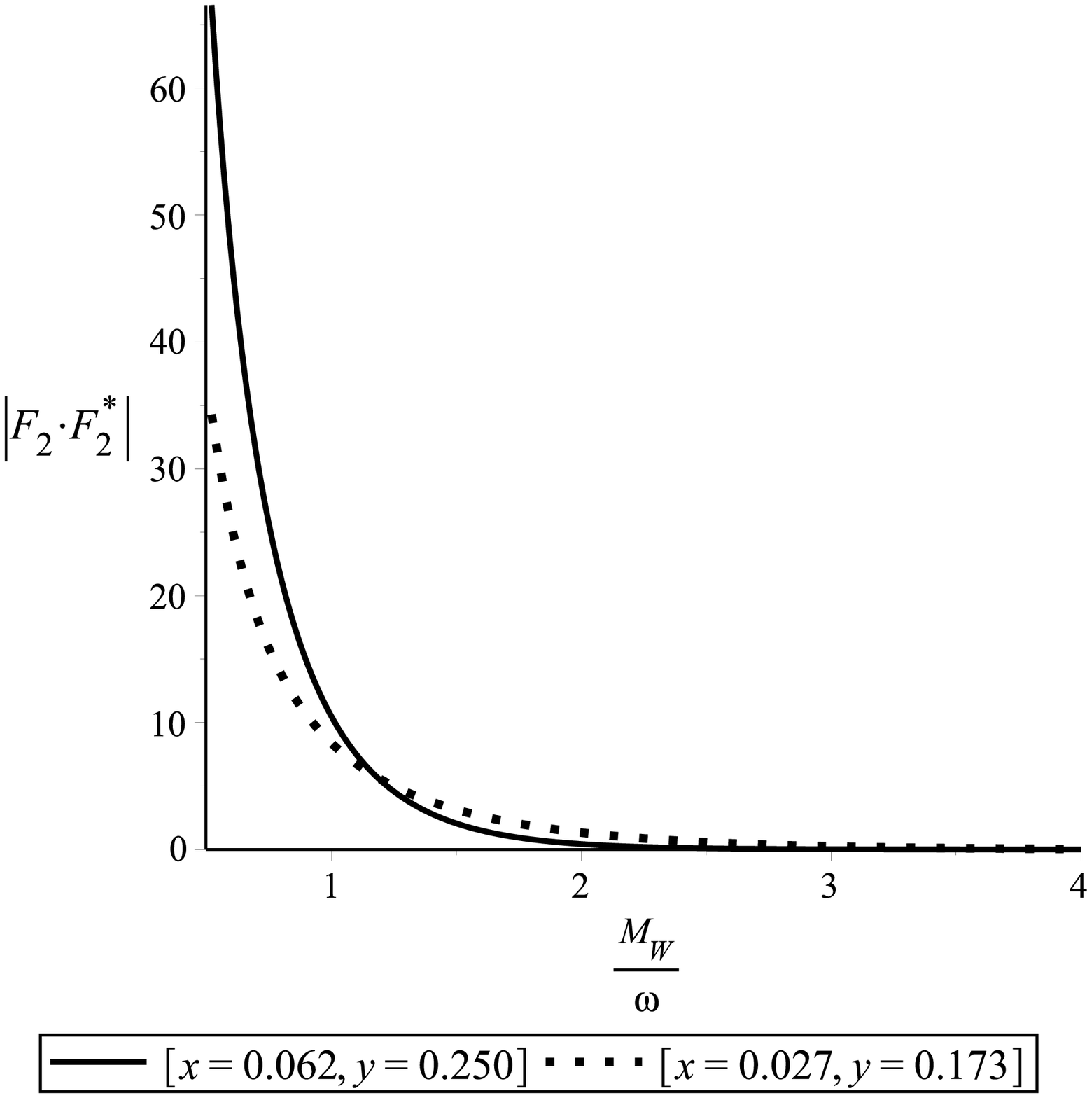}
\includegraphics[scale=0.41]{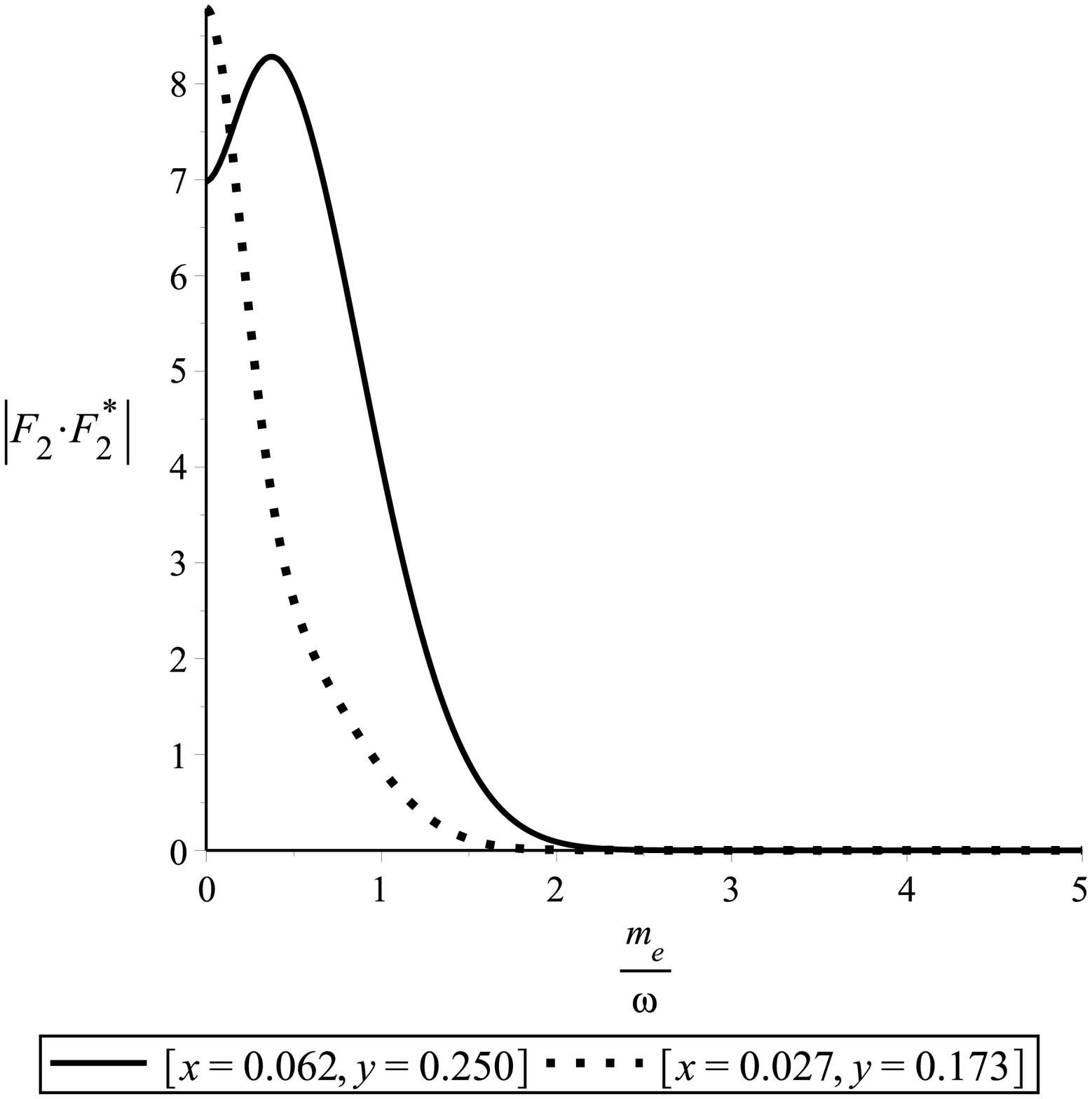}
\caption{The absolute value of $F_{2}$ for $k=0.1$ on the left and $\frac{M_{W}}{\omega}=0.6$ on the right, and momenta values $p' = 0.2, p = 0.1, P = 0.4$ for the straight line and $p' = 0.25, p = 0.1, P = 0.6$ for the dotted line.}\label{F2}
\end{figure}

For the graph in figure (\ref{F2}) with respect to the $W$ boson mass, the starting value is $M_{W}/\omega =0.52$ in order to avoid the divergence which emerges at $M_{W}/\omega=0$ due to a factor of $\omega / M_{W}$ which multiplies all terms in $F_{2}$. Figures (\ref{F2}) prove that the probability of boson production for modes with $\lambda=0$ was high in the early universe, when the expansion factor was much bigger than the mass of the $W$ bosons. In the Minkowski limit, where $\omega \rightarrow 0$, the probability vanishes as expected.

\section{Total probability for $\lambda=\pm1$ in the case of large expansion }

In the limiting case of large expansion, the ratios between particle masses and expansion factor become very small and it is possible to write $m_e/\omega=M_Z/\omega = 0$. In this limit, the amplitude given in equation (\ref{ampl}) reduces to a more simple form:
\begin{eqnarray}\label{amp1o}
&&A_{i\rightarrow f}= -\frac{g}{\sqrt{2}}\frac{\pi \sqrt{p\,^{\prime}} }{(2\pi)^{3/2}}\delta^{3}(\vec{p}+\vec{p^{\prime}}+\vec{P}) \left(\frac{1}{2}-\sigma\right)\sigma^{\prime}  \nonumber \\ &&\times\int_{0}^{+\infty} dz z e^{-ipz}\mathcal{H}_{\frac{1}{2}}^{(2)}(p^{\prime}z)\mathcal{H}_{\frac{1}{2}}^{(2)}(P z)\,\xi_{\sigma}^{\dag}(\vec{p}\,)\sigma^{i}\eta_{\sigma^{\prime}}(\vec{p^{\prime}}){\epsilon}^{*}_{i}(\vec{P},\lambda=\pm 1).
\end{eqnarray}

Taking into account that the neutrino polarization can take only the value $\sigma=-\frac{1}{2}$, the final result for the amplitude reads as:
\begin{eqnarray}
&&A_{i\rightarrow f}=\frac{ig}{2\sqrt{2P}}\frac{ 1}{(2\pi)^{3/2}}\frac{\delta^{3}(\vec{p}+\vec{p^{\prime}}+\vec{P})}{(p+p'+P)} sgn(\sigma^{\prime})\,\xi_{-\frac{1}{2}}^{\dag}(\vec{p}\,)\sigma^{i}\eta_{\sigma^{\prime}}(\vec{p^{\prime}}){\epsilon}^{*}_{i}(\vec{P},\lambda=\pm 1).
\nonumber\\
\end{eqnarray}

The probability per unit volume takes the form:
\begin{eqnarray}
P(\lambda=\pm1)=\frac{1}{4}\sum_{\lambda\sigma'}\frac{g^2}{8(2\pi)^{3}}\frac{\delta^{3}(\vec{p}+\vec{p^{\prime}}+\vec{P})}{P(p+p'+P)^2} \,|\xi_{-\frac{1}{2}}^{\dag}(\vec{p}\,)\sigma^{i}\eta_{\sigma^{\prime}}(\vec{p^{\prime}}){\epsilon}^{*}_{i}(\vec{P},\lambda=\pm 1)|^2.
\nonumber\\
\end{eqnarray}

We set up both the neutrino and the positron to be emitted in the direction of the third axis, such that the angle between the momenta vectors is $0$, i.e $\vec p=p\,\vec e_3\,,\,\vec p\,'=p'\,\vec e_3$. In this case the bispinor summation is reduced to a number since the helicity bispinors are reduced to a column matrix with elements $0,\,\pm 1$. The momentum of the $W^-$ boson is considered to also be on the third axis such that $\vec P=-P\,\vec e_3$. Since the momenta are on fixed directions, the helicity bispinors are very simple and no longer depend on the momenta. This will facilitate our computations of the momenta integrals:
\begin{equation}
  \xi_{-\frac{1}{2}} (\vec{p}\,) = \begin{pmatrix}
                         0 \\
                         1
                       \end{pmatrix} ; \eta_{\frac{1}{2}} (\vec{p}\,) = \begin{pmatrix}
                                                             0 \\
                                                             -1
                                                           \end{pmatrix};
        \eta_{-\frac{1}{2}} (\vec{p}\,) = \begin{pmatrix}
                                                             1 \\
                                                             0
                                                           \end{pmatrix}.
\end{equation}

In this particular case the bispinor and polarization vector summation gives:
\begin{eqnarray}
&&\sum_{\lambda\sigma'}|\xi_{-\frac{1}{2}}^{\dag}(\vec{p}\,)\sigma^{i}\eta_{\sigma^{\prime}}(\vec{p^{\prime}}){\epsilon}^{*}_{i}(\vec{P},\lambda=\pm 1)|^2\nonumber\\
&&=\sum_{\lambda\sigma'}\xi_{-\frac{1}{2}}^{\dag}(\vec{p}\,)\sigma^{i}
{\epsilon}^{*}_{i}(\vec{P},\lambda=\pm 1){\epsilon}^{*}_{j}(\vec{P},\lambda=\pm 1)\eta_{\sigma^{\prime}}(\vec{p^{\prime}})\eta_{\sigma'}^{\dag}(\vec{p^{\prime}})\sigma^{j}\xi_{-\frac{1}{2}}(\vec{p}\,),
\nonumber\\
\end{eqnarray}
which reduces to unity if we use the above equations for spinors and the following relations:
\begin{eqnarray}
&&\sum_{\lambda}\epsilon_i^*(\vec{n}_{\mathcal{P}},\,\lambda)\epsilon_j(\vec{n}_{\mathcal{P}},\,\lambda)=\delta_{ij},\nonumber\\
&&\sum_{\sigma'}\eta_{\sigma^{\prime}}(\vec{p^{\prime}})\eta_{\sigma'}^{\dag}(\vec{p^{\prime}})=I_{2\times 2}.
\end{eqnarray}

The total probability for the process is obtained by computing the integral with respect to the momenta of the final particles:
\begin{equation}
P_{tot}=\int d^{3}p \int d^{3}P \int d^{3}p^{\prime}\,P(\lambda=\pm1).
\end{equation}

We denote the momenta integrals by $I$ and solve the integral with regards to $P$ by using the properties of the delta Dirac function:
\begin{eqnarray}
I &=& \int d^{3}p \int d^{3}P \int d^{3}p^{\prime}\, \frac{\delta^{3}(\vec{p}+\vec{p^{\prime}}+\vec{P})}{(2\pi)^3P(p^{\prime} + p + P)^{2}}\nonumber \\
&=& \int \frac{d^{3}p}{4(2\pi)^3} \int d^{3}p' \frac{1}{(p'+p)^{3}},
\end{eqnarray}
keeping in mind that $|\vec p\,'+\vec p|=p'+p$, because the particles move in the same direction and the momenta are aligned with $\vec e_3$.

Because our remaining integrals are divergent, we make use of dimensional regularization, a method used for studying the properties of propagators in the de Sitter geometry \cite{PT,PC}. We mention that the problem of Dirac and Proca propagators in the momentum and coordinate representations in the de Sitter geometry was analysed in \cite{WT,PC,CRR,COT}.

The $p'$ integral will be solved using dimensional regularization \cite{WF,GHV,IT,GT} such that the integral in $D$ dimensions becomes:
\begin{equation}
\int \frac{d^{3}p'}{(2\pi)^3(p'+p)^3}  \rightarrow  I(D)=\int \frac{d^{D}p'}{(2\pi)^{D}(p'+p)^3},
\end{equation}

Considering the case when the variable is changed as $p'=py$, the above integral changes into:
\begin{eqnarray}\label{rd1}
I(D) = \frac{2\pi^{D/2}p^{D-3}}{(2\pi)^{D}\Gamma(D/2)}\int_0^\infty\frac{dy \,y^{D-1}}{(1+y)^3}.
\end{eqnarray}

This is just the integral of the Beta Euler function, defined bellow \cite{AS} as:
\begin{equation}
B(\alpha,\gamma) = \frac{\Gamma(\alpha)\Gamma(\gamma)}{\Gamma(\alpha + \gamma)} = \int_{0}^{\infty} dy \frac{y^{\alpha -1}}{(1+y)^{\alpha + \gamma}}.
\end{equation}
which for $\alpha=D,\,\beta=3-D$ becomes:
\begin{equation}
I(D) =  \frac{2\pi^{D/2}p^{D-3}}{(2\pi)^{D}}\frac{\Gamma(D)\Gamma(3-D)}{\Gamma(3)\Gamma\left(\frac{D}{2}\right)}.
\end{equation}

The above result contains ultraviolet divergences for $D\geq3$ and infrared divergences for $D\leq0$. These divergences are contained in the poles of the gamma Euler functions. The poles of the gamma functions can be extracted by introducing an arbitrary mass parameter denoted by $\mu$, and an arbitrary coupling dimensionless constant denoted by $\gamma=\lambda \mu^{D-3}=\lambda \mu^{-\varepsilon}$.

We can then write the result of integral (\ref{rd1}) as a function of $\gamma$ and $\varepsilon$ for $D=3-\varepsilon$:
\begin{equation}
 I(D)_r =\lambda\int\frac{d^D p'}{(2\pi)^D}\frac{1}{(p'+ p)^3}=\frac{\gamma }{(4\pi)^{3/2}}\left(\frac{\mu^{2}}{p^{2}}4\pi\right)^{\epsilon/2}\frac{\Gamma(3-\epsilon)\Gamma(\epsilon-1)}{\Gamma\left(\frac{3-\epsilon}{2}\right)}.
\end{equation}

The series expansion in terms of $\epsilon$ of the last two factors gives:
\begin{equation}
\left(\frac{\mu^{2}}{P^{2}}4\pi\right)^{\epsilon/2} = 1 + \frac{\epsilon}{2}\ln\left(\frac{\mu^{2}}{p^{2}}4\pi\right) + \mathcal{O}(\epsilon^{2}),
\end{equation}

\begin{eqnarray}
\frac{\Gamma(3-\epsilon)\Gamma(\epsilon)}{\Gamma\left(\frac{3-\epsilon}{2}\right)}=\frac{4}{\sqrt{\pi}\,\epsilon}+\frac{2(\psi(3/2)-3)}{\sqrt{\pi}}\nonumber\\
+\frac{\epsilon\,(5\pi^2+48-36\psi(3/2)+6\psi^2(3/2))}{12\sqrt{\pi}}+\mathcal{O}(\epsilon^2),
\end{eqnarray}
where $\psi(a)$ are the digamma Euler functions defined in Appendix.

Collecting the above expansions we obtain the final result for the regularized integral in the case of small momenta $p\sim M_W$:
\begin{eqnarray}\label{dr2}
&&I_r(D) = \frac{\gamma }{(4\pi)^{3/2}}\biggl\{ \frac{4}{\sqrt{\pi}\,\epsilon} + \frac{2\left(\psi(3/2)+\ln\left(\frac{4\pi\mu^{2}}{M_W^{2}}\right)-3\right)}{\sqrt{\pi}}\nonumber\\
&&+\epsilon\,\left[\frac{5\pi^2+48-36\psi(3/2)+6\psi^2(3/2)}{12\sqrt{\pi}}+\frac{(\psi(3/2)-3)}{\sqrt{\pi}}\ln\left(\frac{4\pi\mu^{2}}{M_W^{2}}\right)\right]+ \mathcal{O}(\epsilon^{2})\biggl\}.
\nonumber\\
\end{eqnarray}

The result is finite for arbitrary small values of $\epsilon$, but becomes infinite for $\epsilon=0$. In order to replace the term corresponding to the $1/\epsilon$ divergence, we will solve the integral with respect to $p^{\prime}$ by using the Pauli-Villars regularization method \cite{PV}. At the end of the computations we will find the exact substitute for the term proportional with $1/\epsilon$ from the previous expression.

Let us write the integral with $p\sim M_W$
\begin{equation}
\int \frac{d^{3}p'}{(2\pi)^3(p'+M_W)^3}=4\pi\int_0^\infty \frac{dp'\,p'^2}{(2\pi)^3(p'+M_W)^3},
\end{equation}
and expand the integrand
\begin{eqnarray}
\frac{1}{(p'+M_W)^3}\rightarrow \frac{1}{(p'+M_W)^3}+\frac{a}{(p'+\Lambda)^3}
\end{eqnarray}

$\Lambda$ is a bosonic mass parameter and $a$ will be determined as function of $\Lambda$.

The integrand can be rewritten in terms of a common denominator:
\begin{eqnarray}
\frac{p'^3+\Lambda^3+3p'\Lambda^2+3p'^2\Lambda+a(p'^3+M_W^3+3M_Wp'^2+3M_W^2p')}{(p'+M_W)^3(p'+\Lambda)^3}.
\end{eqnarray}

The ultra-violet divergences can be removed from the integral by demanding that terms proportional with $\,p'^3$ vanish. This gives rise to the following condition:
\begin{eqnarray}
&&1+a=0,
\end{eqnarray}

And we obtain $a=-1$, which cancel the terms in $p'^3$.

The final form of the integral is:
\begin{eqnarray}
&&\int_0^\infty dp'\,\frac{p'^2(\Lambda^3-M_W^3+3p'\Lambda^2+3p'^2\Lambda-3p'M_W^2-3p'^2M_W)}{(p'+M_W)^3(p'+\Lambda)^3}
=\ln\left(\frac{\Lambda}{M_W}\right).
\nonumber\\
\end{eqnarray}

Then the final result of our integral obtained with the Pauli-Villars method \cite{PV} reads
\begin{equation}\label{pv1}
\int \frac{d^{3}p'}{(2\pi)^3(p'+M_W)^2}=\frac{1}{2\pi^2}\ln\left(\frac{\Lambda}{M_W}\right).
\end{equation}

By comparing equations (\ref{dr2}) and (\ref{pv1}) one can see that the first terms from both expressions are similar and we can write:
\begin{equation}
\frac{1}{\epsilon}=\frac{1}{2}\ln\left(\frac{\Lambda}{M_W}\right).
\end{equation}

The above result proves that the divergences manifest as logarithmic functions of $\Lambda$. The last integral that needs to be solved is the one with respect to the momentum of the neutrino.

We must specify that this first order process with three particles in the final state is not encountered in the Minkowski field theory and for that reason the regularization of the third momenta integral is hard to accomplish by using the known regularization methods. For this reason we will make a cutoff of the integral up to the mass of the $W$ boson such that the upper limit of the $p$ integral is fixed to $M_W$. This is equivalent to the computation of total probability for a fixed value of the neutrino moment.

Our analysis is important since the regularization of the probability helps us understand the nature of the divergences that appear in our exact perturbative calculations and could help one develop a complete renormalized theory by introducing the necessary counter-terms in the lagrangian density. In order to do that, however, further study of particle production processes which are forbidden in Minkowski space-time is required. It may also be possible that new regularization methods will be needed for the regularization of these types of integrals, since in the Minkowski theory we do not encounter such processes.

Collecting our prior results, we can write down the total probability in the case of large expansion:
\begin{eqnarray}
&&P_{tot}=\frac{g^2\gamma M_W^3}{96\pi^2}\biggl\{\frac{1}{2}\ln\left(\frac{\Lambda}{M_W}\right) + 2\left(\psi(3/2)+\ln\left(\frac{4\pi\mu^{2}}{M_W^{2}}\right)-3\right)\nonumber\\
&&+\epsilon\,\left[\frac{5\pi^2+48-36\psi(3/2)+6\psi^2(3/2)}{12}+(\psi(3/2)-3)\ln\left(\frac{4\pi\mu^{2}}{M_W^{2}}\right)\right]+ \mathcal{O}(\epsilon^{2})\biggl\}.
\nonumber\\
\end{eqnarray}

The above total probability is finite for small values of $\epsilon$, as well as in the case of $\epsilon=0$.
\begin{eqnarray}
&&P_{tot}=\frac{g^2\gamma M_W^3}{48\pi^2}\biggl\{\frac{1}{4}\ln\left(\frac{\Lambda}{M_W}\right) + \psi(3/2)+\ln\left(\frac{4\pi\mu^{2}}{M_W^{2}}\right)-3\biggl\}.
\end{eqnarray}

In this section we have shown that some of the infrared and ultraviolet divergences can be removed from the momenta integrals by applying a combination of methods used in the flat space-time theory. Still, in order to achieve a complete renormalizable theory, one needs to write down the renormalized amplitudes by adding the necessary counter-terms to the lagrangian density.

\section{The transition rate for $\lambda=\pm 1$}

The probability of the particle production process suggests that this type of boson emission was highly likely for $\omega \gg M_{W}$, however one must find measurable quantities to validate such a process. This leads to the problem of computing the transition rate, i.e. the probability per unit time.

In flat space-time, energy-momentum conservation allows us to find a transition rate which depends on the volume $V$ and interaction time $T$. But energy and momentum are not simultaneously conserved in de Sitter space-time, and the amplitude is also time-dependant.

We can see this better when writing the amplitude as such:
\begin{eqnarray}\label{amp1rat}
&&A_{i\rightarrow f}= -\frac{ig}{2\sqrt{2}}\frac{ \sqrt{p^{\prime}} e^{-\pi k/2}}{(2\pi)^{3/2}}\delta^{3}(\vec{p}+\vec{p^{\prime}}+\vec{P}) \left(\frac{1}{2}-\sigma\right)\sigma^{\prime} \cdot \nonumber \\
 &&\times\int_{0}^{+\infty} dz  z e^{-ipz}\mathcal{H}_{\nu_{-}}^{(2)}(p^{\prime}z)\mathcal{K}_{-iK}(iP z)\xi_{\sigma}^{\dag}(\vec{p})\sigma^{i}\eta_{\sigma^{\prime}}(\vec{p^{\prime}}){\epsilon}^{*}_{i}(\vec{P},\lambda=\pm 1).
\end{eqnarray}

The previous expression for amplitude can be rewritten as
\begin{equation}
A_{i\rightarrow f} = M_{i\rightarrow f}I_{i\rightarrow f}\delta^{3}(\vec{p_{f}}-\vec{p_{i}}),
\end{equation}
where $I_{i\rightarrow f}$ represents the time-dependent integral:
\begin{equation}\label{arat}
I_{i\rightarrow f} =  e^{-\pi k/2}\int_{0}^{\infty}  K_{i\rightarrow f} dz,
\end{equation}

The integrand in (\ref{arat}) is given by :
\begin{equation}
K_{i\rightarrow f} = z\,e^{-ipz} \mathcal{H}_{\nu_{-}}^{(2)}(p^{\prime}z)\mathcal{K}_{-iK}(P z),
\end{equation}
while
\begin{eqnarray}\label{mif}
M_{i\rightarrow f} = -\frac{ig}{2\sqrt{2}}\frac{\pi\sqrt{p^{\prime}}}{(2\pi)^{3/2}} \left(\frac{1}{2} -\sigma\right)\sigma^{\prime}\xi_{\sigma}^{\dag}(\vec{p})\sigma^{i}\eta_{\sigma^{\prime}}(\vec{p^{\prime}}){\epsilon}^{*}_{i}(\vec{P},\lambda= \pm1).\nonumber \\
\end{eqnarray}

The transition rate with respect to the conformal time is defined by \cite{cpc}:
\begin{equation}
R_{i\rightarrow f}=\lim_{t_c\rightarrow 0}\frac{1}{2V}\frac{d}{dt_c}|A_{if}|^2=\lim_{t\rightarrow \infty}\frac{e^{\omega t}}{2V}\frac{d}{dt}| A_{if}|^2.
\end{equation}

This expression can be brought to the form:
\begin{equation}\label{rate}
R_{i\rightarrow f} = \frac{1}{(2\pi)^{3}}\delta^{3}(\vec{p_{f}}-\vec{p_{i}})|M_{i\rightarrow f}|^{2}|I_{i\rightarrow f}|\lim_{t\rightarrow \infty}|e^{\omega t}K_{i\rightarrow f}|.
\end{equation}

In our case we must take into account the summations after $\sigma',\lambda$ since $\sigma=-1/2$. We also specify that the limit in equation (\ref{rate}) will be taken up to a sufficiently large time after the interaction, denoted by $t_{\infty}$.

With these things in mind, the formula of the transition rate in our case is:
\begin{equation}\label{rates}
R_{i\rightarrow f} = \frac{1}{4}\sum_{\sigma'\lambda}\frac{1}{(2\pi)^{3}}\delta^{3}(\vec{p_{f}}-\vec{p_{i}})|M_{i\rightarrow f}|^{2}|I_{i\rightarrow f}|\lim_{t\rightarrow t_{\infty}}|e^{\omega t}K_{i\rightarrow f}|.
\end{equation}

Because we are working in the limit $t\rightarrow t_{\infty}$, the Hankel functions may be approximated using the formula for small arguments \cite{AS}:
\begin{equation}
\mathcal{H}_{\nu}^{(2)}(z) \approx i\left(\frac{2}{z}\right)^{\nu}\frac{\Gamma(\nu)}{\pi},
\end{equation}
which works for $z\rightarrow 0$.

Thus the first Hankel function becomes:
\begin{equation}
\mathcal{H}_{\frac{1}{2}-ik}^{(2)}\left(\frac{p^{\prime}}{\omega}e^{-\omega t}\right) \simeq i \left(\frac{2}{\frac{p^{\prime}}{\omega}e^{-\omega t}}\right)^{\frac{1}{2}-ik}\frac{\Gamma(\frac{1}{2}-ik)}{\pi}.
\end{equation}

The transition rate is relevant when $\omega\gg M_{W}$ and $\omega\gg m_{e}$, meaning that we can further approximate the index of the modified Bessel function $\mathcal{K}_{-iK}$:
\begin{equation}
-iK = -i\sqrt{\left(\frac{M_{W}}{\omega}\right)^{2}-\frac{1}{4}} = \frac{1}{2}\sqrt{1 - 4\left(\frac{M_{W}}{\omega}\right)^{2}}\approx \frac{1}{2}.
\end{equation}

Consequently, the behaviour of the modified Bessel function for small arguments is:
\begin{equation}
\mathcal{K}_{-iK} \left(\frac{P}{\omega}e^{-\omega t}\right) \simeq \frac{\Gamma(1/2)}{\sqrt{2i\frac{P}{\omega}e^{-\omega t}}}.
\end{equation}

This leads to the limit:
\begin{eqnarray}\label{liim}
\lim_{t\rightarrow t_{\infty} } |e^{\omega t}K_{i\rightarrow f}| &=&\lim_{t\rightarrow t_{\infty} }\bigg|i\frac{e^{-\omega t}}{\omega}e^{-ip\frac{e^{-\omega t}}{\omega}} \left(\frac{2}{\frac{p^{\prime}}{\omega}e^{-\omega t}}\right)^{\frac{1}{2}-ik}\frac{\Gamma(\frac{1}{2}-ik)}{\pi}\frac{\Gamma(1/2)}{\sqrt{2i\frac{P}{\omega}e^{-\omega t}}}\bigg|\nonumber\\
&=& \frac{1}{\sqrt{Pp^{\prime}\cosh(\pi k)}},
\end{eqnarray}
where we have used \cite{AS}:
\begin{equation}
\Gamma\left(\frac{1}{2}-ik\right) \Gamma\left(\frac{1}{2}+ik\right) = \frac{\pi}{\cosh(\pi k)}.
\end{equation}

By substituting the limit of the temporal integral in the transition rate formula (\ref{rate}) we get:
\begin{equation}
R_{i\rightarrow f} =\sum_{\sigma'\lambda} \frac{g^2\pi^2pp'}{32(2\pi)^{3}}\delta^{3}(\vec{p}+\vec{p^{\prime}} + \vec{P}) |\xi_{-1/2}^{\dag}(\vec{p})\sigma^{i}\eta_{\sigma^{\prime}}(\vec{p^{\prime}}){\epsilon}^{*}_{i}(\vec{P},\lambda= \pm1)|^{2}\frac{|I_{i\rightarrow f}| }{\sqrt{Pp^{\prime}\cosh(\pi k)}},
\end{equation}
where $|I_{i\rightarrow f}| = \sqrt{I_{i\rightarrow f}I_{i\rightarrow f}^{*}}$.

The explicit form of $I_{i\rightarrow f}$ is:
\begin{equation}
I_{i\rightarrow f}=\frac{e^{-i\pi/4}e^{-\pi k/2}}{i\sqrt{p}\cosh(\pi k)}(T_1-T_2-T_3+T_4),
\end{equation}
where $T_1,T_2,T_3,T_4$ are defined in equations (\ref{ttt}).

The final result for the transition rate in the general case is:
\begin{equation}
R_{i\rightarrow f} = \frac{g^2\pi^2\sqrt{pp'}}{32(2\pi)^{3}}\delta^{3}(\vec{p}+\vec{p^{\prime}} + \vec{P})\frac{e^{-\pi k/2}}{\cosh(\pi k)} \sqrt{\frac{(T_1-T_2-T_3+T_4)(T_1-T_2-T_3+T_4)^* }{P\cosh(\pi k)}}.
\end{equation}

For the sake of examining the behaviour of the transition rate it is useful to perform a graphical analysis of the terms that depend on $\frac{m}{\omega}$ and $\frac{M_{W}}{\omega}$:
\begin{equation}
R_{1}= \frac{\sqrt{pp'}e^{-\pi k/2}}{\cosh(\pi k)} \sqrt{\frac{(T_1-T_2-T_3+T_4)(T_1-T_2-T_3+T_4)^* }{P\cosh(\pi k)}}.
\end{equation}

\begin{figure}[H]
\includegraphics[scale=0.4]{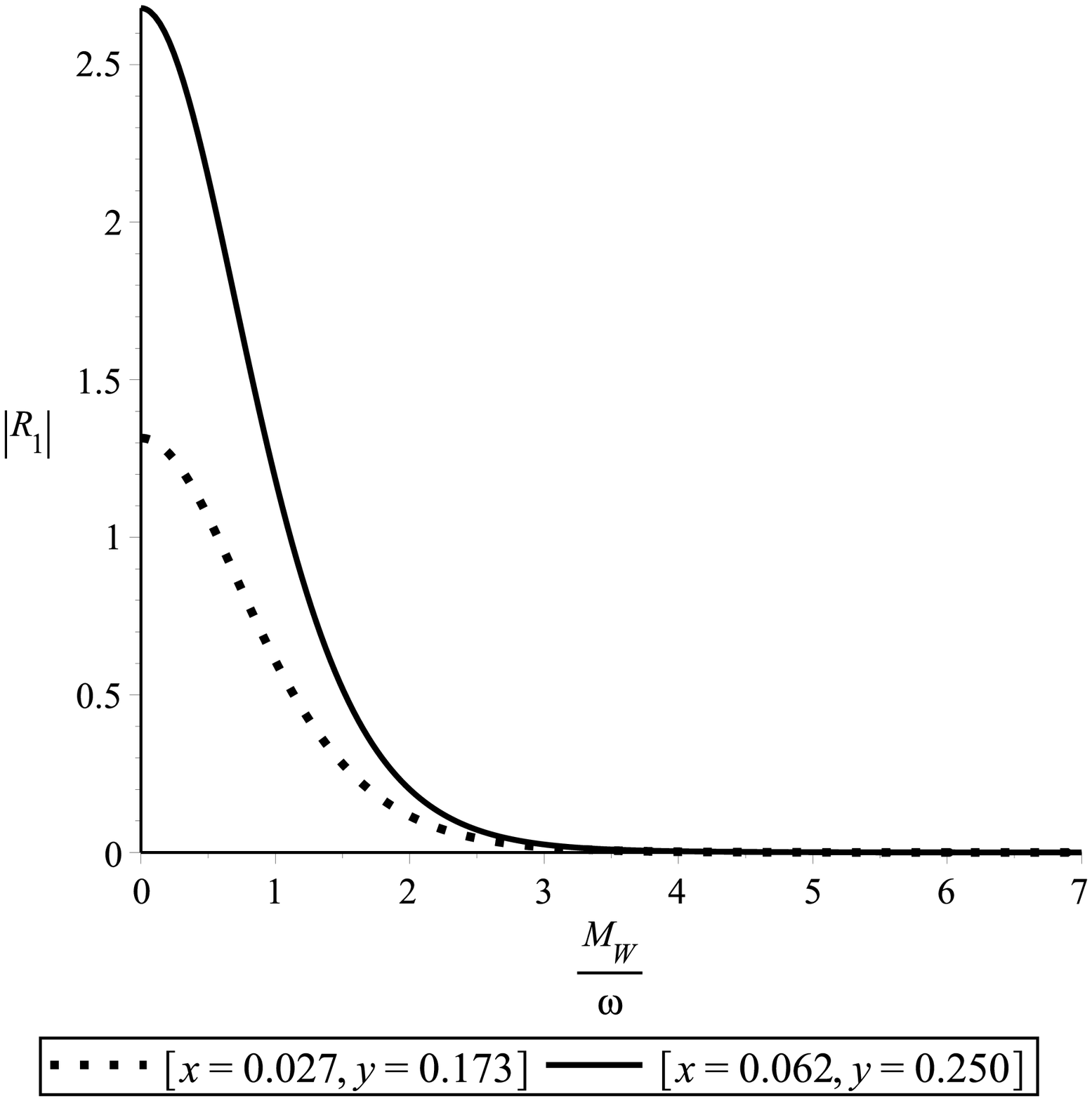}
\includegraphics[scale=0.4]{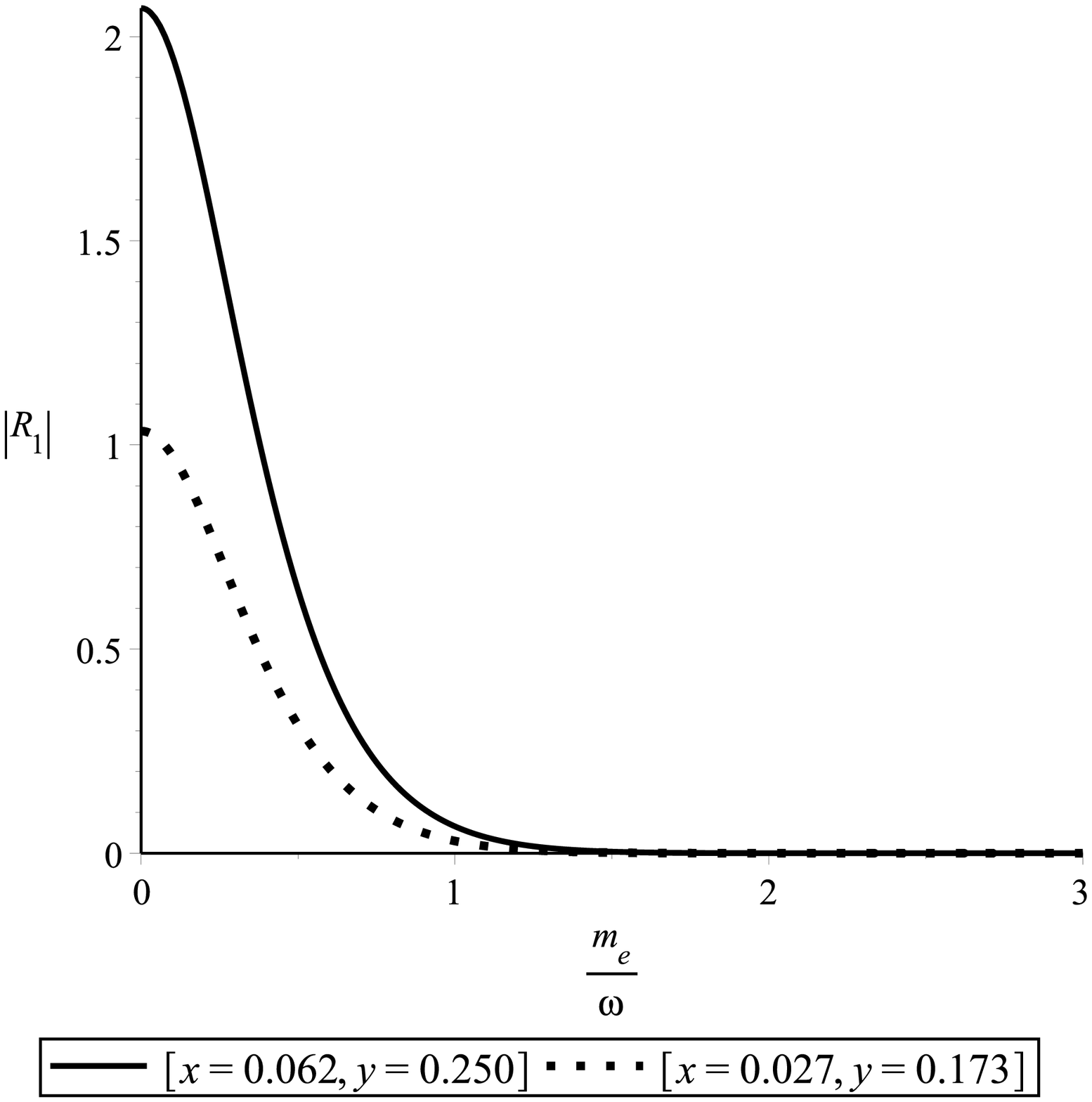}
\caption{The absolute value of $R_{1}$ for $\frac{m_e}{\omega}=0.1$ on the left and $\frac{M_{W}}{\omega}=0.6$ on the right, and momenta values $p' = 0.2, p = 0.1, P = 0.4$ for the straight line and $p' = 0.25, p = 0.1, P = 0.6$ for the dotted line.}
\label{R12}
\end{figure}

The transition rate variation with respect to $\frac{m_e}{\omega},\frac{M_{W}}{\omega}$, presented in the above figures, shows that the spontaneous vacuum generation of the $W$ bosons and fermions is a process possible only when the expansion parameter is far larger than the particle masses, which corresponds to the early universe period. In the Minkowski limit, which corresponds to infinite values of parameters $k,K$, the rate vanishes. It is also worth mentioning that massive bosons can exist as stable particles only as long as the background energy is larger than their rest energy. When this condition is no longer valid, the massive bosons decay into more stable particles with smaller mass.

\subsection{Total rate in the limit of large expansion}

In this section we will discuss the problem of the total transition rate in the limit of large expansion, where the functions that define the amplitude and transition rate from the previous section become simpler and depend only on the momenta of the emitted particles.

The first task is to establish the expression of the transition rate in the limit $\frac{m_e}{\omega}=\frac{M_W}{\omega}=0$. In order to do this we first need to write the amplitude in this limit:
\begin{eqnarray}
&&A_{i\rightarrow f}= -\frac{ig}{2\sqrt{2}}\frac{ \sqrt{p^{\prime}} }{(2\pi)^{3/2}}\delta^{3}(\vec{p}+\vec{p^{\prime}}+\vec{P}) \left(\frac{1}{2}-\sigma\right)\sigma^{\prime}  \nonumber \\ &&\times\int_{0}^{+\infty} dz  z e^{-ipz}\mathcal{H}_{1/2}^{(2)}(p^{\prime}z)\mathcal{K}_{1/2}(iP z)\xi_{\sigma}^{\dag}(\vec{p}\,)\sigma^{i}\eta_{\sigma^{\prime}}(\vec{p^{\prime}}){\epsilon}^{*}_{i}(\vec{P},\lambda=\pm 1)\nonumber\\
&&=M_{i\rightarrow f}I_{i\rightarrow f}\delta^{3}(\vec{p_{f}}-\vec{p_{i}}),
\end{eqnarray}
where $M_{i\rightarrow f}$ is defined in equation (\ref{mif}).

The temporal integral is solvable for the Bessel and Hankel functions of index $1/2$:
\begin{equation}
I_{i\rightarrow f}=\lim_{\varepsilon\rightarrow0}\frac{i}{\sqrt{ip'P}}\int_{0}^{+\infty} dz\,e^{-i(p+p'+P)z-\varepsilon z}=\frac{1}{\sqrt{ip'P}(p+p'+P)}.
\end{equation}

The integrand is of the form $K_{i\rightarrow f}=e^{-i(p+p'+P)z}$, while the limit that needs to be done for computing the rate reads:
\begin{equation}
\lim_{t\rightarrow t_{\infty} } |e^{\omega t}K_{i\rightarrow f}|=\lim_{t\rightarrow t_{\infty} }\bigg|\frac{1}{\sqrt{ip'P}}e^{-i(p+p'+P)\frac{e^{-\omega t}}{\omega}}\bigg|=\frac{1}{\sqrt{p'P}},
\end{equation}
which is precisely the limit obtained directly from equation (\ref{liim}) for $k=0$.

Then the rate in the limit $\frac{m_e}{\omega}=\frac{M_W}{\omega}=0$ can be written in final form as:
\begin{eqnarray}
R=\frac{g^2}{8(2\pi)^{3}}\delta^{3}(\vec{p}+\vec{p^{\prime}}+\vec{P})\frac{1}{P(p+p'+P)},
\end{eqnarray}
where the results of the $\sigma',\lambda$ summations were included.

The total rate of transition can be obtained by integrating over all possible values of the momenta:
\begin{equation}
R_{vac\rightarrow W^-\nu e^+}=\int d^{3}p\,d^{3}p^{\prime}\,d^{3}P\, R_{i\rightarrow f}.
\end{equation}

We assume that the particles are emitted in the same direction as in the case of the total probability, with the momenta of the neutrino and positron being oriented in the direction of the positive third axis, while the boson momentum is oriented in the direction of the negative third axis.

We solve the $P$ integral and obtain:
\begin{eqnarray}
I = \int d^{3}p \int d^{3}P \int d^{3}p^{\prime} \frac{\delta^{3}(\vec{p}+\vec{p^{\prime}} +\vec{P})}{(2\pi)^3P(p+p^{\prime}+P)}=\int d^{3}p \int d^{3}p^{\prime} \frac{1}{2(2\pi)^3(p^{\prime}+p)^2}.
\nonumber\\
\end{eqnarray}

The $p'$ integral is divergent, therefore we have to make use of dimensional regularization \cite{WF,GHV,IT,GT} in order to solve it. The integral in $D$ dimensions becomes:
\begin{equation}
\int \frac{d^{3}p'}{(2\pi)^3(p'+p)^2}  \rightarrow  I(D)=\int \frac{d^{D}p'}{(2\pi)^{D}(p'+p)^2},
\end{equation}

By performing the change of variables $p'=py$, one obtains:
\begin{eqnarray}\label{rdo}
I(D) = \frac{2\pi^{D/2}p^{D-2}}{(2\pi)^{D}\Gamma(D/2)}\int_0^\infty\frac{dy \,y^{D-1}}{(1+y)^2}.
\end{eqnarray}

This is just the integral of the Beta Euler function defined for $\alpha=D,\,\beta=2-D$:
\begin{equation}
I(D) =  \frac{2\pi^{D/2}p^{D-2}}{(2\pi)^{D}}\frac{\Gamma(D)\Gamma(2-D)}{\Gamma(2)\Gamma\left(\frac{D}{2}\right)}.
\end{equation}

The above result contains ultraviolet divergences for $D\geq2$ and infrared divergences for $D\leq0$. These divergences are contained in the poles of the gamma Euler functions. The poles of the gamma functions can be removed by introducing an arbitrary mass parameter denoted by $\mu$ and an arbitrary coupling dimensionless constant denoted by $\gamma=\lambda \mu^{D-3}=\lambda \mu^{-\varepsilon}$.

The final result of the integral (\ref{rdo}) as a function of $\gamma$ and $\varepsilon$ for $D=3-\varepsilon$ is:
\begin{equation}
 I(D)_r =\frac{-2\gamma p}{(4\pi)^{3/2}}\left(\frac{\mu^{2}}{p^{2}}4\pi\right)^{\epsilon/2}\frac{\Gamma(3-\epsilon)\Gamma(\epsilon-1)}{\Gamma\left(\frac{3-\epsilon}{2}\right)}.
\end{equation}

The series expansion of the Gamma Euler functions in terms of $\epsilon$ is:
\begin{eqnarray}
\frac{\Gamma(3-\epsilon)\Gamma(\epsilon-1)}{\Gamma\left(\frac{3-\epsilon}{2}\right)}=-\frac{4}{\sqrt{\pi}\,\epsilon}+\frac{2(\psi(3/2)-1)}{\sqrt{\pi}}\nonumber\\
+\frac{\epsilon\,(-5\pi^2-24+12\psi(3/2)-6\psi^2(3/2))}{12\sqrt{\pi}}+\mathcal{O}(\epsilon^2),
\end{eqnarray}
where $\psi(a)$ are the Digamma Euler functions defined in Appendix.

Finally we obtain for the regularized integral:
\begin{eqnarray}\label{rrdd}
&&I(D)_r = \frac{2\gamma p}{(4\pi)^{3/2}}\bigg\{ \frac{4}{\epsilon\sqrt{\pi}}+\frac{2\psi(3/2)}{\sqrt{\pi}}-\frac{2}{\sqrt{\pi}}\left[\ln\left(4\pi\frac{\mu^{2}}{p^{2}}\right)+1\right]\nonumber\\
&&+\epsilon\,\left[\frac{-5\pi^2-24+12\psi(3/2)-6\psi^2(3/2)}{24\sqrt{\pi}}+ \frac{(\psi(3/2)-1)}{\sqrt{\pi}}\ln\left(4\pi\frac{\mu^{2}}{p^{2}}\right)\right] + \mathcal{O}(\epsilon^{2}) \bigg\}.
\nonumber\\
\end{eqnarray}

In the limit where $\epsilon=0$, the first term of the regularized integral diverges. We will replace this term by the result obtained when we apply the Pauli-Villars regularization for the $p'$ integral:
\begin{equation}
\int\frac{d^3p'}{(2\pi)^3(p'+p)^2}\rightarrow \int \frac{d^3p'}{(2\pi)^3}\left[\frac{1}{(p'+p)^2}+\frac{a_1}{(p'+\Lambda_1)^2}+\frac{a_2}{(p'+\Lambda_2)^2}\right].
\end{equation}
where $\Lambda_1,\Lambda_2$ are fermionic masses.

The integrand can be written as:
\begin{eqnarray}
\frac{(p'+\Lambda_1)^2(p'+\Lambda_2)^2+a_1(p'+p)^2(p'+\Lambda_2)^2+a_2(p'+p)^2(p'+\Lambda_1)^2}{(p'+p)^2(p'+\Lambda_1)^2(p'+\Lambda_2)^2}
\nonumber\\
\end{eqnarray}

In order to avoid the ultra-violet divergences in the integral, we impose the condition that all terms proportional with $p'^4$ and $p'^3$ should vanish. The cancellation of the coefficients of  $p'^4,\,p'^3$ leads to the following equations:
\begin{eqnarray}
&&1+a_1+a_2=0;\nonumber\\
&&\Lambda_1+\Lambda_2+a_1\Lambda_2+a_2\Lambda_1+a_1p+a_2p=0;\nonumber\\
\end{eqnarray}

The parameters $a_1,\,a_2$ are determined to be:
\begin{eqnarray}\label{a12}
&&a_1=\frac{\Lambda_2-p}{\Lambda_1-\Lambda_2},\,\,a_2=\frac{p-\Lambda_1}{\Lambda_1-\Lambda_2}.
\end{eqnarray}

If we replace $a_1,a_2$ in the numerator all terms proportional with momentum $p'^4,p'^3$ vanish and we obtain for the rest of the terms from the nominator:
\begin{eqnarray}
D&=&\Lambda_1^2\Lambda_2^2 +p^3(\Lambda_1+\Lambda_2)-p^2(\Lambda_1^2+\Lambda_2^2+\Lambda_1\Lambda_2)+3p'^2(\Lambda_1\Lambda_2+p^2-p(\Lambda_1+\Lambda_2))\nonumber\\
&&+2p'[\Lambda_1^2\Lambda_2+\Lambda_1\Lambda_2^2+p^3-p(\Lambda_1^2+\Lambda_2^2+\Lambda_1\Lambda_2)]\nonumber\\
&&=\Lambda^4+2p^3\Lambda-3p^2\Lambda^2+2p'(2\Lambda^3+p^3-3p\Lambda^2)+3p'^2(p^2+\Lambda^2-2p\Lambda),
\end{eqnarray}
where the last equality is obtained by making $\Lambda_1=\Lambda_2=\Lambda$.

The momentum integral becomes:
\begin{eqnarray}
&&\int_0^\infty \frac{dp'}{2\pi^2}\frac{p'^2[\Lambda^4+2p^3\Lambda-3p^2\Lambda^2+2p'(2\Lambda^3+p^3-3p\Lambda^2)+3p'^2(p^2+\Lambda^2-2p\Lambda)]}{(p'+p)^2(p'+\Lambda)^4}
\nonumber\\
&&= \frac{p}{\pi^2}\left[\frac{\Lambda}{p}-\ln\left(\frac{\Lambda}{p}\right)-1\right].
\end{eqnarray}

Now we can identify the term containing $1/\epsilon$ from the dimensional regularization as the logarithmic and linear term obtained by using the Pauli-Villars method:
\begin{equation}
\frac{1}{\varepsilon}=-\ln\left(\frac{\Lambda}{p}\right)+\frac{\Lambda}{p},
\end{equation}
which is exactly the first term from equation (\ref{rrdd}).

The final formula for the total transition rate can be established by performing the integral over the momenta of the $W$ boson. We also specify that we take the limit $\epsilon=0$ in the result of the regularized integral given in equation (\ref{rrdd}). The upper limit of the third momentum integral is set up to be $p\simeq M_W$ and we obtain:
\begin{eqnarray}
&&R_{vac\rightarrow W^-\nu e^+}=\int d^3p\,\frac{g^2\gamma p}{64\pi^2}\bigg\{ \frac{\Lambda}{p}-\ln\left(\frac{\Lambda}{p}\right)+2\left[\psi(3/2)-\ln\left(4\pi\frac{\mu^{2}}{p^{2}}\right)-1\right]\bigg\}.
\nonumber\\
\end{eqnarray}

The final result after performing the momentum integrals with the above cutoff is:
\begin{eqnarray}\label{radivv}
&&R_{vac\rightarrow W^-\nu e^+}=\frac{g^2\gamma M_W^4}{16\pi}\left[\frac{\psi(3/2)}{2}+\frac{1}{3}\frac{\Lambda}{M_W}-\frac{1}{4}\ln\left(\frac{\Lambda}{M_W}\right)-2-\frac{1}{2}\ln\left(\frac{\mu^2}{M_W^2}\right)\right]
\nonumber\\
\end{eqnarray}

In the present paper we have managed to obtain the form of the divergences which appear in the probabilities and rates corresponding to the spontaneous emission from vacuum of bosons and fermions on de Sitter space-time. However, for a complete picture, one needs to take into account the possible processes of emission of massive bosons by fermions, in which case we expect to encounter also divergences. Then it may be possible that the divergences obtained in our computations to be cancelled by adding counter terms in the lagrangian density.

Let us recall the results from the Minkowski electro-weak theory. In the Minkowski field theory, in order to compute the transition rate for the $W^-$ boson decay into an electron and antineutrino, one considers that the massive boson is at rest. Then energy-momentum conservation takes care of the momenta integrals in a simpler manner by setting up the momentum of the electron to be $p=\frac{M_W}{2}$. In the de Sitter case the energy is not conserved. In addition, the modes have a well defined momentum, but the energy is not specified since the energy and momentum operators do not commute. The vacuum transitions are forbidden in the Minkowski theory and we do not have at our disposal methods for obtaining finite general results from the momenta integrals encountered in our study. Moreover if one want to study the decay of a massive boson in de Sitter geometry one has to take into account that the rest frame solutions for Dirac and Proca equations \cite{rfv,rfv1}, define a new vacuum and their expressions are different from the solutions used in our computations. Thus it seems that the problem could be related to the limits of a perturbative treatment of the production process, or one can think that new regularization methods need to be used in the field theory on curved space-times. Then is seems that the finite results for total transition rates can be obtained for the moment in particular cases.

The logarithmic divergences from rate equation (\ref{radivv}) can be removed by considering a particular case. Let us recall the situation when the particles are on the same direction i.e. the third axis, $\vec p=p \vec e_3,\,\,\vec p\,'=p' \vec e_3$ and $\vec P=-P\vec e_3$. We perform the integral over $P$ :
\begin{eqnarray}
I &=& \int d^{3}p \int d^{3}P \int d^{3}p^{\prime} \frac{\delta^{3}(\vec{p}+\vec{p^{\prime}} +\vec{P})}{(2\pi)^3P(p+p^{\prime}+P)}\nonumber\\
&=&\int d^{3}p \int d^{3}p^{\prime} \frac{1}{(2\pi)^3|\vec p+\vec p\,'|(p+p^{\prime}+|\vec p+\vec p'|)}
=\int d^{3}p \int d^{3}p^{\prime} \frac{1}{2(2\pi)^3(p+p')^2}.
\nonumber\\
\end{eqnarray}
We choose the situation when the momentum modulus of the positron and neutrino are close in value, such that we can approximate the $(p+p')^2\simeq 2(p^2+p'^2)$. With this supposition the momenta integrals can be solved by using the dimensional regularization:
\begin{eqnarray}
\int d^{3}p \int d^{3}p^{\prime} \frac{1}{4(2\pi)^3(p^2+p'^2)}
\end{eqnarray}
The $p'$ integral in $D$ dimensions reads:
\begin{eqnarray}
\int d^{3}p^{\prime} \frac{1}{(2\pi)^3(p^2+p'^2)}\rightarrow \int d^{D}p^{\prime} \frac{1}{(2\pi)^D(p^2+p'^2)}=\frac{2\pi^{D/2}}{(2\pi)^D\Gamma(D/2)}\int_0^{\infty}dp' \frac{p'^{D-1}}{(p^2+ p'^2)}.
\nonumber\\
\end{eqnarray}
The substitution $p'=p\sqrt{y}$ allows one to bring the integral in the form of a Beta Euler function
\begin{eqnarray}
I(D)=\frac{2\pi^{D/2}(p^2)^{D/2-1}}{2(2\pi)^D\Gamma(D/2)}\int_0^{\infty}dy \frac{y^{D/2-1}}{1+y}=\frac{(p^2)^{D/2-1}\Gamma(1-D/2)}{(4\pi)^{D/2}}.
\end{eqnarray}
Following the steps from previous sections we write down the result for the regularized integral, by introducing the dimensionless constant $\gamma$ and taking $D=3-\epsilon$:
\begin{equation}
I(D)_r=\frac{-\gamma p}{(4\pi)^{3/2}}\left(\frac{4\pi \mu^2}{p^2}\right)^{\epsilon/2}\Gamma\left(\frac{\epsilon}{2}-\frac{1}{2}\right)
\end{equation}
Further we expand the power function and gamma Euler function:
\begin{equation}
\left(\frac{4\pi \mu^2}{p^2}\right)^{\epsilon/2}=1+\frac{\epsilon}{2}\ln\left(\frac{4\pi \mu^2}{p^2}\right)+o(\epsilon^2),
\end{equation}
\begin{equation}
\Gamma\left(\frac{\epsilon}{2}-\frac{1}{2}\right)=-2\sqrt{\pi}-\epsilon\sqrt{\pi}\psi(-1/2)+o(\epsilon^2).
\end{equation}
At the end of our computations we take the limit $\epsilon \rightarrow 0$ and obtain:
\begin{eqnarray}
\lim_{\epsilon \rightarrow 0}I(D)_r=\frac{\gamma p}{4\pi},
\end{eqnarray}
This a finite result, which proves that the $p'$ integral can be regularized in some cases, all divergences being removed by the dimensional regularization procedure. The $p$ integral can be solved by using the fact that the momentum is conserved in this process $\vec{p}+\vec{p^{\prime}} +\vec{P}=0$. Taking the boson momenta $P\simeq M_W$, we can choose for fermion  momenta $p'\simeq p\simeq M_W/2$. Our approximation for the momenta module represents the case of small momenta. In this particular case the upper limit of the $p$ integral can be taken as $M_W/2$, and the final result for the momenta integrals reads:
\begin{eqnarray}
I=\frac{\gamma M_W^4}{128}.
\end{eqnarray}
The final finite result for the transition rate is:
\begin{equation}
R_{vac\rightarrow W^-\nu e^+}=\frac{g^2\gamma M_W^4}{8\cdot128}=\frac{\gamma G_F M_W^6}{128\sqrt{2}},
\end{equation}
where the last equality is obtained when we introduce the Fermi constant $G_F$. With this approach we obtain the rate equation when the neutrino momentum $p$ is specified.

The density number of $W$ bosons must be proportional with the ratio between the rate of production from vacuum and the decay rate of the $W$ boson:
\begin{equation}
n_{W}\sim\frac{R_{vac\rightarrow W^-\nu e^+}}{R_{des}}.
\end{equation}

For a complete picture of boson densities due to perturbative production it will be necessary to analyse all first order processes which give contribution to the density number of $W$ bosons, including those processes in which massive bosons are emitted by fermions.

We must also mention results obtained by other ways \cite{37}, where the authors consider nonperturbative methods for studying the production of massive bosons in gravitational fields. In our opinion, both perturbative and nonperturbative methods should be taken into account when the problem of particle production in the early universe is approached. An analysis of the both perturbative and nonperturbative methods for particle production in gravitational fields can be found in \cite{17,18}.

Our analysis presents a microscopic process which is produced by a large scale event, namely space-time expansion, which could play a similar role to the thermal bath. The two effects must be taken into account when we analyse  the mechanisms that generate matter and anti-matter in the early universe.

\section{Concluding remarks}
The present study focuses on the study of the interaction between charged $W^{\pm}$ bosons and fermions in the de Sitter geometry by using perturbative methods. The equations of interaction between massive charged vector fields and fermions are established and their solutions are written with the help of the usual method of the Green functions. With this approach one can introduce the $IN/OUT$ fields and develop the formalism of reduction for the Proca field in the de Sitter geometry. The definition of the transition amplitudes in the first order of perturbation theory is also established. This is a powerful tool for the study of the phenomenon of particle production in the early stages of the universe.

As an application of our formalism, we have studied the problem of generating the triplet $W^{-}$ boson, positron and neutrino from the de Sitter vacuum. We established the transition amplitudes for both the transversal and longitudinal modes and defined the transition probabilities per unit volume. From our analytical and graphical results we prove that that the phenomenon of massive boson production is present during the inflation period. From our general results one recovers the Minkowski limit, where the amplitude and probability of this process vanish. We also study the interesting limiting case where $m_{e}/\omega=M_{W}/\omega=0$, which corresponds to large values of the expansion parameter $\omega$. Under these limiting conditions we obtain the total probability and rate of transition by using dimensional regularization and Pauli-Villars regularization for computing the momenta integrals. We conclude that the process of spontaneous particle emission from vacuum is possible only for a large expansion factor, which corresponds to the early universe period.

For further analysis, the study of all possible processes which generate charged $W^{\pm}$ bosons would be of interest, since all possible transitions in the first order of perturbation theory have nonvanishing amplitudes for the large expansion conditions of the early universe. Our results aim to open the way for the study of electro-weak interactions in curved backgrounds and for applying perturbative methods to the understanding of mechanisms which generate massive bosons in the early universe.

\section{Appendix}

We make use of the formula for solving the temporal integrals in our amplitudes \cite{AS,21}:
\begin{eqnarray}\label{appell}
&&\int_{0}^{\infty} x^{Q-1} \mathcal{J}_{\lambda}(ax)\mathcal{J}_{\mu}(bx)\mathcal{K}_{\nu}(cx) dx =\frac{2^{Q-2}a^{\lambda}b^{\mu}c^{-Q-\lambda-\mu}}
{\Gamma(\lambda+1)\Gamma(\mu+1)}\nonumber\\
&&\times \Gamma\left(\frac{Q+\lambda+\mu-\nu}{2}\right) \Gamma\left(\frac{Q+\lambda+\mu+\nu}{2}\right)\nonumber\\
&&\times F_{4} \left(\frac{Q+\lambda+\mu-\nu}{2},\frac{Q+\lambda+\mu+\nu}{2},\lambda+1,\mu+1,-\frac{a^{2}}{c^{2}},-\frac{b^{2}}{c^{2}}\right),
\end{eqnarray}
where $\mathcal{J}$ is the Bessel function of the first kind, $\mathcal{K}$ is the modified Bessel function and $\mathcal{F}_{4}$ is Appell$^{\prime}$s function.
The gamma Euler function can be expanded as \cite{HGF}
\begin{equation}
\Gamma(\epsilon - n) = \frac{(-1)^{n}}{n!}\left\{ \frac{1}{\epsilon} + \psi(n+1) + \frac{\epsilon}{2} \left[\frac{\pi^{2}}{3} + \psi^{2}(n+1) -\psi^{\prime}(n+1)\right]+\mathcal{O}(\epsilon^{2})\right\}
\end{equation}
where $\psi$ are the digamma Euler functions defined below \cite{AS}
\begin{eqnarray}
\psi(n) = -\gamma + \sum_{l=1}^{n-1}\frac{1}{2}; \\
\psi^{\prime}(n) = \frac{\pi^{2}}{6} - \sum_{l=1}^{n-1}\frac{1}{l^{2}},
\end{eqnarray}
while the Euler constant is
\begin{equation}
\gamma=-0,5772.
\end{equation}

\textbf{Acknowledgements}

Amalia Dariana Fodor was supported by a grant of  the Romanian Ministry of Research and Innovation and West University of Timi\c soara, CCCDI-UEFISCDI, under project "VESS, 18PCCDI/2018",  within PNCDI III.


\begin{thebibliography}{99}
\bibitem{1}
C. W. Misner, K. S. Thorne and J. A. Wheleer, {\em Gravitation}
(W. H. Freeman and Company New York, 1973).
\bibitem{PR1}
A. Proca, {\em J. Phys. Radium} \textbf{7},
347–353 (1936);A. Proca, {\em C. R. Acad. Sci. Paris}, \textbf{202}, 1366 (1936); A. Proca, {\em C. R. Acad. Sci. Paris} \textbf{202}, 1490
(1936); A. Proca, {\em  J. Phys. Radium} \textbf{9}, 61 (1938).
\bibitem{2}
Ion I. Cot\u{a}escu, {\em Gen.Rel.Grav.} \textbf{42},861-876,2010.
\bibitem{w1}
S. Weinberg, The First Three Minutes: A Modern View of the Origin of the Universe (Basic Books,New York, 1977).
\bibitem{w2}
S. Weinberg, {\em Phys. Scr.} \textbf{21}, 773 (1979).
\bibitem{3}
S. Weinberg, {\em Phys. Rev. Lett.} \textbf{19}, 1264 (1967).
\bibitem{4}
S. Weinberg, {\em Phys. Rev. Lett.} \textbf{27}, 1688 (1971).
\bibitem{5}
S. Weinberg, {\em Phys. Rev. D} \textbf{5}, 1412 (1972).
\bibitem{6}
S. Weinberg, {\em Phys. Rev. D} \textbf{7}, 1068 (1973).
\bibitem{7}
S. Weinberg, {\em Phys. Rev. D} \textbf{8}, 605 (1973).
\bibitem{8}
S. Weinberg, {\em Rev. Mod. Phys.} \textbf{46}, 255 (1974).
\bibitem{9}
S. L. Glasshow and S. Weinberg, {\em Phys. Rev. D} \textbf{15}, 1958 (1977).
\bibitem{10}
J. D. Bjorken, K. Lane and S. Weinberg, {\em Phys. Rev. D} \textbf{5}, 1474 (1977).
\bibitem{11}
B. W. Lee and S. Weinberg, {\em Phys. Rev. D} \textbf{38}, 1237 (1977).
\bibitem{cr}
C. Rubbia, {\em Rev. Mod. Phys.} \textbf{57}, 699 (1985).
\bibitem{12}
S. Weinberg, {\em The Quantum Theory of Fields}  (Cambridge University Press, Cambridge, 1995).
\bibitem{15}
J. Lankinen and I. Vilja, {\em Phys. Rev. D} \textbf{96}, 105026-1 (2017).
\bibitem{17}
N. D. Birrel and P. C. W. Davies,  {\em Quantum Fields in Curved Space} (Cambridge University Press, Cambridge 1982).
\bibitem{18}
N. D. Birrel, P. C. W. Davies and L. H. Ford, {\em J. Phys. A} \textbf{13}, 961 (1980).
\bibitem{19}
S. Drell and J. D. Bjorken, {\em Relativistic Quantum Fields} (Mc Graw-Hill Book Co., New York 1965).
\bibitem{20}
L. Landau and E. M. Lifsit, {\em Theorie Quantique Relativiste} (Mir Moscou 1972).
\bibitem{AS}
M. Abramowitz and I. A. Stegun, {\em Handbook of Mathematical Functions} (Dover, New York, 1964).
\bibitem{21}
I. S. Gradshteyn and I. M. Ryzhik {\em Table of integrals, series and products} (Academic Press, 2007).
\bibitem{PT}
T. Prokopec, N. C. Tsamis and R. P. Woodard, {\em AnnalsPhys.} \textbf{323}, 1324-1360,(2008).
\bibitem{WT}
N. C. Tsamis and R. P. Woodard, {\em J.Math.Phys.} \textbf{48}, 052306, (2007).
\bibitem{PC}
J. F. Koksma, T. Prokopec, {\em Class.Quant.Grav.} \textbf{26}, 125003, (2009).
\bibitem{CRR}
P. Candelas and D. J . Raine, {\em Phys. Rev. D} \textbf{12}, 965, (1975).
\bibitem{COT}
Ion I. Cot\u{a}escu, {\em Eur. Phys. J. C} \textbf{78:769}, (2018).
\bibitem{WF}
M.E. Fisher and K.G. Wilson, {\em Phys. Rev. Lett.} \textbf{28}, 240 (1972);
K.G. Wilson, {\em Phys. Rev. D} \textbf{7}, 2911 (1973).
\bibitem{GHV}
G.’t Hooft and M. Veltman, {\em Nucl. Phys. B} \textbf{44}, 189 (1972).
\bibitem{IT}
C.G. Bollini, J.J. Giambiagi, {\em Nuovo Cimento B} \textbf{12}, 20 (1972).
\bibitem{GT}
G.’t Hooft, {\em Nucl. Phys. B} \textbf{61}, 455 (1973).
\bibitem{PV}
W. Pauli and F. Villars, {\em Rev. Mod. Phys.} \textbf{21}, 434 (1949).
\bibitem{HGF}
H. Kleinert and V. Schulte-Frohlinde, {\em Critical properties of $\phi^4$-theories} (World Scientific 2001).
\bibitem{22}
Ion I. Cot\u{a}escu, {\em Phys. Rev. D} \textbf{65}, 084008 (2002).
\bibitem{LL}
J. Lankinen and I. Vilja, {\em Phys. Rev. D} \textbf{96}, 105026-1 (2017)
\bibitem{LL1}
J. Lankinen, J. Malmi and I. Vilja, {\em Eur. Phys. J. C} \textbf{80:502}, (2020).
\bibitem{CML}
B. Carter and R. G. McLenaghan, {\em Phys. Rev. D}, {\bf 19} (1979) 1093.
\bibitem{GH}
G. W. Gibbons and S. W. Hawking, {\em Phys. Rev. D} {\bf 15} (1977) 2738.
\bibitem{DW}
B. S. DeWitt, {\em Quantum gravity: the new synthesis} in {\em General Relativity an Einstein centenary} survey ed. S. W. Hawking, W. Israel, (Cambridge University Press, Cambridge, 1979).
\bibitem{23}
Crucean Cosmin, {\em Phys. Rev. D} \textbf{85}, 084036 (2012).
\bibitem{rc}
C. Crucean and R. Racoceanu, {\em Int. J. Mod. Phys. A} \textbf{23}, (2008).
\bibitem{24}
Ion I. Cot\u{a}escu, C. Crucean, {\em Phys. Rev. D} \textbf{87}, 044016 (2013).
\bibitem{25}
Ion I. Cot\u{a}escu, C. Crucean, {\em Progress of Theor. Phys.} \textbf{124}, 1051 (2010).
\bibitem{26}
C. Crucean and M. A. B\u{a}loi {\em Phys. Rev. D} \textbf{93}, 044070 (2016).
\bibitem{27}
C. Crucean, {\em Mod. Phys. Lett. A} \textbf{22}, 2573 (2007).
\bibitem{28}
C. Crucean and M. A. B\u aloi, {\em Int. J. Mod. Phys. A} \textbf{30}, 1550088 (2015).
\bibitem{29}
Ion I. Cotaescu, R. Racoceanu, Radu and C. Crucean, {\em Mod. Phys. Lett. A} \textbf{21}, 1313 (2006).
\bibitem{30}
Ion I. Cot\u{a}escu, C. Crucean, {\em Int. J. Mod. Phys. A} \textbf{23}, 3707 (2008).
\bibitem{31}
M. A. B\u{a}loi, {\em Mod. Phys. Lett. A} \textbf{29}, 1450138 (2014).
\bibitem{b1}
M. A. B\u{a}loi,{\em Int. J. Mod. Phys. A} \textbf{31}, 1650081 (2016).
\bibitem{b2}
M. A. B\u{a}loi, C. Crucean and D. Popescu, {\em Eur. Phys. J. C} \textbf{78:398}, (2018).
\bibitem{cc}
C. Crucean, {\em Eur. Phys. J. C} \textbf{79:483}, (2019).
\bibitem{32}
E. Schr\" odinger, {\em Physica} \textbf{6}, 899 (1939).
\bibitem{33}
L. Parker, {\em Phys. Rev. Lett.} \textbf{21}, 562 (1968).
\bibitem{34}
L. Parker, {\em Phys. Rev.} \textbf{183}, 1057 (1969).
\bibitem{35}
L. Parker, {\em Phys. Rev. D} \textbf{3}, 346 (1971).
\bibitem{cpc}
Ion I. Cot\u{a}escu, D. Popescu, {\em Chinese Phys. C} \textbf{44}, (2020).
\bibitem{37}
Y. Ema, K. Nakayama and Y. Tang, {\em JHEP} \textbf{60} (2019).
\bibitem{rat}
W. N. Cottingham, D. A. Greenwood, {\em An introduction to standard model of particle physics}, (Cambridge University Press, Cambridge 2007).
\bibitem{rfv}
Ion I. Cot\u{a}escu, {\em Eur. Phys. J. C} \textbf{79:696}, (2019).
\bibitem{rfv1}
Ion I. Cot\u{a}escu, {\em Eur. Phys. J. C} \textbf{80:535}, (2020).

\end{thebibliography}
\end{document}